\pdfoutput=1

\documentclass[10pt]{amsart} 

\usepackage{
  amssymb, 
  latexsym, 
  graphics, 
  verbatim, 
  harvard, 
  layout, 
  amsbsy, 
  pstricks,
  pst-plot, 
  pst-tree,
  pstricks-add, 
  pst-pdf,
  accents 
  }
\usepackage[bottom]{footmisc} 
\usepackage[inline]{enumitem}

\newtheorem{lemma}{\indent Lemma}
\newtheorem{nthm}[lemma]{\indent Theorem}

\newtheorem{ncrly}[lemma]{\indent Corollary}
\newtheorem{prop}[lemma]{\indent Proposition}

\theoremstyle{definition} 
\newtheorem{pfa}[lemma]{\indent Proof}
\newenvironment{npf}  
  { \begin{pfa} }
  { \hfill $\Box$ \end{pfa} }

\theoremstyle{remark} 
\newtheorem*{pfb}{\indent Proof}
\newenvironment{pf} 
  { \begin{pfb} }
  { \hfill $\Box$ \end{pfb}  }

\newcounter{tlistno} 
\newenvironment{tlist} 
  { \begin{list} 
      {\indent(\alph{tlistno}) } 
      { \setlength{\leftmargin}{0mm} 
        \setlength{\rightmargin}{0mm} 
        \setlength{\itemindent}{-4.3mm}  
        \setlength{\labelwidth}{0mm} 
        \setlength{\labelsep}{0mm} 
        \setlength{\itemsep}{0mm} 
        \setlength{\topsep}{0mm} 
        \usecounter{tlistno} } } 
  { \end{list} }

\newcommand{\itemi}{ \begin{itemize} }
\newcommand{\itemo}{ \end{itemize} }
\newcommand{\y}{ \item } 
\newcommand{\yl}[1]{\item\label{#1}}

\newcommand{\zz}{}
\newcommand{\subi}{\begin{subequations}}
\newcommand{\subo}{\end{subequations}}
\newcommand{\nt}{\notag \\ }

\newcommand{\mi}{ \left[ \begin{matrix} }
\newcommand{\mo}{ \end{matrix} \right] }
\newcommand{\miz}{ \left. \begin{matrix} }
\newcommand{\moz}{ \end{matrix} \right. }
\newcommand{\mip}{ \left( { \ } \begin{matrix} }
\newcommand{\mop}{ \end{matrix} { \ } \right) }
\newcommand{\mib}{ \left[ { \ } \begin{matrix} }
\newcommand{\mob}{ \end{matrix} { \ } \right] }
\newcommand{\si}{ \left[ \begin{smallmatrix} }
\newcommand{\so}{ \end{smallmatrix} \right] }
\newcommand{\casei}{ \left\{ \begin{matrix} } 
\newcommand{\caseo}{ \end{matrix} \right. }
\newcommand{\mpi}{ \begin{minipage} }
\newcommand{\mpo}{ \end{minipage} }

\newcommand{\nichts}[1]{} 
\newcommand{\showbib}{ \bibliographystyle{econometrica}\bibliography{mrmagoo}}  
\newcommand{\markb}[1]{\markboth{#1}{#1}}
\newcommand{\ssec}[1]{\vspace{0.9mm}\subsection{}{\sc #1}\vspace{1.6mm}}

\newcommand{\myskip}{\vspace{3mm}}

\newcommand{\lstep}[1]{\vspace{1mm} {\em #1}}

\allowdisplaybreaks

\newcommand{\nnexists}{\raisebox{0.15ex}{/}\hspace{-1.15ex}\exists}

\usepackage[utf8]{inputenc}
\usepackage{newunicodechar}
\newcommand{\nuc}[2]{\newunicodechar{#1}{#2}} 

\nuc{α}{{ \alpha}} 
\nuc{β}{{ \beta}}
\nuc{γ}{{ \gamma}} \nuc{Γ}{{ \varGamma}}
\nuc{δ}{{ \delta}} \nuc{Δ}{{ \varDelta}} 
\nuc{ε}{{ \varepsilon}}
\nuc{ζ}{{ \zeta}}
\nuc{η}{{ \eta}}
\nuc{θ}{{ \theta}} \nuc{Θ}{{ \varTheta}}
\nuc{ι}{{ \iota}}
\nuc{κ}{{ \kappa}}
\nuc{λ}{{ \lambda}} \nuc{Λ}{{ \varLambda}}
\nuc{μ}{{ \mu}}
\nuc{ν}{{ \nu}}
\nuc{ξ}{{ \xi}} \nuc{Ξ}{{ \varXi}}
\nuc{π}{{ \pi}} \nuc{Π}{{ \varPi}}
\nuc{ρ}{{ \rho}} 
\nuc{σ}{{ \sigma}} \nuc{Σ}{{ \varSigma}}
\nuc{τ}{{ \tau}}
\nuc{υ}{{ \upsilon}} \nuc{Υ}{{ \varUpsilon}}
\nuc{φ}{{ \phi}} \nuc{Φ}{{ \varPhi}}
\nuc{χ}{{ \chi}} 
\nuc{ψ}{{ \psi}} \nuc{Ψ}{{ \varPsi}}
\nuc{ω}{{ \omega}} \nuc{Ω}{{ \varOmega}}

\nuc{≠}{{ \ne}}
\nuc{∾}{{ \sim}} 
\nuc{≈}{{ \approx}}
\nuc{≡}{{ \equiv}} 
\nuc{∈}{{ \in}} \nuc{∉}{{ \notin}}
\nuc{∋}{{ \ni}} \nuc{∌}{{ \not\ni}}   
\nuc{⊊}{{ \subset}} \nuc{⊆}{{ \subseteq}}
\nuc{⊋}{{ \supset}} \nuc{⊇}{{ \supseteq}} 
\nuc{≤}{{ \leq}}
\nuc{≥}{{ \geq}}
\nuc{≪}{{ \ll}}
\nuc{≫}{{ \gg}}
\nuc{≺}{{ \prec}} \nuc{≼}{{ \preccurlyeq}}
\nuc{≻}{{ \succ}} \nuc{≽}{{ \succcurlyeq}}
\nuc{‖}{{ \parallel}}

\nuc{∑}{{ \Sigma}} 
\nuc{∫}{{ \textstyle \int\nolimits}}
\nuc{±}{{ \pm}}
\nuc{×}{{ \times}} \nuc{∏}{{ \Pi}} 
\nuc{÷}{{ \div}} 
\nuc{○}{{ \circ}}
\nuc{∩}{{ \cap}} \nuc{⋂}{{ \textstyle \bigcap\nolimits}}
\nuc{∪}{{ \cup}} \nuc{⋃}{{ \textstyle \bigcup\nolimits}}
\nuc{⧷}{{ \smallsetminus}} 
\nuc{∧}{{ \wedge}} \nuc{⋀}{{ \textstyle \bigwedge\nolimits}}
\nuc{∨}{{ \vee}} \nuc{⋁}{{ \textstyle \bigvee\nolimits}}
\nuc{±}{{ \oplus}} 
\nuc{⊗}{{ \otimes}}
\nuc{⊙}{{ \odot}} \nuc{⊛}{{ \textstyle \bigodot\nolimits}}
\nuc{•}{{ \bullet}} 
\nuc{∗}{{ \star}} 
\nuc{·}{{ \cdot}}

\nuc{→}{{ \rightarrow}}  \nuc{⇉}{{ \twoheadrightarrow}} 
\nuc{←}{{ \leftarrow}}
\nuc{⟷}{{ \leftrightarrow}}
\nuc{⇒}{{ \Rightarrow}}  
\nuc{⇐}{{ \Leftarrow}} 
\nuc{⟺}{{ \Leftrightarrow}}

\nuc{ℓ}{{ \ell}}
\nuc{∂}{{ \partial}}
\nuc{∞}{{ \infty}}
\nuc{‴}{^{ \prime\prime\prime}} 
\nuc{″}{^{ \prime\prime}}
\nuc{′}{^{ \prime}}
\nuc{¬}{{ \neg}}
\nuc{∀}{{ \forall}}
\nuc{∃}{{ \exists}}
\nuc{∆}{{ \triangle}}
\nuc{∄}{{ \nnexists}}
\nuc{∅}{{ \varnothing}}
\nuc{□}{{ \square}}

\nuc{⎨}{{ \lbrace}}
\nuc{⎬}{{ \rbrace}}
\nuc{⁅}{{ \langle}}
\nuc{⁆}{{ \rangle}}

\nuc{Ṛ}{{ \mathbb R}}
\nuc{Ẓ}{{ \mathbb Z}}
\nuc{⋅}{{ \ }}

\newcounter{cllistno} 
\newenvironment{cllist} 
  { \begin{list} 
      {\emph{Claim \arabic{cllistno}:} } 
      { \setlength{\leftmargin}{0mm} 
        \setlength{\rightmargin}{0mm} 
        \setlength{\itemindent}{\parindent} 
        \setlength{\labelwidth}{0mm} 
        \setlength{\labelsep}{0mm} 
        \setlength{\itemsep}{1.2mm}  
        \setlength{\topsep}{1.2mm} 
        \usecounter{cllistno} } } 
  { \end{list} }

\newcommand{\qedup}{\vspace{-5.2mm}} 

\newcommand{\ili}{\begin{enumerate*}[label=$\lbrack$\alph*$\rbrack$]}
\newcommand{\ilr}{\begin{enumerate*}[resume*]}
\newcommand{\ilo}{\end{enumerate*}}
\newcommand{\ilitem}[1]{\item\label{#1}\ilo\hspace{-1mm}}
\newcommand{\ilc}[1]{\ili\ilitem{#1}}
\newcommand{\il}[1]{\ilr\ilitem{#1}}

\newcommand{\lii}{\begin{enumerate*}[label=$\lbrack$\arabic*$\rbrack$]}
\newcommand{\lir}{\begin{enumerate*}[resume*]}
\newcommand{\lio}{\end{enumerate*}}
\newcommand{\liitem}[1]{\item\label{#1}\lio\hspace{-1mm}}
\newcommand{\lic}[1]{\lii\liitem{#1}}
\newcommand{\li}[1]{\lir\liitem{#1}}

\newcommand{\ci}{\par\begin{center}}
\newcommand{\co}{\end{center}\par\noindent}

\newcommand{\mpic}{\begin{minipage}{4cm}\begin{center}}
\newcommand{\mpoc}{\end{center}\end{minipage}}

\newenvironment{maneq}{%
  \setlength{\arraycolsep}{.1ex}
  
  \begin{array}{lc} \rule{5ex}{0ex} & \rule{79ex}{0ex} \\[-3ex] }{%
  \end{array}}

  \newcommand{\nl}{\newline}
  \newcommand{\ct}[1]{$\mathbf{#1}$}
  \newcommand{\cy}[1]{\text{\ct{#1}}}
  \newcommand{\ex}[1]{\mathsf{#1}}
  \newcommand{\ud}[1]{{\text{\d{$#1$}}}}

  \newcommand{\mapsfrom}{\text{\reflectbox{$\mapsto$}}}

  \newcommand{\notenclosedby}{{\not\mspace{-5mu}\ud{→}}}
  \newcommand{\sencloses}{\ud{←}\mspace{0mu}{\notenclosedby}}

  \newcommand{\sh}{^{\sharp}}

  \newcommand{\ps}{^{\prime\sharp}}
  \newcommand{\po}{^{\prime o}}
  
  \newcommand{\st}{^{\star}}

  \newcommand{\dC}{\bar{C}}
  \newcommand{\dcc}{\bar{c}}

  \newcommand{\dI}{{\bar{I}}}
  \newcommand{\di}{{\bar{i}}}

  \newcommand{\dT}{\bar{T}} 
  \newcommand{\dt}{\bar{t}}

  \newcommand{\ga}{{\tilde{a}}}

  \newcommand{\HH}{{ \mathcal H}}
  \newcommand{\PP}{{\mathcal P}}

  \newcommand{\ZZ}{{ \mathcal Z}}
  \newcommand{\ZZf}{\mathcal Z_{\mathsf{ft}}}
  \newcommand{\ZZi}{\mathcal Z_{\mathsf{inft}}}

  \newcommand{\tNa}{{{ \mathbb N}_1}}
  \newcommand{\tNz}{{{ \mathbb N}_0}}

  \newcommand{\set}{\ct{Set}}   
  \newcommand{\Set}{\set}

  \newcommand{\src}{\mathsf{src}}
  \newcommand{\trg}{\mathsf{trg}}
  \newcommand{\id}{\mathsf{id}}
  \newcommand{\gr}{\mspace{1mu}^{\mathsf{gr}}}

  \newcommand{\fa}{{\mathsf{a}}}
  \newcommand{\fb}{{\mathsf{b}}}
  \newcommand{\fc}{{\mathsf{c}}}
  \newcommand{\fd}{{\mathsf{d}}}
  
  \newcommand{\ff}{{\mathsf{f}}}
  \newcommand{\fg}{{\mathsf{g}}}

  \newcommand{\f}[1]{\mathsf{#1}}

  \newcommand{\FB}{\mathsf{F}}
  \newcommand{\FO}{\mathsf{F_0}}
  \newcommand{\FA}{\mathsf{F_1}}

  \newcommand{\PB}{\mathsf{P}}
  \newcommand{\PO}{\mathsf{P_0}}
  \newcommand{\PA}{\mathsf{P_1}}

\begin{document}

\numberwithin{lemma}{section}
\numberwithin{figure}{section}
\newcommand{\pt}[2]{{\em (\ref{#1}) holds.}}
\newcommand{\ptrf}[1]{Claim \ref{#1p} (\ref{#1})}
\newcommand{\xrf}[1]{$^{\text{\normalfont\ref{#1}}}$} 
\newcommand{\xin}[1]{#1} 
\newcommand{\captionall}{}

\title{The Category of Node-and-Choice Forms, \\
with Subcategories for \\
Choice-Sequence Forms and Choice-Set Forms \\
}

\date{\today.  This paper is very similar to Western University Department of Economics Research Report 2018-6.  Only the introduction and some expository remarks have changed.  {\em Keywords:} extensive form, game form, isomorphic enclosure. {\em JEL Classification:} C73.  {\em AMS Classification:} 91A70. {\em Contact information:} pstreuf@uwo.ca, 519-661-2111x85384, Economics Department, University of Western Ontario, London, Ontario, N6A 5C2, Canada.}  

\thanks{I thank Deanna Walker for many valuable suggestions.} 

\maketitle 

\vspace{-.2cm}
\begin{centering}
Peter A. Streufert \\[-.5mm]
Economics Department \\[-.5mm]
Western University \\
\end{centering} 
\vspace{0.35cm}

\begin{abstract} The literature specifies extensive-form games in many styles, and eventually I hope to formally translate games across those styles.  Toward that end, this paper defines \ct{NCF}, the category of node-and-choice forms.  The category's objects are extensive forms in essentially any style, and the category's isomorphisms are made to accord with the literature's small handful of ad hoc style equivalences.

Further, this paper develops two full subcategories: \ct{CsqF} for forms whose nodes are choice-sequences, and \ct{CsetF} for forms whose nodes are choice-sets.  I show that \ct{NCF} is ``isomorphically enclosed'' in \ct{CsqF} in the sense that each \ct{NCF} form is isomorphic to a \ct{CsqF} form.  Similarly, I show that \ct{CsqF_\ga} is isomorphically enclosed in \ct{CsetF} in the sense that each \ct{CsqF} form with no-absentmindedness is isomorphic to a \ct{CsetF} form.  The converses are found to be almost immediate, and the resulting equivalences unify and simplify two ad hoc style equivalences in Kline and Luckraz 2016 and Streufert 2019.

Aside from the larger agenda, this paper already makes three practical contributions.  Style equivalences are made easier to derive by [1] a natural concept of isomorphic invariance and [2] the composability of isomorphic enclosures.  In addition, [3] some new consequences of equivalence are systematically deduced.  \end{abstract}

\vspace{5mm} %%%%<<<<>>>>
\section{Introduction} \markb{\sc 1. Introduction}

\ssec{Specification styles}

To set the stage, this subsection recalls that there are many styles in which to specify an extensive-form game.  All styles must specify [a] nodes, which are variously called ``histories'', ``vertices'', or ``states''; and [b] choices, which are variously called ``actions'', ``alternatives'', ``labels'', or ``programs''.  The following paragraphs arrange the styles into five broad groups according to how the styles specify nodes and choices.

[1] Some styles specify nodes and choices abstractly without restriction.  Classic examples from economics include the style of Kuhn 1953\nocite{Kuhn5397} and the style of Selten 1975\nocite{Selt75}.  Examples from computer science and/or logic include the style of Shoham and Leyton-Brown 2009, page 125;\nocite{ShohLeyt09} the ``labeled transition system'' style\footnote{Note~\ref{8275} more precisely links ``labeled transition systems'' with ``node-and-choice forms''.} in Blackburn, de Rijke, and Venema 2002, page 3;\nocite{BlacDeriV01} and the ``epistemic process graph'' style of van Benthem 2014, page 70.  A final example is the ``node-and-choice'' style of this paper (see Figure~\ref{8219}).  Because each of these styles specifies nodes and choices abstractly without restriction, each can be roughly understood to encompass all other styles as special cases.\footnote{\label{8282}Accordingly, this paper's ``node-and-choice'' style essentially encompasses all other extensive-form styles.  Several aspects of this claim should be clarified.  [1] An extensive-form game specifies a tree.  This feature excludes recursively specified stochastic games such as those of Mertens 2002\nocite{Merte02}.  [2] A node-and-choice form is assumed to be discrete in the sense that every node has a finite number of predecessors.  This assumption excludes non-discrete extensive-form games such as those of Dockner, J{\o}rgensen, Long, and Sorger 2000, \nocite{DocknJLS00} and Al\'os-Ferrer and Ritzberger 2016.  [3] A node-and-choice form assumes that information sets do not share alternatives.  This assumption is insubstantial in the sense of note~\ref{8155} below.  [4] A node-and-choice form assumes that exactly one player moves at each information set.  Accordingly, simultaneous moves by several players are specified by several information sets, as in Osborne and Rubinstein, 1994, page 202.}

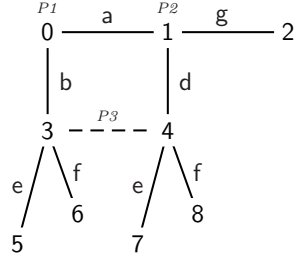
\renewcommand{\captionall}{A node-and-choice form (later called an ``\ct{NCF} form'').  Player {\Small ${\mathit P3}$} selects choice $\ex{e}$ or choice $\ex{f}$ without knowing whether she is at node $\ex{3}$ or node~$\ex{4}$.}
\begin{figure}[h]
  \newcommand{\hgth}{95}  
  \begin{picture}(0,\hgth) 
  \put(-64,-12){\scalebox{.96}{ 
    \begin{pspicture}(-1,1)(5,-5) 
      \end{pspicture}
    }} \end{picture}
  \caption{\small \captionall} \label{8219} 
  \end{figure} 

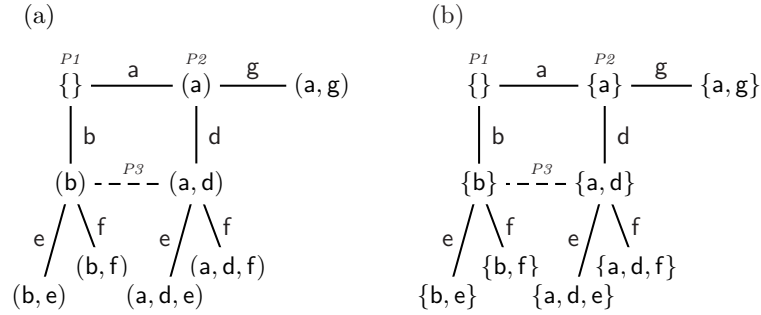
\renewcommand{\captionall}{(a) A choice-sequence form (later called a ``\ct{CsqF} form'').  (b) A choice-set form (later called a ``\ct{CsetF} form'').  These  special kinds of node-and-choice forms are developed further in this paper.}
\begin{figure}[b!]
  \newcommand{\hgth}{112}  
  \begin{picture}(0,\hgth) 
  \put(-172,-11){\scalebox{.96}{ 
    \begin{pspicture}(4,2)(17,-5) 
      \end{pspicture}
    }} \end{picture}
  \caption{\small \captionall} \label{8205} 
  \end{figure}  

[2] Other styles specify nodes as sequences of choices.  A popular example in economics is the style of Osborne and Rubinstein 1994, page 200.\nocite{OsRu94}  Examples from logic include the ``logical game'' style of Hodges 2013, Section 2,\nocite{Hodge13} and the ``epistemic forest model'' style of van Benthem 2014, page 130.\nocite{Vben14}  Examples from computer science include the ``protocol'' style of Parikh and Ramanujam 1985,\nocite{PariRama85} the ``history-based multi-agent structure'' style of Pacuit 2007,\nocite{Pacu07} and the ``sequence-form representation'' style of Shoham and Leyton-Brown 2009, page 129.  A final example is the ``choice-sequence'' style of this paper (see Figure~\ref{8205}(a)).  [3] Other styles specify nodes as sets of choices.  Examples include the ``choice-set'' style of Streufert 2019 (henceforth ``SE''), and also the ``choice-set'' style of this paper (see Figure~\ref{8205}(b)).\nocite{five-1811-as-SE}

There are still other possibilities.  [4] Some styles specify choices as sets of nodes, as in the ``simple'' style of Al\'{o}s-Ferrer and Ritzberger 2016, Section~6.3 (see Figure~\ref{8206}(a)).\nocite{AlRi16}  [5] Other styles express both nodes and choices as sets of outcomes, as in the style of von Neumann and Morgenstern 1944, Section~10.\nocite{vNMo53} and the style of Al\'{o}s-Ferrer and Ritzberger 2016, Section~6.2 (see Figure~\ref{8206}(b)).  Possibilities [1]--[5] are arranged in a spectrum by SE (Streufert 2019), Figure~2.  Further, SE Section 7 explains how each possibility has its own advantages and disadvantages.

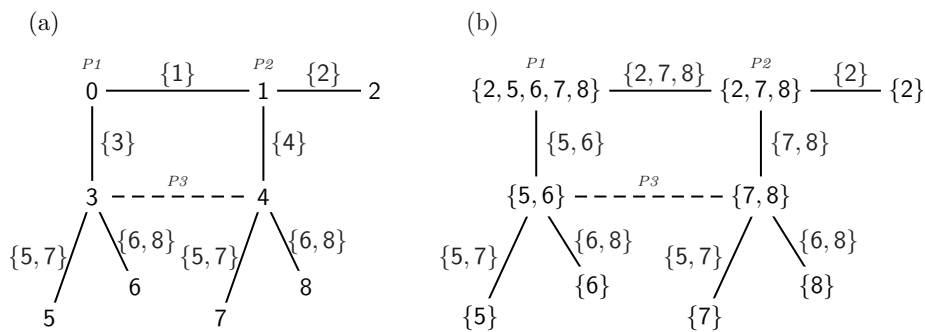
\renewcommand{\captionall}{In (a), choices are node sets.  In (b), both nodes and choices are outcome sets.  These special kinds of node-and-choice forms are not developed further in this paper.}
\begin{figure}[h]
  \newcommand{\hgth}{122}  
  \begin{picture}(0,\hgth) 
  \put(-178,-4){\scalebox{.94}{ 
    \begin{pspicture}(0,2)(17,-5) 
      \end{pspicture}
    }} \end{picture}
  \caption{\small \captionall} \label{8206} 
\end{figure} 

\ssec{General motivation}

It is difficult to formally compare the different styles.  Indeed, the first such results have only recently appeared in Al\'{o}s-Ferrer and Ritzberger 2016 Section~6.3, in Kline and Luckraz 2016, \nocite{KlinLuck16-as-KL16} and in SE (whose Figure~2 provides an overview of all these results).  These contributions show, by ad hoc constructions, that the five styles in the above figures are of roughly equal generality.  To be somewhat more precise, these papers argue that one style is at least as general as another style, by showing that each game\footnote{To be meticulous, these papers concern forms rather than games.  In other words, they stop before specifying player preferences.} in the first style can be reasonably mapped to a game in the second style.  Then two styles are regarded as equivalent if such an argument can be made in both directions.  Notice that each such argument hinges upon an ad hoc mapping linking games in one style to games in another style.  Lacking is a way to compare styles that is based on a systematic way of comparing games.  I hope to provide that systematization in a fashion that is compatible with the prior style equivalences.

Further, I have a larger agenda in mind.  Suppose that two styles have been compared and found to be equivalent.  Then I hope to do more than merely translate each game in one style to an equivalent game in the other style.  I hope to translate properties, defined for games, from one style to the other.  I hope to translate equilibrium concepts from one style to the other.  And ultimately, I hope to translate theorems from one style to the other.  In other words, I hope to formally translate game theory from one style to another.

Such an overarching theory promises to deliver large conceptual benefits.  Foremost among the benefits is the synthesis of results and questions from the many disciplines and subdisciplines which are currently studying some version of game theory.  There is much to be gained because there is so much diversity.  Further, I believe that, fundamentally, we should make the focus of our thinking an equivalence class of games, rather than an individual game.  Such an equivalence class will typically contain games in many styles.  If we can easily translate across those styles, the essence of the equivalence class can emerge.\footnote{Although it lies outside my current expertise, there appears to be a further conceptual benefit, namely, that categorical translations between games may allow for syntactic translations between the logical languages that are interpreted in those games.  This would accord with the correspondence theory of van Benthem 2001, and Conradie, Ghilardi, and Palmigiano 2014.\nocite{Benth01} \nocite{ConraGP14}}

Formal translation is a daunting task.  Fortunately, category theory promises to be a powerful and natural tool.  In order to gain access to this tool, my intermediate-range objective has been to construct a category [a] whose objects are extensive-form games in any style, and [b] whose isomorphisms accord with the handful of style equivalences already in the literature.  My first step was Streufert 2018 (henceforth ``SP'').\nocite{ncp-1809-as-SP} That paper defined \ct{NCP}, which is the category of node-and-choice ``preforms'', where a preform is a rooted tree with choices and information sets.  My second step is the present paper.  Here I define \ct{NCF}, which is the category of node-and-choice ``forms'', where a form augments a preform with players.  Later, a third paper will augment \ct{NCF} forms with preferences in order to define extensive-form games.  

Elsewhere there is little categorical work on game theory.  Lapitsky 1999\nocite{Lapi99} and Jim\'{e}nez 2014\nocite{Jime14} define categories of simultaneous-move games.  Machover and Terrington 2014\nocite{MachTerr14} defines a category of simple voting games.  Vannucci 2007\nocite{Vann07} defines categories of various games, but in its category of extensive-form games, every morphism merely maps a game to itself.  Finally, Hedges 2017 develops morphisms for open games, which resemble extensive-form games but which do not appear to accommodate players with different information.\nocite{Hedg17}%
\footnote{In addition, there have been categories developed for some relatively specialized games within the theoretical computer-science literature.  Examples include, for example, Abramsky, Jagadeesan, and Malacaria 2000,\nocite{AbraJagaM00} Hyland and Ong 2000,\nocite{HylaOngL00} and McCusker 2000.\nocite{McCu00}}

\ssec{Categorical Investments}

As explained two paragraphs ago, this paper constructs a category of forms [a] whose objects are forms in any style, and [b] whose isomorphisms accord with the style equivalences already in the literature.  Goals [a] and [b] are discussed in the next two paragraphs.

Section 2 introduces \ct{NCF}, which is the category of node-and-choice forms, in which both nodes and choices are specified abstractly without restriction.  Thereby goal [a] is achieved.  Further, one special kind of node-and-choice form is a choice-sequence form, in which nodes are choice-sequences.  Correspondingly, Section 3 introduces \ct{CsqF}, which is the full \ct{NCF} subcategory for choice-sequence forms.  Similarly, another special kind of node-and-choice form is a choice-set form, in which nodes are choice-sets.  Correspondingly, Section 4 introduces \ct{CsetF}, which is the full \ct{NCF} subcategory for choice-set forms.  Finally, consider again the five styles in Section~1.1.  \ct{NCF} itself corresponds to style [1], \ct{CsqF} corresponds to style [2], and \ct{CsetF} corresponds to style [3].  Left for future research are style [4] with its node-set choices, and style [5] with its outcome-set nodes and outcome-set choices.  These two additional styles will correspond to two additional subcategories of \ct{NCF}, as suggested in Section~5.2's discussion of future research.

To achieve goal [b], Section 2 defines \ct{NCF}'s morphisms in such a way that the category's isomorphisms accord with the style equivalences in the literature.  Since this paper does not build subcategories for the node-set and outcome-set styles, only two of the literature's style equivalences remain: [i] Kline and Luckraz 2016 Theorems 1 and 2, which are essentially an equivalence between node-and-choice forms and choice-sequence forms, and [ii] SE Theorems 3.1 and 3.2, which are essentially an equivalence between (no-absentminded) choice-sequence forms and choice-set forms.  As discussed earlier, each of these two equivalences is a matching pair of results, in which each result states that each form in one style can be reasonably mapped to a form in the other style.  Section 3.2 proposes to strengthen each such result by requiring that each form in one style is \ct{NCF} isomorphic to a form in the other style.  This new kind of result is called an ``isomorphic enclosure'', and a matching pair of isomorphic enclosures is called an ``isomorphic equivalence''.  Equivalence [i] accords with Corollary~\ref{8052}(b), which states that \ct{NCF} and \ct{CsqF} are isomorphically equivalent.  Similarly, equivalence [ii] accords with Corollary~\ref{8082}(b), which states that \ct{CsqF_\ga} and \ct{CsetF} are isomorphically equivalent.  The paragraphs after these two corollaries provide historical context, more details, and more senses in which the two corollaries accord with literature's equivalences [i] and [ii].

Other results show that \ct{NCF} is pleasant in other ways.  Theorem~\ref{6316} shows that \ct{NCF} is a well-defined category.  Theorem~\ref{6315} shows that an \ct{NCF} isomorphism can be characterized by bijections for nodes, choices, and players.  Theorem~\ref{6318} shows that there is a forgetful functor from \ct{NCF} to \ct{NCP}, which is SP's category of node-and-choice preforms.  In addition, various results in Sections~2.1--2.3 show that the category interacts naturally with game-theoretic concepts like the assignment of information sets to players.  Also, Section 2.4 shows that the properties of no-absentmindedness and perfect-information are invariant to \ct{NCF} isomorphisms.  Finally, the paragraph after Corollary~\ref{8089} shows how the negation of isomorphic enclosure formalizes the notion that a property is truly ``restrictive'' and ``substantial'' as opposed to merely ``notational''.

\ssec{Categorical Dividends}

Section 1.3 above argues that \ct{NCF} systematizes prior style equivalences and that it is a pleasant category in a variety of other ways.  Also, Sections 1.2 and 5.2 argue that \ct{NCF} promises to be of practical importance in the larger agenda of translating game theory across styles.  Further, the following three paragraphs identify three practical ways that \ct{NCF} directly contributes to game theory.

First, isomorphic invariance is a natural and powerful concept.  For example, two elementary propositions in Section 3.3 use isomorphic invariance to find [1] general circumstances in which one subcategory is strictly isomorphically enclosed by another and [2] general circumstances in which an isomorphic enclosure can be restricted to smaller subcategories.  The latter proposition is used by Corollary~\ref{8083}(b) to easily construct an isomorphic enclosure for the proof highlighted in the next paragraph.  Further, both propositions are used by Section 4.3 to easily derive new results about perfect-information.

Second, isomorphic enclosures can be composed (note~\ref{8154}).  Such compositions can make it much easier to derive other isomorphic enclosures.  For example, the proof of Corollary~\ref{8082}(b)'s reverse direction is just six lines long, and the third paragraph following the corollary's proof explains how this simple argument replaces six difficult pages in SE's proof of its Theorem 3.2.  Thus the isomorphic equivalence of Corollary~\ref{8082}(b) is much easier to prove than the corresponding ad hoc equivalence of SE Theorems 3.1 and 3.2 (this was called equivalence [ii] in Section~1.3).

Third, isomorphic enclosures have consequences for form derivatives, and Section 5.1 deduces them simultaneously for all isomorphic enclosures.  More specifically, each isomorphic enclosure is defined via  isomorphisms, and Proposition~\ref{6346} implies that each such isomorphism has consequences not only for form components (such as nodes, choices, and players) but also for form derivatives (such as the precedence relation among nodes, and each player's collection of information sets).  In contrast, the literature's ad hoc style equivalences concern only form components.

\ssec{Organization}

Section~2 develops \ct{NCF}, the category of node-and-choice forms.  Less generally, Section 3 develops the subcategory \ct{CsqF} for choice-sequence forms, and Section 4 develops the subcategory \ct{CsetF} for choice-set forms.  Sections 3.2 and 3.3 use the context of \ct{CsqF} to introduce the general concept of isomorphic enclosure, and to introduce general propositions about isomorphic invariance.  Further, Section 5.1 uses parts of Sections 3 and 4 to illustrate some general consequences of isomorphic enclosure.  Finally, Section 5.2 discusses future research.

Although many proofs appear within the text, twelve lengthy proofs and their associated lemmas are relegated to the appendices.  Appendix~A concerns \ct{NCF}, Appendix~B concerns \ct{CsqF}, and Appendix~C concerns \ct{CsetF}.

\vspace{3mm} %<<<<<<
\section{The Category of Node-and-Choice Forms} 
\markb{\sc 2. The Category of Node-and-Choice Forms}

\ssec{Objects}

Let $T$ be a set of elements $t$ called {\em nodes}.  As in SP Section 2.1 (where ``SP'' abbreviates Streufert 2018), a pair $(T,p)$ is a {\em functioned tree} iff there are $t^o⋅∈⋅T$ and $X⋅⊆⋅T$ such that [T1] $p$ is a nonempty function from $T⧷⎨t^o⎬$ onto $X$ and [T2] $(∀t∈T⧷⎨t^o⎬)(∃m∈\tNa)$ $p^m(t) = t^o$.\footnote{\label{8097}I adopt the conventions that $\tNz$ is $⎨0,1,2,...⎬$, that $\tNa$ is $⎨1,2,...⎬$, and that, for any function $f$, $f^0$ is the identity function.}  Call $p$ the {\em (immediate) predecessor} function.

A functioned tree (uniquely) determines many entities beyond $T$ and $p$.  First, it determines its {\em root} node $t^o$ and its set $X$ of {\em decision} nodes.  Second, it determines its {\em stage} function $k{:}T→\tNz$ by [a] $k(t^o) = 0$ and [b] $(∀t∈T⧷⎨t^o⎬)$ $p^{k(t)}(t) = t^o$.  Further, it determines its {\em (strict) precedence relation} $≺$ on $T$ by $(∀t^1∈T,t^2∈T)$ $t^1⋅≺⋅t^2$ iff $(∃m∈\tNa)$ $t^1 = p^m(t^2)$.  Relatedly, it determines its {\em weak precedence relation} $≼$ on $T$ by $(∀t^1∈T,t^2∈T)$ $t^1⋅≼⋅t^2$ iff $(∃m∈\tNz)$ $t^1 = p^m(t^2)$.  Finally, it determines the set $\ZZ$ of maximal chains in $(T,≼)$.  This can be split into the set $\ZZf$ of finite maximal chains and the (possibly empty) set $\ZZi$ of infinite maximal chains.  These derived entities and their basic properties are developed in SP Sections 2.1 and 2.2.

\newcommand{\noteFinv}{\footnotetext{\label{8128}To be clear, let $F{:}T⇉C$ mean that $F$ is a correspondence from $T$ to $C$, which means that $(∀t∈T)$ $F(t)⋅⊆⋅C$.  Also, for $c⋅∈⋅C$, let $F^{-1}(c) = ⎨t∈T|c∈F(t)⎬$.  Also, let $F^{-1}(C) = ∪_{c∈C}F^{-1}(c)$.}}

\newcommand{\notewellp}{\footnotetext{SP Lemma C.1(a) shows that [P1] implies the well-definition and surjectivity of $p$.}}

\newcommand{\notegraph}{\footnotetext{\label{8189}In contrast to SP, the present paper notationally distinguishes between a correspondence and its graph, between a function and its graph, and between a binary relation and its graph.  Thus [P1] distinguishes between the correspondence $F$ and its graph $F\gr⋅⊆⋅T×C$.  Also, [P2] distinguishes between the function $p$ and its graph $p\gr⋅⊆⋅T×T$, and between the function $⊗$ and its graph $⊗\gr⋅⊆⋅T×C×T$.  Also, for example, Proposition~\ref{6309}(\ref{8166m}) distinguishes between a relation $≺$ and its graph $≺\gr⋅⊆⋅T×T$.}}

\myskip Let $C$ be a set of elements $c$ called {\em choices}.  A triple $Π = (T,C,⊗)$ is a {\em (node-and-choice) preform} (SP Section 3.1) iff \begin{gather}
\zz
\begin{maneq} \text{[P1]} & \text{there is a correspondence}\footnotemark⋅F{:}T⇉C⋅\text{and a}⋅t^o∈T \\   &\text{such that}⋅⊗⋅\text{is a bijection from}\footnotemark⋅F\gr⋅\text{onto}⋅T⧷⎨t^o⎬, \end{maneq} \nt 
\begin{maneq} \text{[P2]} & (T,p)⋅\text{is a functioned tree where}⋅p{:}T⧷⎨t^o⎬→F^{-1}(C) \\
  & \text{is defined}\footnotemark⋅\text{by}⋅p\gr = ⎨(t\sh,t)∈T^2|(∃c∈C)(t,c,t\sh)∈⊗\gr⎬,⋅\text{and} \end{maneq} \nt
\begin{maneq} \text{[P3]} & \HH⋅\text{partitions}⋅F^{-1}(C) \\
  & \text{where}⋅\HH⋅⊆⋅\PP(T)⋅\text{is defined by}⋅\HH = ⎨F^{-1}(c)|c∈C⎬.\end{maneq}\notag
\zz
\end{gather}%
\addtocounter{footnote}{-2}\noteFinv%
\addtocounter{footnote}{1}\notegraph%
\addtocounter{footnote}{1}\notewellp%
Call $⊗$ the {\em node-and-choice} operator,\footnote{\label{8275}A preform's node-and-choice operator $⊗$ can be regarded as a special kind of labeled transition system (e.g.\ Blackburn, de Rijke, and Venema 2001, page 3; van Benthem 2014, page 36).  More precisely, a {\em labeled transition system} is a pair $(S,(R_a)_{a∈A})$ consisting of [1] a set $S$ of states $s$ and [2] a collection of binary relations $R_a$, each defined over $S$, which is indexed by a set $A$ of labels $a$.  A preform's node-and-choice operator $⊗⋅⊆⋅T×C×T$ determines a labeled transition system $(S,(R_a)_{a∈A})$ by setting $S = T$, setting $A = C$, and setting each $R_c = ⎨(t,t\sh)|(t,c,t\sh)∈⊗⎬$.   Conversely, a labeled transition system $(S,(R_a)_{a∈A})$ determines a node-and-choice operator $⊗⋅⊆⋅T×C×T$ by setting $T = S$, setting $C = A$, and setting $⊗ = ⎨(s,a,s′)|(s,s′)∈R_a⎬$.  [P2] restricts this construction by requiring that $⎨(s′,s)|(∃a∈A)(s,s′)∈R_a⎬$ is a functioned tree, and [P3] further restricts the construction by requiring that the labels $a⋅∈⋅A$ serve to specify information sets (these two restrictions concern [1] and [3] in note~\ref{8282}).} %
and let $t⊗c$ denote its value at $(t,c)⋅∈⋅F\gr$.  Call $F$ the {\em feasibility} correspondence, call $t^o$ the {\em root} node, call $p$ the {\em immediate-predecessor} function, and call $\HH$ the collection of {\em information sets}.  In addition, let $X$ equal $F^{-1}(C)$ (inconsequentially, SP uses $F^{-1}(C)$ rather than $X$).  Call $X$ the {\em decision-node set}.\footnote{SP Lemma~C.1(b,c) implies that a preform's $t^o$ and $X$ coincide with the underlying tree's $t^o$ and $X$.  Hence the symbols $t^o$ and $X$ are unambiguous.}   

A node-and-choice preform $Π$ (uniquely) determines many entities.  First, it determines its components $T$, $C$, and $⊗$.  Second, it determines its $F$, $t^o$, $p$, $\HH$, and $X$, as discussed in the previous paragraph.  Third, [P2] determines the functioned tree $(T,p)$, which in turn determines $k$, $≺$, $≼$, $\ZZf$, and $\ZZi$, as discussed in the second-previous paragraph.  Finally, define the preform's {\em previous-choice} function $q{:}T⧷⎨t^o⎬→C$ by $q\gr = ⎨(t\sh,c)∈T×C|(∃t∈T)(t,c,t\sh)∈⊗\gr⎬$.  All these entities and their basic properties are developed in SP Sections 3.1 and 3.2.  Among the basic properties is the convenient fact that $(p,q) = ⊗^{-1}$.  Further properties appear in SP Lemmas A.1, C.1, and C.2, and also in Lemma~\ref{7646} here. 

\myskip Let $I$ be a set of elements $i$ called {\em players}.  A quadruple $Φ = (I,T,(C_i)_{i∈I},⊗)$ is a {\em (node-and-choice) form} iff \begin{gather}
\zz
\begin{maneq} \text{[F1]} & (T,C,⊗)⋅\text{is a preform where}⋅C = ∪_{i∈I}C_i,
  \end{maneq} \notag \\[-1mm]
\begin{maneq} \text{[F2]} & (∀i∈I,j∈I⧷⎨i⎬)⋅C_i∩C_j = ∅,⋅\text{and} \end{maneq} \notag \\[-1mm]
\begin{maneq} \text{[F3]} & (∀t∈T)(∃i∈I)⋅F(t)⋅⊆⋅C_i. \end{maneq} \notag
\zz
\end{gather} Each $C_i$ is the set of choices that are assigned to player $i$.  The definitions in this paragraph are new to this paper (and an earlier version, Streufert 2016).\nocite{ncf-1610}

A node-and-choice form $Φ$ (uniquely) determines many entities.  First, it determines its components $I$, $T$, $(C_i)_{i∈I}$, and $⊗$.  Second, [F1] determines $C$ and the preform $(T,C,⊗)$, which in turn determines $F$, $t^o$, $p$, $q$, $\HH$, $X$, $k$, $≺$, $≼$, $\ZZf$, and $\ZZi$, as discussed in the second-previous paragraph.  In addition, define $(X_i)_{i∈I}$ at each $i$ by $X_i = ∪_{c∈C_i}F^{-1}(c)$.  $X_i$ is the set of decision nodes that are assigned to player $i$.  Further, define $(\HH_i)_{i∈I}$ at each $i$ by $\HH_i = ⎨F^{-1}(c)|c∈C_i⎬$.  $\HH_i$ is the collection of information sets that are assigned to player $i$.

\begin{prop}\label{6295} Suppose $(I,T,(C_i)_i,⊗)$ is a node-and-choice form with its $X$, $\HH$, $(X_i)_{i∈I}$, and $(\HH_i)_{i∈I}$.  Then the following hold. \begin{tlist}
\y $∪_{i∈I}X_i = X$ and $(∀i∈I,j∈I⧷⎨i⎬)⋅X_i∩X_j = ∅$.
\y $(∀i∈I)⋅\HH_i$ partitions $X_i$.
\y $∪_{i∈I}\HH_i = \HH$ and $(∀i∈I,j∈I⧷⎨i⎬)⋅\HH_i∩\HH_j = ∅$.
(Proof~\ref{6295p}.) \end{tlist} \end{prop}

Here are two minor remarks.  [1] A preform can be understood as a one-player form.  Specifically, $(T,C,⊗)$ is a preform iff $(⎨1⎬,T,(C),⊗)$ is a form, where $(C_i)_i = (C)$ is taken to mean $C_1 = C$.  [2] A player $i$ in a form is said to be {\em vacuous} iff $C_i = ∅$.  A vacuous player $i$ necessarily has $X_i = ∅$ and $\HH_i = ∅$.  Vacuous players can be convenient.  For example, one can posit the existence of a chance player, and yet create a game without chance nodes by letting the chance player be vacuous.

\ssec{Morphisms}

A (node-and-choice) {\em preform morphism} (SP Section~3.3) is a quadruple $α = [Π,Π′,τ,δ]$ such that $Π = (T,C,⊗)$ and $Π′ = (T′,C′,⊗′)$ are preforms,\begin{gather}
\zz
\begin{maneq} \text{[PM1]} & τ{:}T→T′, \end{maneq} \notag \\[-1mm]
\begin{maneq} \text{[PM2]} & δ{:}C→C′,⋅\text{and} \end{maneq} \notag \\[-1mm]
\begin{maneq} \text{[PM3]} & ⎨\,(τ(t),δ(c),τ(t\sh))\,|\,(t,c,t\sh)∈⊗\gr\,⎬⋅⊆⋅⊗′\gr. \end{maneq} \notag
\zz
\end{gather} SP Propositions 3.3 and 3.4 give two characterizations of preform morphisms which feel more category-theoretic.  A (node-and-choice) {\em form morphism} is a quintuple $β = [Φ,Φ′,ι,τ,δ]$ s.t.\ $Φ = (I,T,(C_i)_{i∈I},⊗)$ and $Φ′ = (I′,T′,(C′_{i′})_{i′∈I′},⊗′)$ are forms, \begin{gather}
\zz
\begin{maneq} \text{[FM1]} & [Π,Π′,τ,δ]⋅\text{is a preform morphism where} \\
  & Π = (T,C,⊗),⋅C = ∪_{i∈I}C_i,⋅Π′ = (T′,C′,⊗′),⋅\text{and}⋅C′ = ∪_{i′∈I′}C′_{i′}, \end{maneq} \notag \\[-.5mm]
\begin{maneq} \text{[FM2]} & ι{:}I→I′,⋅\text{and} \end{maneq} \notag \\[-.5mm]
\begin{maneq} \text{[FM3]} & (∀i∈I)⋅δ(C_i)⋅⊆⋅C′_{ι(i)}. \end{maneq} \notag
\zz
\end{gather} 

The first paragraph of Proposition~\ref{6309} rearranges the definition of a morphism.  Meanwhile, the second and third paragraphs concern the many derivatives which can be constructed, via Section 2.1, from the source and target forms.  Parts (\ref{8041m}) and (\ref{8042m}) are new, while the remainder are obtained by combining [FM1] with various SP results for preforms and trees.\xin{}

\begin{prop}\label{6309} Suppose $Φ = (I,T,(C_i)_{i∈I},⊗)$ and $Φ′ = (I′,T′,(C′_{i′})_{i′∈I′},⊗′)$ are forms.  Let $C = ∪_{i∈I}C_i$ and $C′ = ∪_{i′∈I′}C′_{i′}$.  Then $[Φ,Φ′,ι,τ,δ]$ is a morphism iff the following hold. \begin{tlist} 
\yl{8173m} $ι\,{:}\,I\,→\,I′$\,.
\yl{8172m} $τ\,{:}\,T\,→\,T′$\,.
\yl{8169m} $δ\,{:}\,C\,→\,C′$\,.
\yl{8040m} $(∀i∈I)⋅δ(C_i)⋅⊆⋅C′_{ι(i)}$.
\yl{8159m} $⎨\,(τ(t),δ(c),τ(t\sh))\,|\,(t,c,t\sh)∈⊗\gr\,⎬⋅⊆⋅⊗′\gr$. 
\end{tlist} 

\vspace{.6mm}Further, suppose $[Φ,Φ′,ι,τ,δ]$ is a morphism.  Let $Π = (T,C,⊗)$ and $Π′ = (T′,C′,⊗′)$.  Also, derive $F$, $t^o$, $p$, $q$, $X$, $(X_i)_{i∈I}$, $\HH$, and $(\HH_i)_{i∈I}$ from $Π$ and $Φ$.  Also, derive $F′$, $t\po$, $p′$, $q′$, $X′$, $(X′_{i′})_{i′∈I′}$, $\HH′$, and $(\HH′_{i′})_{i′∈I′}$ from $Π′$ and $Φ′$.  Then the following hold. \begin{tlist} \setcounter{tlistno}{5}
\yl{8160m} $⎨\,(τ(t),δ(c))\,|\,(t,c)∈F\gr\,⎬⋅⊆⋅F′\gr$.
\yl{8161m} $t\po⋅≼′⋅τ(t^o)$. 
\yl{8162m} $⎨\,(τ(t\sh),τ(t))\,|\,(t\sh,t)∈p\gr\,⎬⋅⊆⋅p′\gr$.
\yl{8163m} $⎨\,(τ(t\sh),δ(c))\,|\,(t\sh,c)∈q\gr\,⎬⋅⊆⋅q′\gr$.
\yl{8164m} $τ(X)⋅⊆⋅X′$.
\yl{8041m} $(∀i∈I)⋅τ(X_i)⋅⊆⋅X′_{ι(i)}$.
\yl{8171m} $(∀H∈\HH)(∃H′∈\HH′)⋅τ(H)⋅⊆⋅H′$.
\yl{8042m} $(∀i∈I,H∈\HH_i)(∃H′∈\HH′_{ι(i)})⋅τ(H)⋅⊆⋅H′$. \end{tlist}

Finally, derive $k$, $≺$, $≼$, $\ZZf$, and $\ZZi$ from $(T,p)$.  Also, derive $k′$, $≺′$, $≼′$, $\ZZf′$, and $\ZZi′$ from $(T′,p′)$. Then the following hold. \begin{tlist} \setcounter{tlistno}{13}
\yl{8165m} $(∀t∈T)⋅k′(τ(t)) = k(t) + k′(τ(t^o))$. 
\yl{8166m} $⎨\,(τ(t^1),τ(t^2))\,|\,(t^1,t^2)∈≺\gr\,⎬⋅⊆⋅≺′\gr$. 
\yl{8167m} $⎨\,(τ(t^1),τ(t^2))\,|\,(t^1,t^2)∈≼\gr\,⎬⋅⊆⋅≼′\gr$. 
\yl{8168m} $(∀Z∈\ZZf)(∃Z′∈\ZZf′∪\ZZi′)⋅τ(Z)⋅⊆⋅Z′$. 
\yl{8170m} $(∀Z∈\ZZi)(∃Z′∈\ZZi′)⋅τ(Z)⋅⊆⋅Z′$. (Proof~\ref{6309p}.) 
\end{tlist}
\end{prop}

\ssec{The category \ct{NCF}}

This paragraph and Theorem~\ref{6316} define the category \ct{NCF}, which is called the {\em category of node-and-choice forms}.  Let an object be a (node-and-choice) form $Φ = (I,T,(C_i)_{i∈I},⊗)$.  Let an arrow be a (node-and-choice) form morphism $β = [Φ,Φ′,ι,τ,δ]$.  Let source, target, identity, and composition be \begin{gather}
\zz
β^\src = [Φ,Φ′,ι,τ,δ]^\src = Φ, \notag \\
β^\trg = [Φ,Φ′,ι,τ,δ]^\trg = Φ′, \notag \\[-0.2mm]
\id_Φ = \id_{(I,T,(C_i)_{i∈I},⊗)} = [Φ,Φ,\id_{I},\id_{T},\id_{∪_{i∈I}C_i}],⋅\text{and} \notag \\[0.4mm]
β′○β = [Φ′,Φ″,ι′,τ′,δ′]○[Φ,Φ′,ι,τ,δ] = [Φ,Φ″,ι′○ι,τ′○τ,δ′○δ],\, \notag 
\zz
\end{gather} where $\id_I$, $\id_T$, and $\id_{∪_{i∈I}C_i}$ are identities in \Set. 

\begin{nthm}\label{6316} \ct{NCF} is a category. (Proof~\ref{6316p}.) \end{nthm}

\begin{nthm}\label{6315} Suppose $β = [Φ,Φ′,ι,τ,δ]$ is a morphism.  Then (a) $β$ is an isomorphism iff $ι$, $τ$, and $δ$ are bijections.  Further (b) if $β$ is an isomorphism, then $β^{-1} = [Φ′,Φ,ι^{-1},τ^{-1},δ^{-1}]$.  (Proof~\ref{6315p}.) \end{nthm}

\begin{ncrly}\label{6333}  Suppose $[Φ,Φ′,ι,τ,δ]$ is a morphism.  Let $Π$ be the preform in $Φ$, and let $Π′$ be the preform in $Φ′$.  Then $[Φ,Φ′,ι,τ,δ]$ is an isomorphism iff [1] $[Π,Π′,τ,δ]$ is a preform isomorphism and [2] $ι$ is a bijection.  (Proof here.) \end{ncrly}

\begin{pf} Note $[Π,Π′,τ,δ]$ is a preform morphism by [FM1] for $[Φ,Φ′,ι,τ,δ]$.  Thus SP Theorem 3.7(a) shows that [1] is equivalent to the bijectivity of $τ$ and $δ$.  Therefore [1] and [2] together are equivalent to the bijectivity of $ι$, $τ$, and $δ$.  By Theorem~\ref{6315}(a), this is equivalent to $[Φ,Φ′,ι,τ,δ]$ being an isomorphism.  \end{pf}

Proposition~\ref{6346} organizes some\footnote{The proposition's list of consequences is far from exhaustive.  For example, in the notation of the proposition's second paragraph, Lemma~\ref{6349}(b) deduces that $(∀c∈C)$ $τ(F^{-1}(c)) = (F′)^{-1}(δ(c))$.} of the consequences of a form isomorphism.  The proposition's first paragraph concerns form components, while the second and third paragraphs concern form derivatives.  
Consequences (\ref{8173})--(\ref{8169}) repeat the forward direction of Theorem~\ref{6315}(a).  Consequences (\ref{8040}), (\ref{8041}), and (\ref{8042}) are new, while the remainder are obtained by combining the forward direction of Corollary~\ref{6333} with SP results about preforms and trees.  The entire proposition is comparable to Proposition~\ref{6309} for morphisms, and Section 5.1 will discuss how the proposition contributes directly to game theory.

To address a minor technical issue, note that many of the proposition's consequences are formulated by restricting functions.  In each case, the codomain of the restriction is defined so that the restriction is surjective.  Some other minor technical issues are discussed in notes \ref{8189}, \ref{8190}, and \ref{8158}.

\begin{prop}\label{6346} Suppose $[Φ,Φ′,ι,τ,δ]$ is an isomorphism, where $Φ = (I,T,$ $(C_i)_{i∈I},⊗)$ and $Φ′ = (I′,T′,(C′_{i′})_{i′∈I′},⊗′)$.  Let $C = ∪_{i∈I}C_i$ and $C′ = ∪_{i′∈I′}C′_{i′}$.  Then the following hold. \begin{tlist} 
\yl{8173} $ι$ is a bijection from $I$ onto $I′$.
\yl{8172} $τ$ is a bijection from $T$ onto $T′$.
\yl{8169} $δ$ is a bijection from $C$ onto $C′$.
\yl{8040} $(∀i∈I)⋅δ|_{C_i}$ is a bijection from $C_i$ onto $C′_{ι(i)}$.\footnote{\label{8190}To be clear, parts (\ref{8040}), (\ref{8041}), and (\ref{8042}) do hold when there is a vacuous player $i$.  In this case, $C_i$ is empty, and thus, $δ|_{C_i}$, $C′_{ι(i)}$, $X_i$, $τ|_{X_i}$, $X′_{ι(i)}$, $\HH_i$, $τ|_{\HH_i}$, and $\HH′_{ι(i)}$ are all empty as well.}
\yl{8159} $(τ,δ,τ)|_{⊗\gr}$ is a bijection from $⊗\gr$ onto $⊗′\gr$. 
\end{tlist}

\vspace{.6mm}Further, let $Π = (T,C,⊗)$ and $Π′ = (T′,C′,⊗′)$.  Also, derive $F$, $t^o$, $p$, $q$, $X$, $(X_i)_{i∈I}$, $\HH$, and $(\HH_i)_{i∈I}$ from $Π$ and $Φ$.  Also, derive $F′$, $t\po$, $p′$, $q′$, $X′$, $(X′_{i′})_{i′∈I′}$, $\HH′$, and $(\HH′_{i′})_{i′∈I′}$ from $Π′$ and $Φ′$.  Then the following hold. \begin{tlist} \setcounter{tlistno}{5}
\yl{8160} $(τ,δ)|_{F\gr}$ is a bijection from $F\gr$ onto $F′\gr$.
\yl{8161} $τ(t^o) = t\po$.
\yl{8162} $(τ,τ)|_{p\gr}$ is a bijection from $p\gr$ onto $p′\gr$.
\yl{8163} $(τ,δ)|_{q\gr}$ is a bijection from $q\gr$ onto $q′\gr$.
\yl{8164} $τ|_X$ is a bijection from $X$ onto $X′$.
\yl{8041} $(∀i∈I)⋅τ|_{X_i}$ is a bijection from $X_i$ onto $X′_{ι(i)}$.\xrf{8190}
\yl{8171} $τ|_{\HH}$ is a bijection from $\HH$ onto $\HH′$.\footnote{\label{8158}In parts  (\ref{8171}), (\ref{8042}), (\ref{8168}), and (\ref{8170}), $τ$ is understood to be the function $\PP(T)⋅∋⋅S⋅\mapsto⋅⎨τ(t)|t∈S⎬⋅∈⋅\PP(T′)$.  For example, if $H⋅∈⋅\HH$, then $τ(H) = ⎨τ(t)|t∈H⎬$.  Similarly, if $Z⋅∈⋅\ZZf∪\ZZi$, then $τ(Z) = ⎨τ(t)|t∈Z⎬$.}  
\yl{8042} $(∀i∈I)⋅τ|_{\HH_i}$ is a bijection from $\HH_i$ onto $\HH′_{ι(i)}$.\xrf{8190}$^,$\xrf{8158}
 \end{tlist} 

Finally, derive $k$, $≺$, $≼$, $\ZZf$, and $\ZZi$ from $(T,p)$.  Also, derive $k′$, $≺′$, $≼′$, $\ZZf′$, $\ZZi′$ from $(T′,p′)$. Then the following hold. \begin{tlist} \setcounter{tlistno}{13}
\yl{8165} $(∀t∈T)⋅k′(τ(t)) = k(t)$.
\yl{8166} $(τ,τ)|_{≺\gr}$ is a bijection from $≺\gr$ onto $≺′\gr$.
\yl{8167} $(τ,τ)|_{≼\gr}$ is a bijection from $≼\gr$ onto $≼′\gr$.
\yl{8168} $τ|_{\ZZf}$ is a bijection from $\ZZf$ onto $\ZZf′$.\xrf{8158}
\yl{8170} $τ|_{\ZZi}$ is a bijection from $\ZZi$ onto $\ZZi′$.\xrf{8158} (Proof~\ref{6346p}.) 
\end{tlist}
\end{prop}

As already noted, the definition of a form incorporates a preform, and the definition of a form morphism incorporates a preform morphism.  Correspondingly, Theorem~\ref{6318} shows there is a ``forgetful'' functor $\PB$ from \ct{NCF} to \ct{NCP}.  Incidentally, SP Theorem 3.9 shows there is a similar functor $\FB$ from \ct{NCP} to \ct{Tree}.  Hence $\FB○\PB$ is a functor from \ct{NCF} to \ct{Tree}.

\begin{nthm}\label{6318} Define $\PB$ from \ct{NCF} to \ct{NCP} by\begin{gather}
\zz
\PO⋅:⋅(I,T,(C_i)_{i∈I},⊗)⋅\mapsto (T,∪_{i∈I}C_i,⊗)⋅\text{and} \nt
\PA⋅:⋅[Φ,Φ′,ι,τ,δ]⋅\mapsto⋅[\PO(Φ),\PO(Φ′),τ,δ].\notag
\zz
\end{gather} Then $\PB$ is a well-defined functor.  (Proof~\ref{6318p}.) \end{nthm}

\ssec{No-absentmindedness and perfect-information}

Consider an arbitrary category \ct{Z}, and a property which is defined for the objects of \ct{Z}.  The property is said to be {\em isomorphically invariant} iff, for each object, the object satisfies the property iff all of its isomorphs satisfy the property.  This section explores two isomorphically invariant properties: [1] no-absentmindedness and [2] perfect-information.  Both properties restrict information sets.

\myskip No-absentmindedness is a standard property which is widely regarded as being very weak (see, for example, Al\'os-Ferrer and Ritzberger 2016 Section 4.2.3).  To define this property in \ct{NCP}, consider an \ct{NCP} preform with its $≺$ and $\HH$.  Then the preform is said to have {\em no-absentmindedness} iff $(∄H∈\HH,t^A∈H,t^B∈H)⋅t^A⋅≺⋅t^B$.{\footnotemark}  Further, consider an \ct{NCF} form with its preform.  Then the form is said to have {\em no-absentmindedness} iff its preform has no-absentmindedness. 

\footnotetext{\label{8095}Piccione and Rubinstein 1997\nocite{PiRu97a} Figure~1 provides an example of absentmindedness.  A corresponding \ct{NCP} preform $Π = (T,C,⊗)$ can be defined by $T = ⎨⎨⎬,(\fa),(\fb),(\fa,\fa),(\fa,\fb)⎬$, $C = ⎨\fa,\fb⎬$, and $⊗ = ⎨(⎨⎬,\fa,(\fa)),$ $(⎨⎬,\fb,(\fb))$, $((\fa),\fa,(\fa,\fa))$, $((\fa),\fb,(\fa,\fb)⎬$.  No-absentmindedness fails because $\HH$ contains $H = ⎨⎨⎬,(\fa)⎬$ and $⎨⎬⋅≺⋅(\fa)$.  A corresponding \ct{NCF} form $Φ = (I,T,(C_i)_i,⊗)$ can be defined by setting $T$ and $⊗$ as above, setting $I = ⎨\f1⎬$, and setting $C_{\f1} = ⎨\fa,\fb⎬$.  The existence of this example is used in the proof of Corollary~\ref{8089}.}

\begin{prop}\label{7710} (a$^o$) If $[Π,Π′,τ,δ]$ is an \ct{NCP} morphism and $Π′$ has no-absentmindedness, then $Π$ has no-absentmindedness.  (a) No-absentmindedness is isomorphically invariant in \ct{NCP}.  (b$^o$) If $[Φ,Φ′,ι,τ,δ]$ is an \ct{NCF} morphism and $Φ′$ has no-absentmindedness, then $Φ$ has no-absentmindedness.  (b) No-absentminded\-ness is isomorphically invariant in \ct{NCF}.  (Proof~\ref{7710p}.) \end{prop}

Let \ct{NCP_\ga} be the full subcategory of \ct{NCP} whose objects are preforms with no-absentmindedness.  (I am endeavouring to use subscripts for isomorphically invariant properties.)  Similarly, let \ct{NCF_\ga} be the full subcategory of \ct{NCF} whose objects are forms with no-absentmindedness.  No-absentmindedness will appear again in Section~3.3.

\newcommand{\notenapi}{\footnote{\label{8191}To see that perfect-information implies no-absentmindedness, assume no-absentmindedness is violated.  Then there is $H⋅∈⋅\HH$, $t^A⋅∈⋅H$, and $t^B⋅∈⋅H$ such that $t^A⋅≺⋅t^B$.  Thus $t^A⋅≠⋅t^B$.  So $|H| > 1$ and perfect-information is violated.}}

\newcommand{\notehorsegame}{\footnote{\label{8096} A simple example of a form which satisfies no-absentmindedness but not perfect-information is a form corresponding to a two-person simultaneous-move game.  Specifically, define the \ct{NCP} preform $Π = (T,C,⊗)$ by $T = ⎨⎨⎬,(\fa),(\fb),(\fa,\fc),(\fa,\fd),(\fb,\fc),(\fb,\fd)⎬$, $C = ⎨\fa,\fb,\fc,\fd⎬$, and $⊗ = ⎨(⎨⎬,\fa,(\fa)),$ $(⎨⎬,\fb,(\fb))$, $((\fa),\fc,(\fa,\fc))$, $((\fa),\fd,(\fa,\fd))$, $((\fb),\fc,(\fb,\fc))$, $((\fb),\fd,(\fb,\fd))⎬$.  Note that $\HH$ consists of $H = ⎨⎨⎬⎬$ and $H′ = ⎨(\fa),(\fb)⎬$.  No-absentmindedness holds because [i] $H$ is a singleton and [ii] neither $(\fa)⋅≺⋅(\fb)$ nor $(\fa)⋅≻⋅(\fb)$.  Perfect-information fails because $|H′|⋅≠⋅1$.  A corresponding \ct{NCF} form $Φ = (I,T,(C_i)_i,⊗)$ can be defined by setting $T$ and $⊗$ as above, setting $I = ⎨\f1,\f2⎬$, and setting $C_\f1 = ⎨\fa,\fb⎬$ and $C_\f2 = ⎨\fc,\fd⎬$.  The existence of this example is used in the proof of Corollary~\ref{8091}.  [A slightly more complicated example with the same combination of properties can be obtained from any of the five figures in Section 1.1.]}}

\myskip Perfect-information is another standard property.  It is restrictive, and at the same time, there are many interesting games which satisfy it (see, for example, Osborne and Rubinstein 1994 Part II).  As in SP Section 3.5, an \ct{NCP} preform, with its collection $\HH$ of information sets $H$, is said to have {\em perfect-information} iff $(∀H∈\HH)$ $|H| = 1$.  Perfect-information is strictly stronger than no-absentmindedness.{\notenapi}$^,${\notehorsegame}  Further, an \ct{NCF} form is said to have {\em perfect-information} iff the form's preform has perfect-information.  (In spite of Proposition~\ref{8046}, the existence of a morphism does not lead to any logical relationship between the source's perfect-information and the target's perfect-information.)

\begin{prop}\label{8046} (a) Perfect-information is isomorphically invariant in \ct{NCP}.  (b) Perfect-information is isomorphically invariant in \ct{NCF}. (Proof~\ref{8046p}.) \end{prop}

Let \ct{NCP_p} be the full subcategory of \ct{NCP} whose objects are preforms with perfect-information.  (The subscript \ct{_{\ga p}} would be equivalent to the subscript \ct{_p}, because no-absentmindedness is implied by perfect-information, as shown in note~\ref{8191}.)  Further, let \ct{NCF_p} be the full subcategory of \ct{NCF} whose objects are forms with perfect-information.  Perfect-information will appear again in Section~4.3.

\section{The Subcategory of Choice-Sequence Forms}
\markb{\sc 3. The Subcategory of Choice-Sequence Forms}

\ssec{Objects}

Let a {\em (finite) sequence} be a function from $⎨1,2,...\,m⎬$ for some nonnegative integer $m$ (to be clear, the empty sequence\footnote{The empty sequence is the empty set.  Further, $⎨⎬$ and $∅$ are alternative notations for the empty set.  I use $⎨⎬$ for a root node, and use $∅$ for all other purposes.} with empty domain is admitted by $m = 0$).  I will regard a sequence as a set of ordered pairs.  For example, $t^* = ⎨(1,\fg),$ $(2,\ff),$ $(3,\ff)⎬$ is a sequence with domain $⎨1,2,3⎬$.  An alternative notation for the same entity is $t^* = (\ex{g,f,f})$.  Yet another is $t^* = (t^*_n)^3_{n=1}$ where $t^*_1 = \fg$ and $t^*_2 = t^*_3 = \ff$.

Let the {\em length} of a sequence $t$ be $|t|$.  For instance, the length of the example sequence is $|t^*|$ $=$ $|⎨(1,\fg),(2,\ff),(3,\ff)⎬|$ $= 3$, which is consistent with the observation that $(2,\ff)$ $≠$ $(3,\ff)$.  Note that the length of the empty sequence $⎨⎬$ is $|⎨⎬| = 0$.  Next, let the {\em range} of a sequence $t$ be $R(t) = ⎨\,t_n\,|\,n∈⎨1,2,...\,|t|⎬\,⎬$.  For instance, the range of the example sequence is $R(t^*) = ⎨\,t^*_n\,|\,n∈⎨1,2,3⎬\,⎬ = ⎨\fg,\ff,\ff⎬ = ⎨\fg,\ff⎬$.  Note that the range of the empty sequence $⎨⎬$ is $R(⎨⎬) = ∅$. 

Let the {\em concatenation} $t±s$ of two sequences $t$ and $s$ be $⎨(1,t_1),$ $...$ $(|t|,t_{|t|}),$ $(|t|{+}1,s_1),$ $...$ $(|t|{+}|s|,s_{|s|})⎬$.  Thus the concatenation of a sequence $t = (t_1,t_2,...\,t_{|t|})$ with a one-element sequence $(c)$ is $t±(c) = (t_1,t_2,...\,t_{|t|},c)$.  Next, for any sequence $t$ and any $ℓ⋅∈⋅⎨0,1,2,...\,|t|⎬$, let $_1t_ℓ$ denote the {\em initial segment} $(t_1,t_2,...\,t_ℓ)$.  Thus for any sequence $t$, $_1t_0 = ⎨⎬$.  

A {\em choice-sequence} \ct{NCP} preform is an \ct{NCP} preform $(T,C,⊗)$ such that\begin{gather}
\zz
\begin{maneq} \text{[Csq1]} & T⋅\text{is a collection of (finite) sequences which contains}⋅⎨⎬, \end{maneq} 
  \notag \\[-1mm]
\begin{maneq} \text{[Csq2]} & (∀\,(t,c,t\sh)\,∈\,⊗\gr)⋅t{\oplus}(c) = t\sh. \end{maneq} \notag
\zz
\end{gather} Let \ct{CsqP} be the full subcategory of \ct{NCP} whose objects are choice-sequence preforms.  Proposition~\ref{7720} lists some of the special properties of \ct{CsqP} preforms.  Incidentally, property (\ref{7733}) and assumption [Csq1] together imply that each node in a \ct{CsqP} preform is actually a choice sequence, as the terminology suggests.  

\begin{prop}\label{7720} Suppose $(T,C,⊗)$ is a \ct{CsqP} preform.  Derive its $F$, $t^o$, $p$, $q$, $k$, $≺$, and $≼$.  Then the following hold.\begin{tlist}
\yl{7721} $t^o = ⎨⎬$.
\yl{7728} $(∀t\sh∈T⧷⎨⎨⎬⎬)⋅p(t\sh) = {_1t\sh_{|t\sh|-1}}\,\text{and}⋅q(t\sh) = t\sh_{|t\sh|}$.            
\yl{7726} $⊗\gr = ⎨⋅(t,c,t\sh)∈T×C×T⋅|⋅t±(c){=}t\sh⋅⎬$.
\yl{7727} $F\gr = ⎨⋅(t,c)∈T×C⋅|⋅t±(c)∈T⋅⎬$.
\yl{7729} $(∀t∈T,m∈⎨0,1,...\,|t|⎬)⋅p^m(t) = {_1t_{|t|-m}}$.
\yl{7722} $(∀t∈T)$ $k(t) = |t|$. 
\yl{7723} $(∀t∈T)⋅t = (q○p^{|t|-ℓ}(t))^{|t|}_{ℓ=1}$.
\yl{7733} $C = ∪_{t∈T}R(t)$. 
\yl{8110} $(∀t^A∈T,t^B∈T)$ $t^A⋅≺⋅t^B$ iff $(|t^A|\,{<}\,|t^B|⋅\text{and}⋅t^A\,{=}\,{_1t^B_{|t^A|}})$.
\yl{8111} $(∀t^A∈T,t^B∈T)$ $t^A⋅≼⋅t^B$ iff $(|t^A|\,≤\,|t^B|⋅\text{and}⋅t^A\,{=}\,{_1t^B_{|t^A|}})$. (Proof \ref{7720p}.)
\end{tlist} \end{prop}

Finally, let a {\em choice-sequence} \ct{NCF} form be an \ct{NCF} form whose preform is a \ct{CsqP} preform.  Then let \ct{CsqF} be the full subcategory of \ct{NCF} whose objects are choice-sequence \ct{NCF} forms.

\ssec{Isomorphic Enclosure}

Consider two full subcategories \ct{A} and \ct{B} of some overarching category \ct{Z}.  Say that \ct{A} is {\em isomorphically enclosed} in \ct{B} (in symbols, \ct{A} $\ud{→}$ \ct{B}) iff every object of \ct{A} is isomorphic to an object of \ct{B}.  Note that \ct{A} $\ud{→}$ \ct{B} concerns not only the subcategories \ct{A} and \ct{B} but also, implicitly, the overarching category \ct{Z} within which isomorphisms are defined.  Further note that isomorphic enclosures can be composed in the sense that \ct{A} $\ud{→}$ \ct{B} and \ct{B} $\ud{→}$ \ct{C} imply \ct{A} $\ud{→}$ \ct{C}.\footnote{\label{8154}To prove composability, recall \ct{A} $\ud{→}$ \ct{B} means that [a] each \ct{A} form is isomorphic to a \ct{B} form.  Similarly, \ct{B} $\ud{→}$ \ct{C} means that [b] each \ct{B} form is isomorphic to a \ct{C} form.  [a] and [b] imply that each \ct{A} form is isomorphic to a \ct{C} form, and this is what is meant by \ct{A} $\ud{→}$ \ct{C}.}  Finally, let $\cy{A}⋅\ud{⟷}⋅\cy{B}$ mean that both $\cy{A}⋅\ud{→}⋅\cy{B}$ and $\cy{A}⋅\ud{←}⋅\cy{B}$ hold.  Call $\ud{⟷}$ {\em isomorphic equivalence}.  Isomorphic equivalence implies the standard categorical concept of equivalence in MacLane 1998 page~18.\nocite{MacL98}

\begin{nthm}\label{8066} (a) \ct{NCP} $\ud{→}$ \ct{CsqP}.  In particular, suppose $Π = (T,C,⊗)$ is an \ct{NCP} preform with its $p$, $q$, and $k$.   Define $\dT = ⎨\,(q○p^{k(t)-ℓ}(t))^{k(t)}_{ℓ=1}\,|\,t∈T\,⎬$, define $\bar{τ}{:}T→\dT$ by $\bar{τ}(t) = (q○p^{k(t)-ℓ}(t))^{k(t)}_{ℓ=1}$, and define $\bar{⊗}$ by surjectivity and $\bar{⊗}\gr = ⎨\,(\bar{τ}(t),c,\bar{τ}(t\sh))\,|\,(t,c,t\sh)∈⊗\gr\,⎬$.  Then $\bar{Π} = (\dT,C,\bar{⊗})$ is an \ct{CsqP} preform, $\bar{τ}$ is a bijection, and $[Π,\bar{Π},\bar{τ},\id_C]$ is an \ct{NCP} isomorphism.  (b) \ct{NCF} $\ud{→}$ \ct{CsqF}.  In particular, suppose $Φ = (I,T,(C_i)_{i∈I},⊗)$ is an \ct{NCF} form.  Define $\dT$, $\bar{τ}$, and $\bar{⊗}$ as in part (a).  Then $\bar{Φ} = (I,\dT,(C_i)_{i∈I},\bar{⊗})$ is a \ct{CsqF} form and $[Φ,\bar{Φ},\id_I,\bar{τ},\id_{∪_{i∈I}C_i}]$ is an \ct{NCF} isomorphism. (Proof~\ref{8066p}.)\footnote{Theorems~\ref{8066} and \ref{7967} draw upon Lemmas~\ref{6946} and \ref{8058}.  These nontrivial lemmas show how to construct isomorphisms in \ct{NCP} and \ct{NCF} from bijections for nodes, choices, and players.  These lemmas appear to have application beyond this paper.} \end{nthm}

\begin{ncrly}\label{8052} (a) \ct{NCP} $\ud{⟷}$ \ct{CsqP}.  (b) \ct{NCF} $\ud{⟷}$ \ct{CsqF}. (Proof here.) \end{ncrly}

\begin{pf} {\em (a)}. \ct{NCP} $\ud{→}$ \ct{CsqP} by Theorem~\ref{8066}(a).  Conversely, each \ct{CsqP} preform is an \ct{NCP} preform by definition.  {\em (b)}. This is very similar to (a).  Change ``preform'' to ``form'', \ct{P} to \ct{F}, and (a) to (b). \end{pf}

\newcommand{\alfred}{[i]}
\newcommand{\berndt}{[ii]}
\newcommand{\ORna}{OR$\bar{\text{a}}$}

This equivalence has a long history.  In the more distant past, it was informally understood that game trees could be specified in terms of either {\alfred} a collection of nodes and a collection of edges or {\berndt} a collection of sequences.  Harris 1985 page 617\nocite{Harris85} provides an example of this informal understanding.  Specification style {\alfred} uses the nomenclature of graph theory (e.g., Tutte 1984),\nocite{Tutt84} and style-{\alfred} trees were the basis on which Kuhn 1953 and Selten 1975 built game forms.  Later, style-{\berndt} trees became the basis on which Osborne and Rubinstein 1994 built game forms.

Kline and Luckraz 2016\footnote{\label{8155}The terms ``choice'', ``action'', and ``alternative'' are fundamentally synonymous.  However, the literature tends to use ``choice'' when it is assumed that information sets do not share alternatives, and conversely, to use ``action'' when the assumption is relaxed.  The assumption itself is insubstantial in the sense that one can always introduce more alternatives until each information set has its own alternatives (see SE Section~5.2, first paragraph, for more discussion).  This paper makes the assumption for notational convenience, and correspondingly, uses ``choice'' (see SP Proposition 3.2(16b) and the paragraphs beforehand).  In contrast, KL16 relaxes the assumption and uses ``action''.} %
(henceforth ``KL16'') develop this equivalence by a pair of theorems.  In recognition of the above authors, they call style-{\alfred} forms ``KS forms'' and call style-{\berndt} forms ``OR forms''.  Then, one of their theorems (their Theorem~2) shows that a KS form can be derived from each OR form, while the other theorem (their Theorem~1) shows that each KS form can be mapped to an OR form.\footnote{SE Theorems 3.2 and 3.1 adapt and slightly extend KL16 Theorems 2 and 1.}  These two theorems are depicted by the two arrows in Figure~\ref{8146}(a).  The arrows are dashed to convey that the equivalence is ad hoc. 

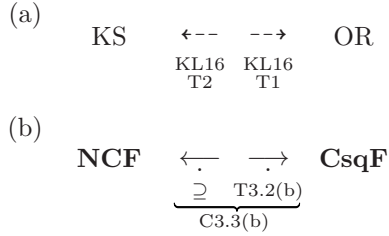
\renewcommand{\captionall}{(a) The ad hoc equivalence of Kline and Luckraz 2016 (KL16).  (b) The isomorphic equivalence of Corollary~\ref{8052}(b).  T\,{=}\,Theorem. C\,{=}\,Corollary.}
\begin{figure}[h]
  \newcommand{\hgth}{90}  
  \begin{picture}(0,\hgth) 
  \put(-195,-9){\scalebox{1}{ 
    \begin{pspicture}(-9,-3)(9,3) 
      \end{pspicture}
    }} \end{picture}
  \caption{\small \captionall} \label{8146} 
  \end{figure} 

Corollary~\ref{8052}(b) develops the equivalence further.  Specification-{\alfred} forms are written as \ct{NCF} forms, and specification-{\berndt} forms are written as \ct{CsqF} forms.  Corollary~\ref{8052}(b) is then a pair of results: one half (the very easy half) shows that an \ct{NCF} form is isomorphic to each \ct{CsqF} form, while the other half (Theorem~\ref{8066}) shows that each \ct{NCF} form is isomorphic to a \ct{CsqF} form.  Thus the corollary's isomorphic equivalence strengthens the KL16 equivalence by introducing isomorphisms.

There are further senses in which the corollary's isomorphic equivalence accords with the KL16 equivalence.  In the backward direction, KL16 Theorem 2 is appealing because the nodes in the constructed KS form are identical to the sequences in the given OR form.  This is possible because KS nodes admit OR sequences as special cases.  Nonetheless KL16 Theorem 2 is nontrivial because KS forms do not admit OR forms as special cases.  Here the analogous result is cleaner: \ct{NCF} forms have been defined so that \ct{NCF} forms admit \ct{CsqF} forms as special cases.  In the forward direction, KL16 Theorem 1 is made appealing by KL16 Lemma~2, which shows that there is a bijection $α$ from the ``vertex histories'' in the given KS form to the nodes in the constructed OR form.  That bijection is closely related to Theorem~\ref{8066}'s bijection $\bar{τ}$, which maps from the nodes of the given \ct{NCF} form to the nodes in the constructed \ct{CsqF} form.

\ssec{More about No-absentmindedness}

\newcommand{\notimply}{\not\mspace{-5mu}{⇒}}
\newcommand{\sweaker}{{⇐}\mspace{0mu}{\notimply}}

3.3.1. Proposition~\ref{8090} describes a general situation in which one subcategory strictly isomorphically encloses another.  In the proposition, $w$ and $s$ are two properties defined for the objects of \ct{Z}.  Further, $w⋅\sweaker⋅s$ means that $w$ is strictly weaker than $s$.  In other words, $w⋅\sweaker⋅s$ means that [a] each object of \ct{Z} satisfies $w$ if it satisfies $s$, and [b] there is an object of \ct{Z} that satisfies $w$ but not $s$.  Corollary~\ref{8089} applies Proposition~\ref{8090} to the nonvacuous property of no-absentmindedness.

\begin{prop}\label{8090} Suppose $w$ and $s$ are properties defined for the objects of \ct{Z}, and that $s$ is isomorphically invariant.  Let \ct{Z_w} be the full subcategory of \ct{Z} whose objects satisfy $w$, and let \ct{Z_s} be the full subcategory of \ct{Z} whose objects satisfy $s$.  Then $w⋅\sweaker⋅s$ implies \ct{Z_w} $\sencloses$ \ct{Z_s}. (Proof here.) \end{prop}  

\begin{pf} Suppose $w⋅\sweaker⋅s$.  To see \ct{Z_w} $\ud{←}$ \ct{Z_s}, take an object of \ct{Z_s}.  Since $w⋅⇐⋅s$, the object is also an object of \ct{Z_w}.  Thus (trivially) the object is isomorphic to an object of \ct{Z_w}.  To see \ct{Z_w} ${\notenclosedby}$ \ct{Z_s}, note the assumption $w⋅\sweaker⋅s$ implies that there is an object of \ct{Z} that satisfies $w$ and violates $s$.  Thus there is an object of \ct{Z_w} that violates $s$.  Thus since $s$ is isomorphically invariant, this object does not have an isomorph that satisfies $s$.  Thus the object does not have an isomorph in \ct{Z_s}. \end{pf}

\begin{ncrly}\label{8089} (a) \ct{NCP} $\sencloses$ \ct{NCP_\ga}.  (b) \ct{NCF} $\sencloses$ \ct{NCF_\ga}. (Proof here.) \end{ncrly}

\begin{pf} {\em (a)}.  Consider Proposition~\ref{8090} at \ct{Z} equal to \ct{NCP}, when $w$ is the vacuous property satisfied by all objects of \ct{NCP}, and $s$ is the property of no-absentmindedness.  No-absentmindedness is invariant by Proposition~\ref{7710}(a).  Further the vacuous property is strictly weaker than no-absentmindedness because there exists an absentminded preform (recall note~\ref{8095}).  Thus Proposition~\ref{8090} implies that \ct{NCP_w} = \ct{NCP} strictly isomorphically encloses \ct{NCP_s} = \ct{NCP_\ga}.  {\em (b)}. This is very similar to (a).  Change ``preform'' to ``form'', \ct{P} to \ct{F}, and (a) to (b). \end{pf}

To better interpret Corollary~\ref{8089}, recall Theorem~\ref{8066}(b) which states \ct{NCF} $\ud{→}$ \ct{CsqF}.  Formally, this means each \ct{NCF} form is isomorphic to a \ct{CsqF} form.  This can be interpreted to mean that the property of having choice-sequence nodes is not ``restrictive''.  In contrast, Corollary~\ref{8089}(b) implies \ct{NCF} $\not\mspace{-5mu}\ud{→}$ \ct{NCF_\ga}.  Formally, this means there is at least one \ct{NCF} form (such as the one in note \ref{8095}) that is not isomorphic to an \ct{NCF_\ga} form.  This can be interpreted to mean that the property of no-absentmindedness is ``restrictive''.  Informally, the first result states that choice-sequence-ness is ``purely notational''.  In contrast, the second result states that no-absentmindedness is ``substantial'', ``significant'', and ``real'', and that it ``limits the range of decision processes and social interactions that can be modelled''.  The categorical concept of isomorphic enclosure ($\ud{→}$) serves to formalize and to standardize these important terms.  Note that both an isomorphic enclosure, and the negation of an isomorphic enclosure, are meaningful.

\myskip  3.3.2. Next, Proposition~\ref{8081} shows that an isomorphic enclosure can be restricted by any isomorphically invariant property.  Corollary~\ref{8083} uses this result to restrict Corollary~\ref{8052} by no-absentmindedness.  Corollary~\ref{8083} will in turn be used in the remarkably quick proof of Corollary~\ref{8082}.

\begin{prop}\label{8081} Suppose that \ct{A} and \ct{B} are full subcategories of \ct{Z}, and that $w$ is an isomorphically invariant property defined for the objects of \ct{Z}.  Let \ct{A_w} be the full subcategory of \ct{A} whose objects satisfy $w$, and let \ct{B_w} be the full subcategory of \ct{B} whose objects satisfy $w$.  Then \ct{A} $\ud{→}$ \ct{B} implies \ct{A_w} $\ud{→}$ \ct{B_w}. (Proof here.) \end{prop}

\begin{pf} Suppose \ct{A} $\ud{→}$ \ct{B}.  To show \ct{A_w} $\ud{→}$ \ct{B_w}, take an object of \ct{A_w}.  Then [1] the object is an object of \ct{A} and [2] the object satisfies $w$.  By [1] and \ct{A} $\ud{→}$ \ct{B}, the object has an isomorph in \ct{B}.  By [2] and the isomorphic invariance of $w$, the isomorph satisfies $w$.  The conclusions of the previous two sentences imply that the isomorph is in \ct{B_w}. \end{pf}

\begin{ncrly}\label{8083} (a) \ct{NCP_\ga} $\ud{⟷}$ \ct{CsqP_\ga}.  (b) \ct{NCF_\ga} $\ud{⟷}$ \ct{CsqF_\ga}. (Proof here.) \end{ncrly}

\begin{pf} (a) follows from Corollary~\ref{8052}(a), Proposition~\ref{8081}, and Proposition~\ref{7710}(a).  (b) is very similar to (a). Just change (a) to (b). \end{pf}

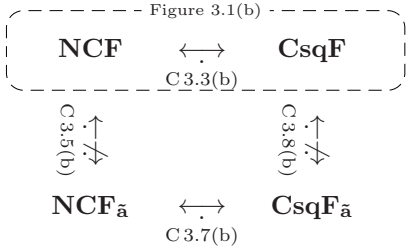
\renewcommand{\captionall}{Half of the previous figure, augmented with some results about no-absentmindedness.  C\,=\,Corollary.}
\begin{figure}[h]
  \newcommand{\hgth}{95}  
  \begin{picture}(0,\hgth)  
  \put(-132,-5){\scalebox{1}{  
    \begin{pspicture}(-6,-2.5)(6,2.5)
      \end{pspicture}
    }} \end{picture}
  \caption{\small \captionall} \label{8147} 
  \end{figure} 

\myskip 3.3.3. Finally, Corollary~\ref{8093} could be proved by mimicking the proof of Corollary~\ref{8089}, in which case Proposition~\ref{8090} would be employed once for part (a) at \ct{Z} = \ct{CsqP}, and again for part (b) at \ct{Z} = \ct{CsqF}.  Instead, Corollary~\ref{8093} is proved by composing isomorphic enclosures (note \ref{8154}), and the proof of the corollary's part (b) is illustrated by Figure~\ref{8147}.  Both proof techniques are straightforward, and a more interesting example of composition will soon appear in the proof of Corollary~\ref{8082}.

\begin{ncrly}\label{8093} (a) \ct{CsqP} $\sencloses$ \ct{CsqP_\ga}.  (b) \ct{CsqF} $\sencloses$ \ct{CsqF_\ga}. (Proof here.) \end{ncrly}

\begin{pf} {\em(a)}.  This is very similar to (b).  Change \ct{F} to \ct{P}, and (b) to (a).  {\em (b)}.  To see \ct{CsqF} $\ud{←}$ \ct{CsqF_\ga}, note that \ct{CsqF} $\ud{←}$ \ct{NCF} $\ud{←}$ \ct{NCF_\ga} $\ud{←}$ \ct{CsqF_\ga} by, respectively, Corollary~\ref{8052}(b), Corollary~\ref{8089}(b), and Corollary~\ref{8083}(b).  To see \ct{CsqF} $\notenclosedby$ \ct{CsqF_\ga}, suppose it were.  Then \ct{NCF} $\ud{→}$ \ct{CsqF} $\ud{→}$ \ct{CsqF_\ga} $\ud{→}$ \ct{NCF_\ga} by, respectively, Corollary~\ref{8052}(b), the supposition of the previous sentence, and Corollary~\ref{8083}(b).  This contradicts Corollary~\ref{8089}(b), which states that \ct{NCF} $\notenclosedby$ \ct{NCF_\ga}.  \end{pf}

\section{The Subcategory of Choice-Set Forms}
\markb{\sc 4. The Subcategory of Choice-Set Forms}

\ssec{Objects}

Let a {\em choice-set} \ct{NCP} preform be an \ct{NCP} preform $(T,C,⊗)$ such that \begin{gather}
\zz
\begin{maneq} \text{[Cset1]} & T⋅\text{is a collection of finite sets which contains}⋅⎨⎬⋅\text{and} \end{maneq}
\notag \\[-1mm]
\begin{maneq} \text{[Cset2]} & (∀(t,c,t\sh)∈⊗\gr)⋅t∪⎨c⎬ = t\sh. \notag \end{maneq}
\zz
\end{gather} Then let \ct{CsetP} be the full subcategory of \ct{NCP} whose objects are choice-set \ct{NCP} preforms.  Proposition~\ref{7568} lists some of the special properties of \ct{CsetP} preforms.\footnote{Almost every \ct{CsetP} property in Proposition~\ref{7568} has a \ct{CsqP} analog in Proposition~\ref{7720}.  The properties are merely presented in different orders because they are proved in different orders.  The exceptions are that properties (\ref{7572})--(\ref{7632}) have no \ct{CsqP} analogs in Proposition~\ref{7720}.}  Incidentally, property (\ref{7915}) and assumption [Cset1] together imply that each node in a \ct{CsetP} preform is actually a choice set, in accord with the terminology.  More significantly, property (\ref{7572}) shows that every \ct{CsetP} preform has no-absentmindedness.  In this sense the combination of [Cset1] and [Cset2] is restrictive.

\newcommand{\noteeqvts}{\footnote{Lemma~\ref{8103} shows the following are equivalent: [a] $c∉t$ and $t∪⎨c⎬{=}t\sh$.  [b] $t≠t\sh$ and $t∪⎨c⎬{=}t\sh$.  [c] $t≠t\sh$ and $t{=}t\sh⧷⎨c⎬$.  [d] $t⊆t\sh$ and $⎨c⎬{=}t\sh⧷t$.}}

\begin{prop}\label{7568} Suppose $(T,C,⊗)$ is a \ct{CsetP} preform with its $F$, $t^o$, $p$, $q$, $k$, $≺$, $≼$, and $\HH$.  Then the following hold.\begin{tlist}
\yl{7511} $t^o = ⎨⎬$.
\yl{7592} $(∀t\sh∈T⧷⎨⎨⎬⎬)$ $q(t\sh)⋅∉⋅p(t\sh)$ and $p(t\sh)∪⎨q(t\sh)⎬ = t\sh$.
\yl{7594} $(∀t∈T)$ $k(t) = |t|$. 
\yl{7596} $(∀t∈T,m∈⎨0,1,...\,|t|⎬)$ $p^m(t)\,⊆\,t$ {\em and} $t⧷p^m(t)$ $=$ $⎨\,q○p^n(t)\,|\,\!$ $m{>}n≥0\,⎬$.
\yl{7644} $(∀t∈T)⋅t = ⎨\,q○p^n(t)\,|\,|t|{>}n≥0\,⎬$.
\yl{7915} $C = ∪T$.
\yl{7572} $(T,C,⊗)$ has no-absentmindedness.
\yl{7645} $(∀t∈T,H∈\HH)$ $|t∩F(H)|⋅≤⋅1$.
\yl{7632} $(∀t^A∈T,t^B∈T)$ $t^A⋅⊆⋅t^B$ implies $t^A = p^{|t^B|-|t^A|}(t^B)$.
\yl{7660} $(∀t^A∈T,t^B∈T)⋅t^A⋅≺⋅t^B$ iff $t^A⋅⊊⋅t^B$.
\yl{7661} $(∀t^A∈T,t^B∈T)⋅t^A⋅≼⋅t^B$ iff $t^A⋅⊆⋅t^B$.
\yl{7567} $⊗\gr = ⎨⋅(t,c,t\sh)∈T×C×T⋅|⋅c∉t,⋅t∪⎨c⎬{=}t\sh⋅⎬$.\noteeqvts
\yl{7667} $F\gr = ⎨⋅(t,c)∈T×C⋅|⋅c∉t,⋅t∪⎨c⎬∈T⋅⎬$. (Proof~\ref{7568p}.)
\end{tlist} \end{prop}

Finally, let a {\em choice-set} \ct{NCF} form be an \ct{NCF} form whose preform is a \ct{CsetP} preform.  Then let \ct{CsetF} be the full subcategory of \ct{NCF} whose objects are choice-set \ct{NCF} forms.

\ssec{Isomorphic Enclosure}

\begin{nthm}\label{7967}  (a) \ct{CsqP_\ga} $\ud{→}$ \ct{CsetP}. In particular, suppose $\bar{Π} = (\dT,\dC,\bar{⊗})$ is a \ct{CsqP_\ga} preform.  Define $T = R(\dT)$, and define $⊗$ by surjectivity and $⊗\gr = ⎨\,(R(\dt),\dcc,R(\dt\sh))\,|\,(\dt,\dcc,\dt\sh)∈\bar{⊗}\gr\,⎬$.  Then $Π = (T,\dC,⊗)$ is a \ct{CsetP} preform, $R|_{\dT}$ is a bijection, and $[\bar{Π},Π,R|_{\dT},\id_{\dC}]$ is an \ct{NCP} isomorphism.  (b) \ct{CsqF_\ga} $\ud{→}$ \ct{CsetF}.  In particular, suppose $\bar{Φ} = (\dI,\dT,(\dC_\di)_{\di∈\dI},\bar{⊗})$ is a \ct{CsqF_\ga} form.  Define $T$ and $⊗$ as in part (a).  Then $Φ = (\dI,T,(\dC_\di)_{\di∈\dI},⊗)$ is a \ct{CsetF} form and $[\bar{Φ},Φ,\id_\dI,R|_{\dT},\id_{∪_{\di∈\dI}\dC_\di}]$ is an \ct{NCF} isomorphism. (Proof~\ref{7967p}.) \end{nthm}

\begin{ncrly}\label{8082} (a) \ct{CsqP_\ga} $\ud{⟷}$ \ct{CsetP}.  (b) \ct{CsqF_\ga} $\ud{⟷}$ \ct{CsetF}. (Proof here.) \end{ncrly}

\begin{pf} {\em (a)}.  This is very similar to (b).  Change ``form'' to ``preform'', \ct{F} to \ct{P}, (b) to (a), and the last phrase to ``because it has no-absentmindedness by Proposition~\ref{7568}(\ref{7572})''.

{\em (b)}.  Theorem~\ref{7967}(b) shows \ct{CsqF_\ga} $\ud{→}$ \ct{CsetF}.  Thus it remains to show \ct{CsqF_\ga} $\ud{←}$ \ct{CsetF}.  Since isomorphic enclosures can be composed, it suffices to show [1] \ct{CsetF} $\ud{→}$ \ct{NCF_\ga} and [2] \ct{NCF_\ga} $\ud{→}$ \ct{CsqF_\ga}.  [2] is the forward direction of Corollary~\ref{8083}(b).  [1] holds simply because any \ct{CsetF} form is a \ct{NCF_\ga} form.  To see this, take a \ct{CsetF} form.  It is an \ct{NCF} form by construction.  It has no-absentmindedness because its preform has no-absentmindedness by Proposition~\ref{7568}(\ref{7572}). \end{pf}

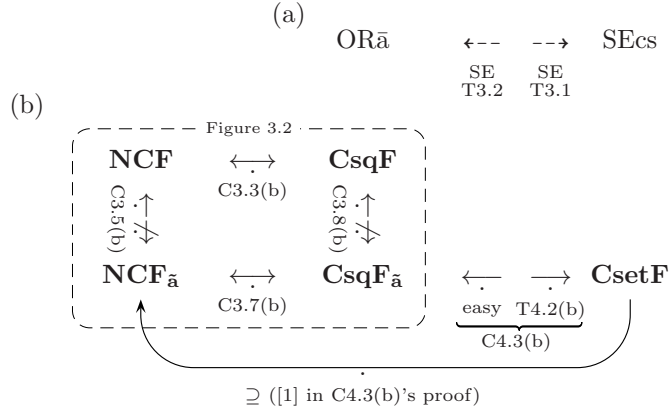
\renewcommand{\captionall}{(a) An ad hoc equivalence from SE.  (b) The previous figure, augmented with Corollary~\ref{8082}(b) and its proof.  T\,{=}\,Theorem. C\,{=}\,Corollary.}
\begin{figure}[h]
  \newcommand{\hgth}{155}  
  \begin{picture}(0,\hgth) 
  \put(-163,-13){\scalebox{1}{  
    \begin{pspicture}(-8,-5)(8,10) 
      \end{pspicture}
    }} \end{picture}
  \caption{\small \captionall} \label{8148} 
  \end{figure}

Corollary~\ref{8082}(b) is analogous to an ad hoc style equivalence in SE.  There, a pair of results argues that no-absentminded OR forms (``{\ORna} forms'' in this subsection) are equivalent to SE-choice-set forms (``SEcs forms'' in this subsection).  One of the results (SE Theorem~3.2) shows that an {\ORna} form can be reasonably derived from each SEcs form, and the other result (SE Theorem~3.1) shows that each {\ORna} form can be reasonably mapped to an SEcs form.  These two theorems are depicted by the two dashed arrows in Figure~\ref{8148}(a).

Corollary~\ref{8082}(b) strengthens this equivalence.  \ct{CsqF_\ga} forms are like {\ORna} forms in that both specify nodes as choice-sequences, and \ct{CsetF} forms are like SEcs forms in that both specify nodes as choice-sets.  Then, Corollary~\ref{8082}(b)'s isomorphic equivalence is a matching pair of results: one half (labelled ``easy'' in Figure~\ref{8148}(b)) shows that a \ct{CsqF_\ga} form is isomorphic to each \ct{CsetF} form, while the other half (Theorem~\ref{7967}) shows that each \ct{CsqF_\ga} form is isomorphic to a \ct{CsetF} form.  Thus Corollary~\ref{8082}(b) strengthens the SE equivalence by introducing isomorphisms.\footnote{There is also another sense in which Corollary~\ref{8082}(b) accords with the SE equivalence.  The forward half of the corollary is Theorem~\ref{7967}, and that theorem transforms choice-sequence nodes to choice-set nodes via the bijection $R|_{\dT}$.  That same bijection is used in SE Theorem~3.1.}

Corollary~\ref{8082}(b)'s proof highlights how useful it is to compose isomorphic enclosures.  In particular, consider the reverse direction of Corollary~\ref{8082}(b), which is \ct{CsqF_\ga} $\ud{←}$ \ct{CsetF} in Figure~\ref{8148}(b), and compare it with SE Theorem 3.2, which is {\ORna} $\dashleftarrow$ SEcs in Figure~\ref{8148}(a).  The lemmas and proof for SE Theorem~3.2 span six difficult pages.  In contrast, the reverse direction of Corollary~\ref{8082}(b) is proved in six lines by composing an easily-proved enclosure (\ct{CsetF} $\ud{→}$ \ct{NCF_\ga} in part [1] of proof) with a previously-proved enclosure (\ct{NCF_\ga} $\ud{→}$ \ct{CsqF_\ga} from the forward half of Corollary~\ref{8083}(b)).  Figure~\ref{8148}(b) shows this composition as the curved arrow followed by the forward direction of Corollary~\ref{8083}(b).

\ssec{More about Perfect-Information}

Corollaries \ref{8091} and \ref{8085} are additional applications of Section 3.3's general propositions using isomorphic invariance.

\begin{ncrly}\label{8091} (a) \ct{NCP_\ga} $\sencloses$ \ct{NCP_p}.  (b) \ct{NCF_\ga} $\sencloses$ \ct{NCF_p}. (Proof here.) \end{ncrly}

\begin{pf} {\em (a)}.  Consider Proposition~\ref{8090} at \ct{Z} equal to \ct{NCP}, when $w$ is the property of no-absentmindedness $\ga$, and $s$ is the property of perfect-information $p$.  Perfect-information is isomorphically invariant by Proposition~\ref{8046}(a).  Further no-absentmindedness is strictly weaker than perfect-information by notes \ref{8191} and \ref{8096}.  Thus Proposition~\ref{8090} implies that \ct{NCP_\ga} strictly isomorphically encloses \ct{NCP_p}.  {\em (b)}. This is very similar to (a).  Change \ct{P} to \ct{F}, and (a) to (b). \end{pf}

\begin{ncrly}\label{8085} (a) \ct{NCP_p} $\ud{⟷}$ \ct{CsqP_p} $\ud{⟷}$ \ct{CsetP_p}.  (b) \ct{NCF_p} $\ud{⟷}$ \ct{CsqF_p} $\ud{⟷}$ \ct{CsetF_p}. (Proof here.) \end{ncrly}

\begin{pf} {\em (a)}. Corollary~\ref{8083}(a) and Corollary~\ref{8082}(a) imply \ct{NCP_\ga} $\ud{⟷}$ \ct{CsqP_\ga} $\ud{⟷}$ \ct{CsetP}.  Thus, Propositions \ref{8081} and \ref{8046}(a) imply that \ct{NCP_{\ga p}} $\ud{⟷}$ \ct{CsqP_{\ga p}} $\ud{⟷}$ \ct{CsetP_p}, where \ct{NCP_{\ga p}} is the full subcategory of \ct{NCP} consisting of those objects that satisfy both no-absentmindedness and perfect-information, and where similarly \ct{CsqP_{\ga p}} is the full subcategory of \ct{CsqP} consisting of those objects that satisfy both no-absentmindedness and perfect-information.  Since no-absentmindedness is weaker than perfect-information (note \ref{8191}), \ct{NCP_{\ga p}} = \ct{NCP_p} and \ct{CsqP_{\ga p}} = \ct{CsqP_p}.  {\em (b)}.  This is very similar to (a).  Change \ct{P} to \ct{F}, and (a) to (b). \end{pf}

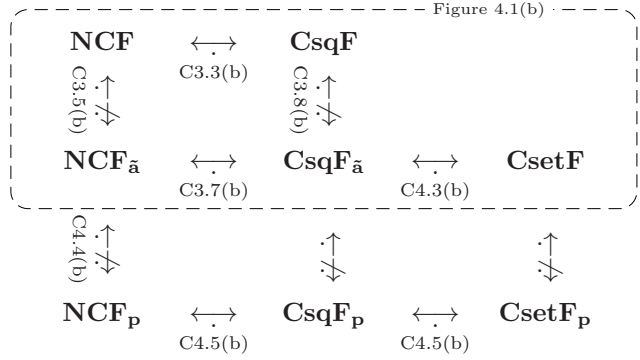
\renewcommand{\captionall}{Most of the previous figure, augmented with some results about perfect-information.  C\,=\,Corollary.}
\begin{figure}[h]
  \newcommand{\hgth}{136}  
  \begin{picture}(0,\hgth) 
  \put(-198,-14){\scalebox{1}{  
    \begin{pspicture}(-9,-4)(9,4) 
      \end{pspicture}
    }} \end{picture}
  \caption{\small \captionall} \label{8149} 
  \end{figure}

Incidentally, since isomorphic equivalence implies categorical equivalence, Corollary~\ref{8085}(a) implies \ct{NCP_p}, \ct{CsqP_p}, and \ct{CsetP_p} are categorically equivalent.  Further, SP Theorem 3.13 and Corollary~3.14 show that \ct{NCP_p}, \ct{Tree}, and \ct{Grph_{ca}} are categorically equivalent, where  \ct{Tree} is the category of functioned trees which SP uses in its development of \ct{NCP}, and where \ct{Grph_{ca}} is the full subcategory of \ct{Grph} whose objects are converging arborescences.  Thus, \ct{NCP_p}, \ct{CsqP_p}, \ct{CsetP_p}, \ct{Tree}, and \ct{Grph_{ca}} are categorically equivalent.

\myskip \pagebreak Figure~\ref{8149}'s arrow diagram illustrates most of the isomorphic-enclosure results from Sections 3.2 and following.  In addition, the diagram has some unlabelled arrows.  They are derived by composing arrows as in the proof of Corollary~\ref{8093}.  Many diagonal arrows could be similarly derived.

\section{Further Remarks} \markb{\sc 5. Further Remarks}

\ssec{Deducing consequences from an isomorphic enclosure}

Consider this paper's first isomorphic enclosure.  Theorem~\ref{8066} shows that each \ct{NCF} form $Φ$ is isomorphic to a \ct{CsqF} form $\bar{Φ}$ by means of an isomorphism which transforms nodes via the bijection $\bar{τ}$.  Proposition~\ref{6346} deduces many consequences from such an isomorphism.  For example, its part (\ref{8166}) implies that $(∀t^1∈T,t^2∈T)$ $t^1⋅≺⋅t^2$ iff $\bar{τ}(t^1)⋅\bar{≺}⋅\bar{τ}(t^2)$, where $T$ is the node set of $Φ$, $≺$ is derived from $Φ$, and $\bar{≺}$ is derived from $\bar{Φ}$. 
Although such consequences about form derivatives like $≺$ and $\bar{≺}$ are tantalizingly natural, the consequences about form derivatives in Proposition~\ref{6346}(\ref{8160})--(\ref{8170}) take about 10 pages to prove.  That work is important because such consequences are fundamental to drawing more conclusions from the isomorphic enclosure of \ct{NCF} in \ct{CsqF}.

As Section~3.2 explained, the isomorphic enclosure of \ct{NCF} in \ct{CsqF} is analogous to KL16 Theorem~1.  No consequences about form derivatives have been deduced from that ad hoc theorem, and an analog of Proposition~\ref{6346}(\ref{8160})--(\ref{8170}) would likely require about 10 pages to prove.  Moreover, like KL16 Theorem~1, no consequences about form derivatives have been deduced from KL16 Theorem~2 or from SE Theorems 3.1 and 3.2.  Each of these ad hoc theorems has its own formulation, so deriving analogs of Proposition~\ref{6346}(\ref{8160})--(\ref{8170}) for the three of them would likely require another $3×10 = 30$ pages.

In contrast, Proposition~\ref{6346}(\ref{8160})--(\ref{8170}) applies not only to the isomorphic enclosure of \ct{NCF} in \ct{CsqF}.  It applies to any isomorphic enclosure.  Thus it applies to all the arrows in Figure~\ref{8149}, as well as to all isomorphic enclosures in the future.

\ssec{Future research}

As discussed in Section 1.2, this paper is part of a larger agenda to translate game theory across specification styles.  In this larger context, isomorphic enclosures can be seen as a way to translate form components from one style to another, and on the basis of these isomorphic enclosures, Proposition~\ref{6346}(\ref{8160})--(\ref{8170}) (discussed just above) can be seen as a way of translating form derivatives from one style to another.  

The \xin{%
\begin{picture}(0,0)
\put(85,56){\color{white} \rule{40ex}{3ex}}
\put(111,60){\sc \SMALL 5. Further Remarks}
\end{picture}}%%%%%%%%%%%%%%%
results of this paper wait to be expanded in three orthogonal directions.

[1] There is more to translate beyond forms and their derivatives.  This would include properties that forms might satisfy, and theorems that might relate these properties to one another.  (This paper makes some limited progress in this direction by exploring the isomorphically invariant properties of no-absentmindedness and perfect-information, and by identifying some special properties of \ct{CsqF} forms and \ct{CsetF} forms via Propositions~\ref{7720} and \ref{7568}.)  Expanding in this direction would correspond to expanding the three substantive sections of this paper.

[2] This paper concerns only three styles: \ct{NCF}, \ct{CsqF}, and \ct{CsetF}.  There are other styles to explore, including the two neglected styles mentioned at the start of this paper, namely, the ``node-set'' style of Al\'os-Ferrer and Ritzberger 2016 Section 6.3, and the ``outcome-set'' style of von Neumann and Morgenstern 1944 and Al\'os-Ferrer and Ritzberger 2016 Section 6.2.  Expanding in this direction will require defining new \ct{NCF} subcategories for ``node-set'' forms and ``outcome-set'' forms, and will correspond to adding, to the present paper, two new sections for the two new subcategories.

[3] This paper concerns only forms, which need to be augmented with preferences in order to define games.  At the higher level of games, many more issues emerge.  To return to [1], there is more to translate, including equilibrium concepts and the theorems which might relate one equilibrium concept to another.  To return to [2], there will be more than five styles because there are alternative ways to specify preferences over the same form.  Expanding in this third direction will require building a new category for games that incorporates this paper's category for forms.

\appendix

\section{\ct{NCF}} \markb{\sc Appendix A. \ct{NCF}} 

\begin{lemma}\label{7646} Suppose $(T,C,⊗)$ is an \ct{NCP} preform with its $F$, $t^o$, $p$, $q$, and $\HH$.  Then the following hold. \begin{tlist}
\yl{7914} $|T|⋅≥⋅2$, $|C|⋅≥⋅1$, $|⊗\gr|⋅≥⋅1$.
\yl{7647} $(∀H∈\HH,c∈C)$ $c⋅∈⋅F(H)$ iff $F^{-1}(c) = H$.
\yl{7648} $(∀H∈\HH,t\sh∈T⧷⎨t^o⎬)$ $q(t\sh)⋅∈⋅F(H)$ iff $p(t\sh)⋅∈⋅H$. 
\end{tlist} \end{lemma} 

\begin{pf} {\em (\ref{7914})}.  In the paragraph after SP equation (1), remark [ii] shows that $(∄t∈T)$ $p(t) = t$.  Thus, since $p$ is nonempty by [T1], there are distinct $t^1⋅∈⋅T$ and $t^2⋅∈⋅T$ such that $t^1 = p(t^2)$.  Thus, by the definition of $p$ in [P2], there is $c⋅∈⋅C$ such that $(t^1,c,t^2)⋅∈⋅⊗\gr$.

\lstep{(\ref{7647})}. (Forward direction). Suppose \ilc{7717} $c⋅∈⋅C$, \il{7715} $H⋅∈⋅\HH$, and \il{7716} $c⋅∈⋅F(H)$.  \ref{7716} implies there is \il{7649} $t⋅∈⋅H$ such that \il{7650} $c⋅∈⋅F(t)$.  \ref{7650} implies \il{7651} $t⋅∈⋅F^{-1}(c)$.  Meanwhile, \ref{7717} and [P3] imply \il{8195} $F^{-1}(c)⋅∈⋅\HH$.  Since $\HH$ is a partition by [P3], \ref{7715} and \ref{8195} imply $H$ and $F^{-1}(c)$ are elements of the same partition.  Hence \ref{7649} and \ref{7651} imply $H = F^{-1}(c)$.

(Reverse direction).  Suppose $c⋅∈⋅C$, \ilc{7718} $H⋅∈⋅\HH$, and \il{7714} $F^{-1}(c) = H$.  Since $H$ belongs to a partition by \ref{7718} and [P3], there is \il{7713} $t⋅∈⋅H$.  \ref{7713} and \ref{7714} implies $t⋅∈⋅F^{-1}(c)$, which implies $c⋅∈⋅F(t)$.  This and \ref{7713} imply $c⋅∈⋅F(H)$.

\lstep{(\ref{7648})}.  (Forward direction).  Suppose $H⋅∈⋅\HH$, \ilc{7654} $t\sh⋅∈⋅T⧷⎨t^o⎬$ and \il{7652} $q(t\sh)⋅∈⋅F(H)$.  \ref{7652} and the forward direction of part (\ref{7647}) imply \il{7653} $F^{-1}(q(t\sh)) = H$.  Meanwhile, \ref{7654} and SP Proposition~3.1(b) imply $p(t\sh)⊗q(t\sh) = t\sh$.  This and [P1] imply $(p(t\sh),q(t\sh))⋅∈⋅F\gr$.  This implies $p(t\sh)⋅∈⋅F^{-1}(q(t\sh))$, which equals $H$ by \ref{7653}. 

(Reverse direction).  Suppose $H⋅∈⋅\HH$, \ilc{7711} $t\sh⋅∈⋅T⧷⎨t^o⎬$ and \il{7712} $p(t\sh)⋅∈⋅H$.  \ref{7711} and SP Proposition~3.1(b) imply $p(t\sh)⊗q(t\sh) = t\sh$.  This and [P1] imply $(p(t\sh),q(t\sh))$ $∈⋅F\gr$.  This implies $q(t\sh)⋅∈⋅F(p(t\sh))$.  This and \ref{7712} imply $q(t\sh)⋅∈⋅F(H)$. \end{pf}

\begin{lemma}\label{6349}%
\footnote{This lemma excerpts parts of proofs from SP.  In particular, the proof of part (a) rearranges part of SP Proof C.12's argument for SP Proposition 3.5, and the proof of the part (b) rearranges part of the argument for SP Lemma~C.17(e).}%
Suppose $α = [Π,Π′,τ,δ]$ is a preform morphism, where $Π = (T,C,⊗)$ determines $F$ and where $Π′ = (T′,C′,⊗′)$ determines $F′$.  Then the following hold.  (a) $(∀c∈C)⋅τ(F^{-1}(c))⋅⊆⋅(F′)^{-1}(δ(c))$.  (b) Suppose $α$ is an isomorphism.  Then $(∀c∈C)⋅τ(F^{-1}(c)) = (F′)^{-1}(δ(c))$. \end{lemma}

\begin{pf} \lstep{(a)}.  Take $c$.  I argue \begin{align}
\zz
τ(F^{-1}(c)) 
=&⋅⎨⋅t′∈T′⋅|⋅(∃t∈T)⋅t′{=}τ(t)⋅\text{and}⋅t∈F^{-1}(c)⋅⎬ \nt
=&⋅⎨⋅t′∈T′⋅|⋅(∃t∈T)⋅t′{=}τ(t)⋅\text{and}⋅(t,c)∈F\gr⋅⎬ \nt
⊆&⋅⎨⋅t′∈T′⋅|⋅(∃t∈T)⋅t′{=}τ(t)⋅\text{and}⋅(τ(t),δ(c))∈F′\gr⋅⎬ \nt
=&⋅⎨⋅t′∈T′⋅|⋅(∃t∈T)⋅t′{=}τ(t)⋅\text{and}⋅(t′,δ(c))∈F′\gr⋅⎬ \nt
⊆&⋅⎨⋅t′∈T′⋅|⋅(t′,δ(c))∈F′\gr⋅⎬ \nt
=&⋅(F′)^{-1}(δ(c)). \notag
\zz
\end{align} The first inclusion holds by (18a) of SP Lemma~C.6.  The second inclusion holds because $τ(T)⋅⊆⋅T′$ by [PM1].  The equalities are rearrangements.

\lstep{(b)}.  Take $c$.  I argue \begin{align}
\zz
τ(F^{-1}(c)) 
=&⋅⎨⋅t′∈T′⋅|⋅(∃t∈T)⋅t′{=}τ(t)⋅\text{and}⋅t∈F^{-1}(c)⋅⎬ \nt
=&⋅⎨⋅t′∈T′⋅|⋅(∃t∈T)⋅t′{=}τ(t)⋅\text{and}⋅(t,c)∈F\gr⋅⎬ \nt
=&⋅⎨⋅t′∈T′⋅|⋅(∃t∈T)⋅t′{=}τ(t)⋅\text{and}⋅(τ(t),δ(c))∈F′\gr⋅⎬ \nt
=&⋅⎨⋅t′∈T′⋅|⋅(∃t∈T)⋅t′{=}τ(t)⋅\text{and}⋅(t′,δ(c))∈F′\gr⋅⎬ \nt
=&⋅⎨⋅t′∈T′⋅|⋅(t′,δ(c))∈F′\gr⋅⎬ \nt
=&⋅(F′)^{-1}(δ(c)). \notag
\zz
\end{align} The third equality holds by SP Proposition~3.8(c).  The fifth holds because $τ$ is a bijection by SP Theorem~3.7 (second sentence).  The remaining equalities are rearrangements. \end{pf}

\begin{npf}[for Proposition~\ref{6295}]\label{6295p} {\em (a)}.  First I show \lic{hh00} $∪_{i∈I}X_i = X$ by arguing, in steps, that $∪_{i∈I}X_i$ by the definition of $(X_i)_{i∈I}$ equals $∪_{i∈I}(∪_{c∈C_i}F^{-1}(c))$; which by rearrangement equals $∪_{c∈∪_{i∈I}C_i}F^{-1}(c)$; which by the definition of $C$ equals $∪_{c∈C}F^{-1}(c)$; which by definition equals $F^{-1}(C)$; which by definition (in Section 2.1) equals $X$.  Thus it remains to show that $(∀i∈I,j∈I⧷⎨i⎬)$ $X_i∩X_j = ∅$.  Toward that end, suppose there are $i^1⋅∈⋅I$ and $i^2⋅∈⋅I$ such that \li{hh11} $i^1⋅≠⋅i^2$ and $X_{i^1}∩X_{i^2}⋅≠⋅∅$.  This nonemptiness and \ref{hh00} imply there is \li{hh33} $t⋅∈⋅X$ such that \li{hh21} $t⋅∈⋅X_{i^1}∩X_{i^2}$.  \ref{hh33} and [F3] imply there is $i^*⋅∈⋅I$ such that \il{hh66} $F(t)⋅⊆⋅C_{i^*}$. %
% For 1
\ref{hh21} implies $t⋅∈⋅X_{i^1}$, which by the definition of $X_{i^1}$ implies there is [6$^1$] $c^1⋅∈⋅C_{i^1}$ such that $t⋅∈⋅F^{-1}(c^1)$.  The previous set membership is equivalent to [7$^1$] $c^1⋅∈⋅F(t)$.  [7$^1$] and \ref{hh66} imply $c^1⋅∈⋅C_{i^*}$, and thus [6$^1$] and [F2] imply [8$^1$] $i^1 = i^*$.  Similarly, %
% For 2
\ref{hh21} implies $t⋅∈⋅X_{i^2}$, which by the definition of $X_{i^2}$ implies there is [6$^2$] $c^2⋅∈⋅C_{i^2}$ such that $t⋅∈⋅F^{-1}(c^2)$.  The previous set membership is equivalent to [7$^2$] $c^2⋅∈⋅F(t)$.  [7$^2$] and \ref{hh66} imply $c^2⋅∈⋅C_{i^*}$, and thus [6$^2$] and [F2] imply [8$^2$] $i^2 = i^*$. %
% Finish
[8$^1$] and [8$^2$] imply $i^1 = i^2$, which contradicts \ref{hh11}.

{\em (b)}.  Take $i$.  First I show \lic{gg11} $\HH_i⋅⊆⋅\HH$.  I do this by arguing, in steps, that $\HH_i$ by definition equals $⎨F^{-1}(c)|c∈C_i⎬$; which by the definition of $C$ is a subset of $⎨F^{-1}(c)|c∈C⎬$; which by definition (in [P3]) equals $\HH$.  Since $\HH$ is a partition by [P3], \ref{gg11} implies that the elements of $\HH_i$ are nonempty and disjoint.  Thus it remains to show that $∪\HH_i = X_i$.  I argue, in steps, that $∪\HH_i$ by the definition of $\HH_i$ equals $∪⎨F^{-1}(c)|c∈C_i⎬$; which by the definition of $X_i$ equals $X_i$.

{\em (c)}.  First I show $∪_{i∈I}\HH_i = \HH$.  I do this by arguing, in steps, that $∪_{i∈I}\HH_i$ by the definition of $(\HH_i)_{i∈I}$ equals $∪_{i∈I}⎨F^{-1}(c)|c∈C_i⎬$; which by rearrangement equals $⎨F^{-1}(c)|c∈∪_{i∈I}C_i⎬$; which by the definition of $C$ equals $⎨F^{-1}(c)|c∈C⎬$; which by definition (in [P3]) equals $\HH$.  Thus it remains to show $(∀i∈I,j∈I⧷⎨i⎬)⋅\HH_i∩\HH_j⋅≠⋅∅$.  Toward that end, suppose $i^1⋅∈⋅I$ and $i^2⋅∈⋅I$ satisfy \lic{ii21} $i^1⋅≠⋅i^2$ and $\HH_{i^1}∩\HH_{i^2}⋅≠⋅∅$.  This nonemptiness implies there is $H⋅∈⋅\HH_{i^1}∩\HH_{i^2}$.  $H⋅∈⋅\HH_{i^1}$ and part (b) implies $H$ is a nonempty subset of $X_{i^1}$.  Similarly, $H⋅∈⋅\HH_{i^2}$ and part (b) implies $H$ is a nonempty subset of $X_{i^2}$.  The previous two sentences imply $X_{i^1}∩X_{x^2}⋅≠⋅∅$.  Hence part (a) implies $i^1 = i^2$, which contradicts \ref{ii21}. \end{npf}

\begin{npf}[for Proposition~\ref{6309}]\label{6309p}  The next two paragraphs prove the first paragraph of the proposition.  In particular, the next two paragraphs show that $[Φ,Φ′,ι,τ,δ]$ is a morphism iff (\ref{8173m})--(\ref{8159m}) hold.

{\em Forward Direction}. Assume $[Φ,Φ′,ι,τ,δ]$ is a morphism.  Then [FM1] implies $[(T,C,⊗),(T′,C′,⊗′),τ,δ]$ is a preform morphism, so [PM1] implies (\ref{8172m}), [PM2] implies (\ref{8169m}), and [PM3] implies (\ref{8159m}).  Further, [FM2] implies (\ref{8173m}), and [FM3] implies (\ref{8040m}).  

{\em Reverse Direction}. Assume (\ref{8173m})--(\ref{8159m}).  Since $Φ$ and $Φ′$ are forms by assumption, it suffices to show [FM1]--[FM3].  [FM3] holds by (\ref{8040m}).  [FM2] holds by (\ref{8173m}).  For [FM1], note that $Π$ and $Π′$ are preforms by [F1] and the assumption that $Φ$ and $Φ′$ are forms.  Thus it suffices to show [PM1]--[PM3].  [PM1] holds by (\ref{8172m}), [PM2] holds by (\ref{8169m}), and [PM3] holds by (\ref{8159m}).

\vspace{1mm} Henceforth assume that $[Φ,Φ′,ι,τ,δ]$ is a morphism.  The remaining two paragraphs of the proposition follow from Claims \ref{8185}, \ref{8184}, \ref{8178}, and \ref{8177} below.

\begin{cllist} \yl{8185} {\em (\ref{8041m}) holds}.  Take $i$.  I argue, in steps, that $τ(X_i)$ by definition equals $τ(∪_{c∈C_i}F^{-1}(c))$, which by rearrangement equals $∪⎨\,τ(F^{-1}(c))\,|$ $c∈C_i\,⎬$, which by Lemma~\ref{6349}(a) is included in $∪⎨\,(F′)^{-1}(δ(c))\,|\,c∈C_i\,⎬$, which by rearrangement is $∪⎨\,(F′)^{-1}(c′)\,|\,c′∈δ(C_i)\,⎬$, which by [FM3] is included in $∪⎨\,(F′)^{-1}(c′)\,|\,c′∈C′_{ι(i)}\,⎬$, which by definition is $X′_{ι(i)}$.

\yl{8184} {\em (\ref{8042m}) holds}.  Take $i$ and $H⋅∈⋅\HH_i$.  By the definition of $\HH_i$, there exists \lic{ii99} $c⋅∈⋅C_i$ such that \li{ii98} $H = F^{-1}(c)$.  Let \li{ii97} $H′ = (F′)^{-1}(δ(c))$.  \ref{ii99} and [FM3] imply $δ(c)⋅∈⋅C′_{ι(i)}$.  Thus the definition of $\HH′_{ι(i)}$ implies $(F′)^{-1}(δ(c))⋅∈⋅\HH′_{ι(i)}$.  This and \ref{ii97} imply $H′⋅∈⋅\HH′_{ι(i)}$.  Thus it remains to show that $τ(H)⋅⊆⋅H′$. I argue, in steps, that $τ(H)$ by \ref{ii98} equals $τ(F^{-1}(c))$, which by Lemma~\ref{6349}(a) is included in $(F′)^{-1}(δ(c))$, which by \ref{ii97} equals $H′$.

\yl{8175} {\em (a)} $[Π,Π,τ,δ]$ {\em is an \ct{NCP} morphism. (b)} $[(T,p),(T′,p′),τ]$ {\em is a \ct{Tree} morphism}.  (a) follows from [FM1].  For (b), note that (a) and SP Theorem 3.9 imply that $\FA([Π,Π′,τ,δ])$ is a \ct{Tree} morphism.  By that theorem's definition of $\FB$, $\FA([Π,Π′,τ,δ]) = [\FO(Π),\FO(Π′),τ] = [(T,p),(T′,p′),τ]$. 

\yl{8178} {\em (\ref{8160m}), (\ref{8162m}), (\ref{8163m}), (\ref{8164m}), and (\ref{8171m}) hold}.  Because of Claim~\ref{8175}(a), these parts follow from various results in SP.  In particular, %
(\ref{8160m}) follows from SP Lemma~C.6(18a). %
(\ref{8162m}) follows from SP Lemma~C.9(20a). %
(\ref{8163m}) follows from SP Lemma~C.9(20b). %
(\ref{8164m}) follows from SP Proposition~3.4(22a) since Section~2.1 defines $X$ equal to $F^{-1}(C)$ and thus $X′$ equal to $(F′)^{-1}(C′)$. %
(\ref{8171m}) follows from SP Proposition~3.5.  

\yl{8177} {\em (\ref{8161m}) and (\ref{8165m})--(\ref{8170m}) hold}.  Because of Claim~\ref{8175}(b), these parts of the proposition follow from various parts of SP Proposition~2.4.  In particular, %
(\ref{8161m}) follows from SP Proposition~2.4(a). %
(\ref{8165m})--(\ref{8167m}) follow from SP Proposition~2.4(c)--(e). %
(\ref{8168m}) follows from SP Proposition~2.4(h). %
(\ref{8170m}) follows from SP Proposition~2.4(g). \end{cllist}\qedup\end{npf}

\begin{npf}[for Theorem~\ref{6316}]\label{6316p} 
 The next two paragraphs draw upon SP Theorem~3.6, which showed that \ct{NCP} is a well-defined category.

This paragraph shows that, for each form $Φ$, $\id_Φ$ is a form morphism.  Toward this end, take a form $Φ = [I,T,(C_i)_{i∈I},⊗]$.  By [F1], let $Π = (T,∪_{i∈I}C_i,⊗)$ be its \ct{NCP} preform.  It must be shown that $\id_Φ = [Φ,Φ,\id_T,\id_I,\id_{∪_{i∈I}C_i}]$ satisfies [FM1]--[FM3].  [FM1] holds because $[Π,Π,\id_T,\id_{∪_{i∈I}C_i}]$ is an \ct{NCP} identity, and hence, an \ct{NCP} morphism.  [FM2] holds because $\id_I{:}I→I$.  [FM3] holds because $(∀j∈I)$ $\id_{∪_{i∈I}C_i}(C_j) = C_j = C_{\id_I(j)}$.

This paragraph shows that, for any two form morphisms $β$ and $β′$, $β′○β$ is a form morphism.  Toward this end, take form morphisms $β = [Φ,Φ′,ι,τ,δ]$ and $β′ = [Φ′,Φ″,ι′,τ′,δ′]$, where $Φ = (I,T,(C_i)_{i∈I},⊗)$, $Φ′ = (I′,T′,(C′_{i′})_{i′∈I′},⊗′)$, and $Φ″ = (I″,T″,(C″_{i″})_{i″∈I″},⊗″)$.  By [F1], let $Π$, $Π′$, and $Π″$ be the NCP preforms underlying $Φ$, $Φ′$, and $Φ″$.  It must be shown that $β′○β = [Φ,Φ″,τ′○ι,τ′○ι,δ′○δ]$ satisfies [FM1]--[FM3].  For [FM1], it must be shown that the quadruple $[Π,Π″,τ′○τ,δ′○δ]$ is an \ct{NCP} morphism.  This holds because [a] the quadruple equals $[Π′,Π″,τ′,δ′]○$ $[Π,Π′,τ,δ]$ in \ct{NCP}, and because [b] $[Π,Π′,τ,δ]$ and $[Π′,Π″,τ′,δ′]$ are \ct{NCP} morphisms by [FM1] for $β$ and $β′$.  For [FM2], it must be shown that $ι′○ι{:}I→I″$.  This holds because $ι{:}I→I′$ by [FM2] for $β$, and because $ι′{:}I′→I″$ by [FM2] for $β′$.  For [FM3], it must be shown that $(∀i∈I)$ $(δ′○δ)(C_i)⋅⊆⋅C″_{ι′○ι(i)}$.  To prove this, take $i$.  I argue $δ′(δ(C_i))$ $⊆$ $δ′(C′_{ι(i)})$ $⊆$ $C″_{ι′○ι(i)}$, where the first inclusion holds because $δ(C_i)⋅⊆⋅C′_{ι(i)}$ by [FM3] for $β$, applied at $i$, and where the second inclusion holds by [FM3] for $β′$, applied at $i′ = ι(i)$. 

The previous two paragraphs have established the well-definition of identity and composition.  The unit and associative laws are immediate.  Thus \ct{NCF} is a category (e.g.\ Mac Lane 1998, page~10).\nocite{MacL98} \end{npf}

\begin{lemma}\label{6348} Suppose $[Φ,Φ′,ι,τ,δ]$ is a morphism, where $Φ = (I,T,(C_i)_{i∈I},⊗)$ and $Φ′ = (I′,T′,(C′_{i′})_{i′∈I′},⊗′)$.  Further suppose that $ι$ and $δ$ are bijections.  Then the following hold. 
\nl(a) $(∀i∈I)⋅δ|_{C_i}$ is a bijection from $C_i$ onto $C′_{ι(i)}$. 
\nl(b) $(∀i′∈I′)⋅δ^{-1}|_{C′_{i′}}$ is a bijection from $C′_{i′}$ onto $C_{ι^{-1}(i′)}$. \end{lemma}

\begin{pf} Define $C = ∪_{i∈I}C_i$ and $C′ = ∪_{i′∈I′}C′_{i′}$.  The lemma follows from Claims~\ref{8130} and \ref{8131}.

\begin{cllist} \yl{8132} $δ$ {\em is a bijection from} $∪_{i∈I}C_i$ {\em onto} $∪_{i′∈I′}C′_{i′}$.  [FM1] implies [PM2], which implies $δ$ is a function from $C$ to $C′$.  Thus the definitions of $C$ and $C′$ imply $δ$ is a function from $∪_{i∈I}C_i$ to $∪_{i′∈I′}C′_{i′}$.  $δ$ is a bijection by assumption.

\yl{8133} $(∀i∈I)⋅δ(C_i) = C′_{ι(i)}$.  Take $i$.  [FM3] implies $δ(C_i)⋅⊆⋅C′_{ι(i)}$.  Thus it remains to show that $C′_{ι(i)}⧷δ(C_i) = ∅$.  Toward that end, suppose contrariwise there is $c′$ such that \ilc{as00} $c′⋅∈⋅C′_{ι(i)}$ and \il{as01} $c′⋅∉⋅δ(C_i)$.  \ref{as00} and Claim~\ref{8132} implies that $δ^{-1}(c′)$ is a well-defined element of $∪_{k∈I}C_k$.  Thus there is $j⋅∈⋅I$ such that $δ^{-1}(c′)⋅∈⋅C_j$.  This implies \il{as66} $c′⋅∈⋅δ(C_j)$.  \ref{as66} and \ref{as01} imply \il{as67} $i⋅≠⋅j$.  Also, \ref{as66} and [FM3] imply $c′⋅∈⋅C′_{ι(j)}$.  This and \ref{as00} imply \il{as11} $c′⋅∈⋅C′_{ι(i)}∩C′_{ι(j)}$.  Meanwhile, \ref{as67} and the bijectivity of $ι$ imply \il{as22} $ι(i)⋅≠⋅ι(j)$.  \ref{as11} and \ref{as22} contradict [F2] for $Φ′$.

\yl{8130} {\em (a) holds}.  This follows from the bijectivity of $δ$ and Claim~\ref{8133}.

\yl{8131} {\em (b) holds}.  Since $ι$ is bijective, it suffices to prove that $(∀i∈I)$ $δ^{-1}|_{C′_{ι(i)}}$ is a bijection from $C′_{ι(i)}$ onto $C_i$.  By Claim~\ref{8133}, this is equivalent to proving that $(∀i∈I)$ $δ^{-1}|_{δ(C_i)}$ is a bijection from $δ(C_i)$ onto $C_i$.  This follows from part (a). \end{cllist}\qedup\end{pf}

\begin{npf}[for Theorem~\ref{6315}]\label{6315p} Let the components of $Φ$ be $(I,T,(C_i)_{i∈I},⊗)$, define $C = ∪_iC_i$, let the components of $Φ′$ be $(I′,T′,(C′_{i′})_{i′∈I′},⊗′)$, and define $C′ = ∪_{i′}C′_{i′}$.

\lstep{The forward half of (a) and all of (b)}.  Suppose that $β$ is an isomorphism (Awodey 2010, page~12, Definition~1.3).  Recall that $β = [Φ,Φ′,ι,τ,δ]$ and let $β^{-1} = [Φ^*,Φ^{**},ι^*,τ^*,δ^*]$.  Then \ci
\zz ⋅\\[-3mm]
\lic{6335} $[Φ^*,Φ^{**},ι^*,τ^*,δ^*]○[Φ,Φ′,ι,τ,δ] = \id_Φ = [Φ,Φ,\id_I,\id_{T},\id_{C}]$ and \\[1mm]
\li{6336} $[Φ,Φ′,ι,τ,δ]○[Φ^*,Φ^{**},ι^*,τ^*,δ^*] = \id_{Φ′} = [Φ′,Φ′,\id_{I′},\id_{T′},\id_{C′}]$, \\[-3mm]⋅
\zz 
\co where the first equality in both lines holds by the definition of $β^{-1}$, and the second equality in both lines holds by the definition of $\id$.  The well definition of $○$ in \ref{6335} implies \ilc{6337} 
$Φ^* = Φ′$.  Analogously, the well definition of $○$ in \ref{6336} implies \il{6338} $Φ^{**} = Φ$.  The third component of \ref{6335} implies $ι^*○ι = \id_{I}$, and the third component of \ref{6336} implies $ι○ι^* = \id_{I′}$.  Thus $ι$ is a bijection from $I$ onto $I′$ and \il{6339} $ι^* = ι^{-1}$.  Similarly, the fourth components of \ref{6335} and \ref{6336} imply $τ$ is a bijection from $T$ onto $T′$ and \il{6340} $τ^* = τ^{-1}$.  Similarly again, the fifth components of \ref{6335} and \ref{6336} imply $δ$ is a bijection from $C$ onto $C′$ and \il{6341} $δ^* = δ^{-1}$.  To conclude, the previous three sentences have shown that $ι$, $τ$, and $δ$ are bijections.  Further, \begin{gather} 
\zz
β^{-1} = [Φ^*,Φ^{**},ι^*,τ^*,δ^*] = [Φ′,Φ,ι^{-1},τ^{-1},δ^{-1}], \notag
\zz
\end{gather} where the first equality follows from the definition of $β^{-1}$, and where the second equality follows from \ref{6337}--\ref{6341}.

\lstep{The reverse half of (a)}. Suppose that $ι$, $τ$, and $δ$ are bijections.  Define $β^* = [Φ′,Φ,ι^{-1},τ^{-1},δ^{-1}]$.  Derive $Π$ from $Φ$ and $Π′$ from $Φ′$.  The remainder of this paragraph will show that $β^*$ is a form morphism by showing that it satisfies \begin{gather}
\zz
\begin{maneq} \text{[FM1$′$]} & [Π′,Π,τ^{-1},δ^{-1}]⋅\text{is a preform morphism}, \end{maneq} \nt
\begin{maneq} \text{[FM2$′$]} & ι^{-1}{:}I′→I, \text{and} \end{maneq} \nt
\begin{maneq} \text{[FM3$′$]} & (∀i′∈I′)⋅δ^{-1}(C′_{i′})⋅⊆⋅C_{ι^{-1}(i′)}. \end{maneq} \notag 
\zz
\end{gather} To see [FM1$′$], first note that $[Π,Π′,τ,δ]$ is a preform morphism by [FM1] for $β$.  Thus the bijectivity of $τ$ and $δ$, together with SP Theorem 3.7(a), imply that $[Π′,Π,τ^{-1},δ^{-1}]$ is an \ct{NCP} isomorphism.  Hence a fortiori, it is a preform morphism.  To see [FM2$′$], first note that $ι{:}I→I′$ by [FM2] for $β$.  Thus the bijectivity of $ι$ implies that $ι^{-1}{:}I′→I$.  Finally, to see [FM3$′$], consider  Lemma~\ref{6348}.  The lemma's assumptions are met because the theorem assumes that $β = [Φ,Φ′,ι,τ,δ]$ is a morphism and because the start of this paragraph assumes that $ι$ and $δ$ are bijections.  Thus the lemma's part (b) implies that $(∀i′∈I′)$ $δ^{-1}(C′_{i′}) = C_{ι^{-1}(i′)}$.  

To conclude, $β^*$ is a form morphism by the previous paragraph.  Further, \begin{gather}
\zz
β^*○β = [Φ′,Φ,ι^{-1},τ^{-1},δ^{-1}]○[Φ,Φ′,ι,τ,δ] = \id_Φ⋅⋅\text{and} \nt 
β○β^* = [Φ,Φ′,ι,τ,δ]○[Φ′,Φ,ι^{-1},τ^{-1},δ^{-1}] = \id_{Φ′}. \notag
\zz
\end{gather} Hence $β$ is an isomorphism (and incidentally, $β^{-1} = β^*$).  \end{npf}

\begin{lemma}\label{6350} Suppose $[Φ,Φ′,ι,τ,δ]$ is a morphism, where $Φ =$ $(I,T,(C_i)_{i∈I},⊗)$ determines $(\HH_i)_{i∈I}$, and where $Φ′ = (I′,T′,(C′_{i′})_{i′∈I′},⊗′)$ determines $(\HH′_{i′})_{i′∈I′}$.  Further suppose that $[Π,Π′,τ,δ]$ is an isomorphism, where $Π = (T,C,⊗)$, $C = ∪_{i∈I}C_i$, $Π′ = (T′,C′,⊗′)$, and $C′ = ∪_{i′∈I′}C′_{i′}$.  Then $(∀i∈I,H∈\HH_i)⋅τ(H)⋅∈⋅\HH′_{ι(i)}$.  \end{lemma}

\begin{pf} Derive $F$ from $Π$, and $F′$ from $Π′$.  Since $[Π,Π′,τ,δ]$ is an isomorphism, Lemma~\ref{6349}(b) implies \lic{cf00} $(∀c∈C)$ $τ(F^{-1}(c)) = (F′)^{-1}(δ(c))$.

Now take $i$ and $H⋅∈⋅\HH_i$.  Then there is \li{cf10} $c^*⋅∈⋅C_i$ such that \li{cf11} $H = F^{-1}(c^*)$.  First I show \li{cf12} $δ(c^*)⋅∈⋅C′_{ι(i)}$ by arguing, in steps, that $δ(c^*)$ by \ref{cf10} belongs to $δ(C_i)$, which by [FM3] is included in $C′_{ι(i)}$.  Finally, I argue, in steps, that $τ(H)$ by \ref{cf11} equals $τ(F^{-1}(c^*))$, which by \ref{cf00} equals $(F′)^{-1}(δ(c^*))$, which by \ref{cf12} belongs to $\HH′_{ι(i)}$. \end{pf}

\begin{npf}[for Proposition~\ref{6346}]\label{6346p} The proposition follows from Claims \ref{8179}--\ref{8182} and \ref{8183}--\ref{8181}. \begin{cllist}

\yl{8179} {\em (\ref{8173})--(\ref{8169}) hold}.  The forward direction of Theorem~\ref{6315}(a) implies that $ι$, $τ$, and $δ$ are bijections.

\yl{8192} {\em (\ref{8040}) holds}.  This follows from Lemma~\ref{6348}(a). 

\yl{8193} {\em (\ref{8041}) holds}.  Take $i$.  Since $τ$ is a bijection by Claim~\ref{8179} (part (\ref{8172})), it suffices to argue that \begin{gather}
\zz
τ(X_i) = ∪⎨⋅τ(F^{-1}(c))⋅|⋅c∈C_i⋅⎬ \nt
= ∪⎨⋅(F′)^{-1}(δ(c))⋅|⋅c∈C_i⋅⎬ \nt
= ∪⎨⋅(F′)^{-1}(c′)⋅|⋅c′∈C′_{ι(i)}⋅⎬ = X′_{ι(i)}. \notag
\zz
\end{gather} The first equality holds by the definition of $X_i$ and a rearrangement.  The second equality follows from Lemma~\ref{6349}(b) because $[Π,Π′,τ,δ]$ is an isomorphism by Corollary~\ref{6333}.  The third equality holds by Claim~\ref{8192} (part (\ref{8040})).  The fourth equality holds by the definition of $X′_{ι(i)}$.

\yl{8182} {\em (\ref{8042}) holds}.  Take $i$.  Since $[Π,Π′,τ,δ]$ is an isomorphism by Corollary~\ref{6333}, Lemma~\ref{6350} implies that $τ|_{\HH_i}$ is a well-defined function from $\HH_i$ into $\HH′_{ι(i)}$.  It is injective because $τ$ is injective by Claim~\ref{8179} (part (\ref{8172})).  To show that it is surjective, take $H′⋅∈⋅\HH_{ι(i)}$.  Since $[Φ′,Φ,ι^{-1},τ^{-1},δ^{-1}]$ is an isomorphism by Theorem~\ref{6315}(b), $[Π′,Π,τ^{-1},δ^{-1}]$ is an isomorphism by Corollary~\ref{6333}.  Thus Lemma~\ref{6350} can be applied to $[Φ′,Φ,τ^{-1},ι^{-1},δ^{-1}]$.  Therefore $H′⋅∈⋅\HH_{ι(i)}$ implies $τ^{-1}(H′)⋅∈⋅\HH_{ι^{-1}○ι(i)}$.  Hence $τ^{-1}(H′)⋅∈⋅\HH_i$.  This implies that $τ(τ^{-1}(H′)) = H′$ is in the range of $τ|_{\HH_i}$. 

\yl{8180} {\em (a). $[Π,Π′,τ,δ]$ is an \ct{NCP} isomorphism, where $Π = (T,C,⊗)$ and $Π = (T′,C′,⊗′)$.}  {\em (b) $[(T,p),(T′,p′),τ]$ is a \ct{Tree} isomorphism.}  (a) holds by Corollary~\ref{6333}.  For (b), note that (a) and SP Theorem 3.9 imply  $\FA([Π,Π′,τ,δ])$ is a \ct{Tree} isomorphism.  By that theorem's definition of $\FB$, $\FA([Π,Π′,τ,δ]) = [\FO(Π),\FO(Π′),τ] = [(T,p),(T′,p′),τ]$.

\yl{8183} {\em (\ref{8159}), (\ref{8160}), (\ref{8163}), (\ref{8164}), and (\ref{8171}) hold}.  These hold by Claim~\ref{8180}(a) and the parts of SP Proposition~3.8.  In particular, %
(\ref{8159}) holds by SP Proposition~3.8(b). %
(\ref{8160}) holds by SP Proposition~3.8(c). %
(\ref{8163}) holds by SP Proposition~3.8(d). %
(\ref{8164}) holds by SP Proposition~3.8(a) since Section 2.1 defines $X$ as $F^{-1}(C)$ and thus $X′$ as $(F′)^{-1}(C′)$. %
(\ref{8171}) holds by SP Proposition~3.8(e). 

\yl{8181} {\em (\ref{8161}), (\ref{8162}), and (\ref{8165})--(\ref{8170}) hold}.  These hold by Claim~\ref{8180}(b) and various parts of SP Proposition~2.7.  In particular, %
(\ref{8161}) holds by SP Proposition~2.7(c). %
(\ref{8162}) holds by SP Proposition~2.7(e). %
(\ref{8165}) holds by SP Proposition~2.7(d). %
(\ref{8166})--(\ref{8170}) hold by SP Proposition~2.7(f)--(i). %
\end{cllist}\qedup\end{npf}

\begin{npf}[for Theorem~\ref{6318}]\label{6318p} By [F1], $\PO$ maps any form into a preform.  By [FM1], $\PA$ maps any form morphism into a preform morphism.  Thus it suffices to show that $\PB$ preserves source, target, identity, and composition (Mac Lane 1998 page 13).  This is done in the following four claims.

\begin{cllist} \yl{8135} $\PA(β)^\src = \PO(β^\src)$.  Take $β = [Φ,Φ′,ι,τ,δ]$.  Then I argue, in steps, that $\PA(β)^\src$ by the definition of $β$ is equal to $\PA([Φ,Φ′,ι,τ,δ])^\src$, which by the definition of $\PA$ is equal to $[\PO(Φ),\PO(Φ′),τ,δ]^\src$, which by the definition of $\src$ in \ct{NCP} is equal to $\PO(Φ)$, which by the definition of $\src$ in \ct{NCF} is equal to $\PO([Φ,Φ′,ι,τ,δ]^\src)$, which by the definition of $β$ is equal to $\PO(β^\src)$.

\yl{8136} $\PA(β)^\trg = \PO(β^\trg)$.  This is very similar to Claim~\ref{8135}.  Simply change $\src$ to $\trg$.

\yl{8137} $\PA(\id_Φ) = \id_{\PO(Φ)}$.  Take $Φ = (I,T,(C_i)_{i∈I},⊗)$ and let $C = ∪_iC_i$.  First I show [a] $\PO(Φ) = (T,C,⊗)$ by arguing, in steps, that $\PO(Φ)$ by the definition of $Φ$ is $\PO(I,T,(C_i)_{i∈I},⊗)$, which by the definition of $\PO$ is $(T,∪_{i∈I}C_i,⊗)$, which by the definition of $C$ is $(T,C,⊗)$.  Then I argue, in steps, that $\PA(\id_Φ)$ by the definition of $\id$ in \ct{NCF} is equal to $\PA([Φ,Φ,\id_{I},\id_{T},\id_{C}])$, which by the definition of $\PA$ is equal to $[\PO(Φ),\PO(Φ),\id_{T},\id_{C}]$, which by [a] is equal to $[(T,C,⊗),(T,C,⊗),\id_{T},\id_{C}]$, which by the definition of $\id$ in \ct{NCP} is equal to $\id_{(T,C,⊗)}$, which by [a] is equal to $\id_{\PO(Φ)}$.

\yl{8138} $\PA(β′○β) = \PA(β′)○\PA(β)$.  Take $β = [Φ,Φ′,ι,τ,δ]$ and $β′ = [Φ′,Φ″,$ $ι′,τ′,δ′]$.  First I note that, since $\PA$ is well-defined by the first paragraph, $\PA([Φ,Φ′,$ $ι,τ,δ]) = [\PO(Φ),\PO(Φ′),τ,δ]$ and $\PA([Φ′,Φ″,ι′,τ′,δ′]) = [\PO(Φ′),\PO(Φ″),τ′,δ′]$ are preform morphisms.  Then I argue that \begin{align}
\zz
\PA(β′○β)
=&⋅\PA([Φ′,Φ″,ι′,τ′,δ′]○[Φ,Φ′,ι,τ,δ]) \nt
=&⋅\PA([Φ,Φ″,ι′○ι,τ′○τ,δ′○δ]) \nt
=&⋅[\PO(Φ),\PO(Φ″),τ′○τ,δ′○δ] \nt
=&⋅[\PO(Φ′),\PO(Φ″),τ′,δ′]○[\PO(Φ),\PO(Φ′),τ,δ] \nt
=&⋅\PA([Φ′,Φ″,ι′,τ′,δ′]○\PA[Φ,Φ′,ι,τ,δ]) \nt
=&⋅\PA(β′)○\PA(β), \notag
\zz
\end{align} where the first equality holds by the definitions of $β$ and $β′$, the second by the definition of $○$ in \ct{NCF}, the third by the definition of $\PA$, the fourth by the previous sentence and by the definition of $○$ in \ct{NCP}, the fifth by the definition of $\PA$, and the sixth by the definitions of $β$ and $β′$. \end{cllist}\qedup\end{npf}

\begin{npf}[for Proposition~\ref{7710}]\label{7710p} {\em (a$^{\text{o}}$)}. Suppose $[Π,Π′,τ,δ]$ is a preform morphism, with $Π =(T,C,⊗)$ determining $≺$ and $\HH$, and with $Π′ = (T′,C′,⊗′)$ determining $≺′$ and $\HH′$.  It suffices to show that the absentmindedness of $Π$ implies the absentmindedness of $Π′$.  Toward that end, suppose $Π$ has absentmindedness.  Then there are \lic{ee11} $H⋅∈⋅\HH$, \li{ee12} $t^A⋅∈⋅H$, and \li{ee13} $t^B⋅∈⋅H$ such that \li{ee14} $t^A⋅≺⋅t^B$.  \ref{ee11} and SP Proposition~3.5 imply there exists \li{ee31} $H′⋅∈⋅\HH′$ such that \li{ee21} $τ(H)⋅⊆⋅H′$.  \ref{ee12} implies $τ(t^A)⋅∈⋅τ(H)$ and thus \ref{ee21} implies \li{ee32} $τ(t^A)⋅∈⋅H′$.  Similarly, \ref{ee13} implies $τ(t^B)⋅∈⋅τ(H)$ and thus \ref{ee21} implies \li{ee33} $τ(t^B)⋅∈⋅H′$.  In addition, \ref{ee14} and SP Proposition~2.4(d) (via SP Corollary~3.10) imply \li{ee24} $τ(t^A)⋅≺′⋅τ(t^B)$.  \ref{ee31}, \ref{ee32}, \ref{ee33}, and \ref{ee24} imply $Π′$ has absentmindedness.

{\em(a)}.  Suppose $Π$ and $Π′$ are isomorphic.  Then (a fortiori) there is a morphism to $Π$ from $Π′$ and also a morphism from $Π$ to $Π′$.  By part (a$^{\text{o}}$) and the first morphism, the no-absentmindedness of $Π$ implies the no-absentmindedness of $Π′$.  Similarly, by part (a$^{\text{o}}$) and the second morphism, the no-absentmindedness of $Π$ is implied by the no-absentmindedness of $Π′$.

{\em(b$^{\text{o}}$)}.  This follows from part (a$^{\text{o}}$) and the definition of no-absentmindedness for forms.

{\em(b)}. This follows from part (b$^{\text{o}}$) just as part (a) follows from part (a$^{\text{o}}$). \end{npf}

\begin{npf}[for Proposition~\ref{8046}]\label{8046p} {\em Claim 1}.  {\em If $[Π,Π′,τ,δ]$ is an isomorphism and $Π′$ has perfect-information, then $Π$ has perfect-information.}  Suppose \linebreak $[Π,Π′,τ,δ]$ is an isomorphism, with $Π = (T,C,⊗)$ determining $\HH$ and $Π′ = (T′,C′,⊗)$ determining $\HH′$.  Further suppose $Π$ does not have perfect-information.  It suffices to show that $Π′$ does not have perfect-information.  Because $Π$ does not have perfect-information, there are $t^1⋅∈⋅T$, $t^2⋅∈⋅T$, and \ilc{ss13} $H⋅∈⋅\HH$ such that \il{ss14} $t^1⋅≠⋅t^2$ and \il{ss15} $⎨t^1,t^2⎬⋅⊆⋅H$.  SP Proposition~3.8(e) implies $τ|_\HH$ is a bijection from $\HH$ onto $\HH′$.  Hence \ref{ss13} implies \il{ss23} $τ(H)⋅∈⋅\HH′$.  Further, SP Theorem~3.7 implies that $τ$ is a bijection from $T$ onto $T′$.  Hence \ref{ss14} implies \il{ss24} $τ(t^1)⋅≠⋅τ(t^2)$.  Yet further, \ref{ss15} implies \il{ss25} $⎨τ(t^1),τ(t^2)⎬⋅⊆⋅τ(H)$.  \ref{ss23}, \ref{ss24}, and \ref{ss25} imply that $Π′$ does not have perfect-information.

{\em(a)}. This follows from Claim~1. 

{\em(b)}. This follows from part (a) and the definition of perfect-information for forms. \end{npf}

\begin{lemma}\label{6967} Suppose that $(T,p)$ is a functioned tree and that $τ{:}T→T′$ is a bijection.  Define the function $p′$ by surjectivity and $p′\gr = ⎨\,(τ(t\sh),τ(t))\,|\,(t\sh,t)∈p\gr\,⎬.$  Then $(T′,p′)$ is a functioned tree. \end{lemma}

\begin{pf} Since $(T,p)$ is a functioned tree, there exist $t^o⋅∈⋅T$ and $X⋅⊆⋅T$ to satisfy [T1]--[T2].  Define $t\po = τ(t^o)$ and $X′ = τ(X)$.  It suffices to show \begin{gather}
\zz
\begin{maneq} \text{[T1$′$]} & p′⋅\text{is a nonempty function from}⋅T′⧷⎨t\po⎬⋅\text{onto}⋅X′,⋅\text{and} \end{maneq} \notag \\[-1mm]
\begin{maneq} \text{[T2$′$]} & (∀t′∈T′⧷⎨t\po⎬)(∃m≥1)⋅(p′)^m(t′) = t\po. \end{maneq} \notag
\zz
\end{gather} These two statements are shown by Claims \ref{8117} and \ref{8118}. \begin{cllist}

\yl{8114} $τ|_{T⧷⎨t^o⎬}{:}T⧷⎨t^o⎬→T′⧷⎨t\po⎬$ {\em is a bijection}.  This follows from the bijectivity of $τ$ and the definition of $t\po$.

\yl{8115} $τ|_X{:}X→X′$ {\em is a bijection}.  This follows from the bijectivity of $τ$ and the definition of $X′$.

\yl{8112} $τ|_X○p○(τ|_{T⧷⎨t^o⎬})^{-1}$ {\em is a nonempty function from} $T′⧷⎨t\po⎬$ {\em onto} $X′$.  The claim follows from composition.  In particular, $(τ|_{T⧷⎨t^o⎬})^{-1}{:}T′⧷⎨t\po⎬→T⧷⎨t^o⎬$ is a bijection by Claim~\ref{8114}, $p{:}T⧷⎨t^o⎬→X$ is nonempty and surjective by [T1], and $τ|_X{:}X→X′$ is a bijection by Claim~\ref{8115}.  These bijections appear on the bottom, left, and top of Figure~\ref{6968}. 

\newcommand{\captionparrows}{\ct{Set} diagram for Claims~\ref{8112} and \ref{8116}.}
\begin{figure}[h]
  \newcommand{\hgth}{72}
  \begin{picture}(0,\hgth) 
  \put(-95,-13){\begin{pspicture}(-4,-3)(4,3) 
    \end{pspicture}}
  \end{picture}
  \caption{\small \captionparrows} \label{6968}
  \end{figure}

\yl{8113} $p′\gr = (τ|_X○p○(τ|_{T⧷⎨t^o⎬})^{-1})\gr$.  I argue\begin{align}
\zz
p′\gr 
=&⋅⎨⋅(τ(t\sh),τ(t))⋅|⋅(t\sh,t)∈p\gr⋅⎬ \nt
=&⋅⎨⋅(τ(t\sh),τ(t))⋅|⋅t\sh∈T⧷⎨t^o⎬,⋅t{=}p(t\sh)⋅⎬ \nt
=&⋅⎨⋅(τ(t\sh),τ○p(t\sh))⋅|⋅t\sh∈T⧷⎨t^o⎬⋅⎬ \nt
=&⋅⎨⋅(τ(t\sh),τ○p(t\sh))⋅|⋅t\ps∈T′⧷⎨t\po⎬,⋅t\sh{=}(τ|_{T⧷⎨t\po⎬})^{-1}(t\ps)⋅⎬ \nt
=&⋅⎨⋅(τ○(τ|_{T⧷⎨t\po⎬})^{-1}(t\ps),τ○p○(τ|_{T⧷⎨t\po⎬})^{-1}(t\ps))⋅|⋅t\ps∈T′⧷⎨t\po⎬⋅⎬ \nt
=&⋅⎨⋅(t\ps, τ○p○(τ|_{T⧷⎨t\po⎬})^{-1} (t\ps))⋅|⋅t\ps∈T′⧷⎨t\po⎬⋅⎬ \nt
=&⋅(τ○p○(τ|_{T⧷⎨t\po⎬})^{-1})\gr.\notag
\zz
\end{align} The first equality holds by the lemma's definition of $p′\gr$.  The second holds since the domain of $p$ is $T⧷⎨t^o⎬$ by [T1].  The third is a rearrangement.  The fourth holds by Claim~\ref{8114}.  The fifth and sixth are rearrangements.  The last holds because the domain of $(τ|_{T⧷⎨t\po⎬})^{-1}$ is $T′⧷⎨t\po⎬$ by Claim~\ref{8114}.  

\yl{8116} $p′ = τ|_X○p○(τ|_{t⧷⎨t^o⎬})^{-1}$, {\em that is, Figure~\ref{6968} commutes}.  This follows from Claim~\ref{8113} because [a] $p′$ is surjective by assumption and [b] $τ|_X○p○(τ|_{t⧷⎨t^o⎬})^{-1}$ is surjective by Claim~\ref{8112}.

\yl{8117} {\em [T1\,$′$] holds}.  This follows from Claims \ref{8112} and \ref{8116}.

\yl{6970} $(∀t∈T⧷⎨t^o⎬)(∃m≥1)⋅t^o = p\mspace{2mu}○[(τ|_{T⧷⎨t^o⎬})^{-1}○τ|_X○p]^{m-1}(t)$.  Take $t\,≠\,t^o$.  By [T2] there exists $m⋅≥⋅1$ such that $t^o = p^m(t)$.  On the one hand, suppose $m = 1$.  Then the claim holds by the definition of $m$.  On the other hand, suppose $m⋅≥⋅2$.  Then proving the claim requires several steps.  First, I show \begin{gather}
\zz
\begin{maneq} \text{(a)} & (∀\,n\,|\,m{-}1\,≥\,n\,≥\,1)⋅⋅p^n(t) = (τ|_{T⧷⎨t^o⎬})^{-1}○τ|_X○p^n(t). \end{maneq} \notag
\zz
\end{gather} Take any such $n$.  Since $τ$ is bijective, it suffices to show that the composition $(τ|_{T⧷⎨t^o⎬})^{-1}○τ|_X○p^n(t)$ is well-defined.  In other words, it suffices to show [i] that $p^n(t)⋅∈⋅X$ and [ii] that $τ|_X○p^n(t)$ is in the domain of $(τ|_{T⧷⎨t^o⎬})^{-1}$.  [i] holds because the codomain of $p$ is $X$ by [T1].  To see [ii], note that $t^o = p^m(t)$ and $m{-}1\,≥\,n\,≥\,1$ imply that $p^n(t)$ is in the domain of $p$.  Thus, since the domain of $p$ is $T⧷⎨t^o⎬$ by [T1], we have $p^n(t)⋅∈⋅T⧷⎨t^o⎬$.  Hence the definition of $t\po$ and the bijectivity of $τ$ imply $τ|_X○p^n(t)⋅∈⋅T′⧷⎨t\po⎬$.  This and Claim~\ref{8114} imply [ii].  Second, I argue\begin{gather}
\zz
\begin{maneq} \text{(b)} & (∀\,n\,|\,m{-}1\,≥\,n\,≥\,1)⋅⋅p^n(t) = [(τ|_{T⧷⎨t^o⎬})^{-1}○τ|_X○p]○p^{n-1}(t). \end{maneq} \notag 
\zz
\end{gather} This holds because the right-hand side of (b) is a rearrangement of the right-hand side of (a).  Third, I argue  \begin{align}
\zz
p^{m-1}(t) =&⋅[(τ_{T⧷⎨t^o⎬})^{-1}○τ|_X○p]○p^{m-2}(t) \tag{c} \\ 
=&⋅[(τ_{T⧷⎨t^o⎬})^{-1}○τ|_X○p]^2○p^{m-3}(t) \notag \\[1.5mm]
...⋅=&⋅[(τ_{T⧷⎨t^o⎬})^{-1}○τ|_X○p]^{m-2}○p(t) \nt
=&⋅[(τ_{T⧷⎨t^o⎬})^{-1}○τ|_X○p]^{m-1}(t), \notag
\zz
\end{align} where the first equality holds by (b) at $n{=}m{-}1$, the second by (b) at $n{=}m{-}2$, ..., and the last by (b) at $n{=}1$.  Finally, I argue the claim holds because \begin{gather}
\zz
t^o = p^m(t) = p○p^{m-1}(t) = p\mspace{2mu}○[(τ|_{T⧷⎨t^o⎬})^{-1}○τ|_X○p]^{m-1}(t), \notag
\zz
\end{gather} where the first equality holds by the definition of $m$, the second is a rearrangement, and the third holds by (c).

\yl{8118} {\em [T2$\,′$] holds}.  Take $t′⋅∈⋅T′⧷⎨t\po⎬$.  Then Claim~\ref{8114} implies $(τ|_{T⧷⎨t^o⎬})^{-1}(t′)$ $∈$ $T⧷⎨t^o⎬$.  Thus by Claim~\ref{6970}, there exists $m⋅≥⋅1$ such that \begin{gather}
\zz
t^o = p\mspace{2mu}○[(τ|_{T⧷⎨t^o⎬})^{-1}○τ|_X○p]^{m-1}○(τ|_{T⧷⎨t^o⎬})^{-1}(t′). \notag 
\zz
\end{gather} I now argue \begin{align}
\zz
t\po =&⋅τ|_X(t^o) \nt
=&⋅τ|_X○p\mspace{2mu}○[(τ|_{T⧷⎨t^o⎬})^{-1}○τ|_X○p]^{m-1}○(τ|_{T⧷⎨t^o⎬})^{-1}(t′) \nt
=&⋅[τ|_X○p○(τ^{-1}|_{T⧷⎨t^o⎬})^{-1}]^m(t′) \nt
=&⋅(p′)^m(t′).
\notag
\zz
\end{align} The first equation holds by the definition of $t\po$ and the fact that $t^o⋅∈⋅X$ in any functioned tree (by remark [iv] in the paragraph following SP equation (1)).  The second equation holds by the definition of $m$, the third is a rearrangement, and the fourth holds by Claim~\ref{8116}. \end{cllist}\qedup\end{pf}

\begin{lemma}\label{6946} Suppose $Π = (T,C,⊗)$ is an \ct{NCP} preform.  Also suppose  $τ{:}T→T′$ and $δ{:}C→C′$ are bijections.  Define $⊗′$ by surjectivity and $⊗′\gr =$ \linebreak $⎨\,(τ(t),δ(c),τ(t\sh)\,|\,$ $\!(t,c,t\sh)∈⊗\gr\,⎬$.  Also define $Π′ = (T′,C′,⊗′)$.  Then (a) $Π′$ is an \ct{NCP} preform and (b) $[Π,Π′,τ,δ]$ is an \ct{NCP} isomorphism.  \end{lemma}

\begin{pf} {\em (a)}.  By [P1] there exist $F{:}T⇉C$ and $t^o⋅∈⋅T$ such that $⊗$ is a bijection from $F\gr$ onto $T⧷⎨t^o⎬$.  Define $F′{:}T′⇉C′$ by $F′\gr =$ $⎨(τ(t),δ(c))|(t,c)∈F\gr⎬$.  Also define $t\po = τ(t^o)$.  It suffices to show that\begin{gather}
\zz
\begin{maneq} \text{[P1$′$]} 
& ⊗′⋅\text{is a bijection from}⋅F′\gr⋅\text{onto}⋅T′⧷⎨t\po⎬, \end{maneq} \nt
\begin{maneq} \text{[P2$′$]}
& (T′,p′)⋅\text{is a functioned tree where}⋅p′{:}T′⧷⎨t\po⎬→(F′)^{-1}(C′) \\
& \text{is defined by}⋅p′\gr = ⎨(t\ps,t′)∈(T′)^2|(∃c′∈C′)(t′,c′,t\ps)∈⊗′\gr⎬,⋅\text{and} \end{maneq} \nt
\begin{maneq} \text{[P3$′$]}
& ⎨(F′)^{-1}(c′)|c′∈C′⎬⋅\text{partitions}⋅(F′)^{-1}(C′). \end{maneq} \notag
\zz
\end{gather} This is done by Claims \ref{8124}, \ref{8125}, and \ref{8127}. 
\begin{cllist}

\yl{8120} $(τ,δ)|_{F\gr}{:}F\gr→F′\gr$ {\em is a bijection}.  This follows from the bijectivity of $τ$, the bijectivity of $δ$, and the definition of $F′$.

\yl{8119} $τ|_{τ⧷⎨t^o⎬}{:}T⧷⎨t^o⎬→T′⧷⎨t\po⎬$ {\em is a bijection}.  This follows from the bijectivity of $τ$ and the definition of $t\po$.

\yl{8121} $τ|_{τ⧷⎨t^o⎬}○⊗○[(τ,δ)|_{F\gr}]^{-1}$ {\em is a bijection from} $F′\gr$ {\em onto} $T′⧷⎨t\po⎬$.  The claim follows from composition.  In particular, $((τ,δ)|_{F\gr})^{-1}{:}F′\gr→F\gr$ is a bijection by Claim~\ref{8120}, $⊗{:}F\gr→T⧷⎨t^o⎬$ is a bijection by the definitions of $F$ and $t^o$, and $τ_{T⧷⎨t^o⎬}{:}T⧷⎨t^o⎬→T′⧷⎨t\po⎬$ is a bijection by Claim~\ref{8119}.  These three functions appear on the top, left, and bottom of Figure~\ref{6974}.

\newcommand{\captionoparrows}{\ct{Set} diagram for Claims~\ref{8121} and \ref{8123}.}
\begin{figure}[h]
  \newcommand{\hgth}{72}
  \begin{picture}(0,\hgth) 
  \put(-104,-13){\begin{pspicture}(-4,-3)(4,3) 
    \end{pspicture}}
  \end{picture}
  \caption{\small \captionoparrows} \label{6974}
  \end{figure}

\yl{8122} $⊗′\gr = (τ|_{T⧷⎨t^o⎬}○⊗○[(τ,δ)|_{F\gr}]^{-1})\gr$.  I argue\begin{align}
\zz
⊗′\gr
&=⋅⎨\,(τ(t),δ(c),τ(t\sh))\,|\,(t,c,t\sh)∈⊗\gr\,⎬ \nt
&=⋅⎨\,(τ(t),δ(c),τ(t\sh))\,|\,(t,c)∈F\gr,\,t\sh{=}⊗(t,c)\,⎬ \nt
&=⋅⎨\,((τ,δ)(t,c),τ|_{T⧷⎨t^o⎬}○⊗(t,c))\,|\,(t,c)∈F\gr\,⎬ \nt
&=⋅⎨\,((τ,δ)(t,c),τ|_{T⧷⎨t^o⎬}○⊗(t,c))\,|\,(t′,c′)∈F′\gr,\,(t,c){=}[(τ,δ)|_{F\gr}]^{-1}(t′,c′)\,⎬ \nt
&=⋅⎨\,((τ,δ)○[(τ,δ)|_{F\gr}]^{-1}(t′,c′),τ|_{T⧷⎨t^o⎬}○⊗○[(τ,δ)|_{F\gr}]^{-1}(t′,c′)\,|\,(t′,c′)∈F′\gr\,⎬ \nt
&=⋅⎨\,((t′,c′), τ|_{T⧷⎨t^o⎬}○⊗○[(τ,δ)|_{F\gr}]^{-1} (t′,c′)\,|\,(t′,c′)∈F′\gr\,⎬ \nt
&=⋅(\,τ|_{T⧷⎨t^o⎬}○⊗○[(τ,δ)|_{F\gr}]^{-1}\,)\gr.
\notag
\zz
\end{align} The first equality holds by the lemma's definition of $⊗′$.  The second holds by the definition of $F$, and the third by the definition of $t^o$.  The fourth holds by Claim~\ref{8120}.  The fifth and sixth are rearrangements.  The seventh holds by Claim~\ref{8120}.

\yl{8123} $⊗′ = τ|_{T⧷⎨t^o⎬}○⊗○[(τ,δ)|_{F\gr}]^{-1}$, {\em that is, Figure~\ref{6974} commutes}.  This follows from Claim~\ref{8122} because [a] $⊗′$ is surjective by definition and [b] $τ|_{T⧷⎨t^o⎬}$ is surjective by Claim~\ref{8119}.

\yl{8124} {\em [P1\,$′$] holds}.  This follows from Claims \ref{8121} and \ref{8123}. 

\yl{8125} {\em [P2\,$′$] holds}.  Define $p$ by [P2].  [P2] implies that \ilc{ce01} $(T,p)$ is a functioned tree.  Define $p′$ by [P2$′$].  Claim~\ref{8124} and SP Lemma~C.1(a) implies \il{ce03} $p′$ is well-defined and \il{ce02} $p′$ is surjective.  Because of \ref{ce03}, it suffices to show that $(T′,p′)$ is a functioned tree.  

Toward that end, consider Lemma~\ref{6967}.  Lemma~\ref{6967}'s assumptions are met by \ref{ce01} and the injectivity of $τ$.  Thus Lemma~\ref{6967} implies that $(T′,p\st)$ is a functioned tree, where the function $p\st$ is defined by \il{ce11} $p\st$ being surjective and \il{ce12} $p\st\gr = ⎨\,(τ(t\sh),τ(t))\,|\,(t\sh,t)∈p\gr\,⎬$.  Thus it suffices to show that $p′ = p\st$.  

Toward that end, note \ref{ce02} and \ref{ce11} imply that both $p′$ and $p\st$ are surjective.  Thus it suffices to show $p′\gr = p\st\gr$.  I argue \begin{align}
\zz
p′\gr =&⋅⎨⋅(t\ps,t′)∈(T′)^2⋅|⋅(∃c′∈C′)(t′,c′,t\ps)∈⊗′\gr⋅⎬ \nt
=&⋅⎨⋅(t\ps,t′)∈(T′)^2⋅|⋅(∃c′∈C′)(∃(t,c,t\sh)∈⊗\gr)⋅(t′,c′,t\ps){=}(τ(t),δ(c),τ(t\sh))⋅⎬ \nt
=&⋅⎨⋅(t\ps,t′)∈(T′)^2⋅|⋅(∃(t,c,t\sh)∈⊗\gr)⋅(t′,t\ps){=}(τ(t),τ(t\sh))⋅⎬ \nt
=&⋅⎨⋅(τ(t\sh),τ(t))⋅|⋅(∃c∈C)(t,c,t\sh)∈⊗\gr⋅⎬ \nt
=&⋅⎨⋅(τ(t\sh),τ(t))⋅|⋅(t\sh,t)∈p\gr⋅⎬ \nt
=&⋅p\st\gr. \notag
\zz
\end{align} The first equality holds by the definition of $p′$ two paragraphs ago, and the second equality holds by the definition of $⊗′$ in the lemma statement.  The $⊆$ direction of the third equality holds simply because the variable $c′$ does not appear in the right-hand side.  The $⊇$ direction follows from $⊗\gr⋅⊆⋅T×C×T$ and $δ{:}C→C′$.  The fourth equality holds because the codomain of $τ$ is $T′$.  The fifth equality follows from the definition of $p$ two paragraphs ago, and the sixth equality follows from \ref{ce12}. 

\yl{6966} $(∀c′∈C′)⋅(F′)^{-1}(c′) = τ(F^{-1}(δ^{-1}(c′)))$.  Take $c′⋅∈⋅C′$.  I argue, in seven steps, that %
$(F′)^{-1}(c′)$ by definition is %
$⎨t′∈T′|c′∈F′(t′)⎬$, which by rearrangement is %
$⎨t′∈T′|(t′,c′)∈F′\gr⎬$, which, by the definition of $F′$, the bijectivity of $τ$, and the bijectivity of $δ$, is %
$⎨t′∈T′|(τ^{-1}(t′),δ^{-1}(c′))∈F\gr⎬$, which by the \linebreak bijectivity of $τ$ is %
$⎨t′|(∃t∈T)\,t′{=}τ(t),\,(τ^{-1}(t′),δ^{-1}(c′))∈F\gr⎬$, which by rearrangement is %
$⎨τ(t)|(∃t∈T)(τ^{-1}○τ(t),δ^{-1}(c′))∈F\gr⎬$, which by rearrangement is \linebreak%
$τ(⎨t∈T|(t,δ^{-1}(c′))∈F\gr⎬$, which by rearrangement is %
$τ(F^{-1}(δ^{-1}(c′)))$.

\newcommand{\pA}{^{\prime A}}
\newcommand{\pB}{^{\prime B}}

\yl{8127} {\em [P3\,$′$] holds}. It must be shown that \begin{gather}
\zz
\begin{maneq} \text{[a]} 
& (∀c′∈C′)⋅(F′)^{-1}(c′)⋅≠⋅∅,  \end{maneq} \nt
\begin{maneq} \text{[b]}
& (∀c\pA∈C′,c\pB∈C′)⋅(F′)^{-1}(c\pA)∩(F′)^{-1}(c\pB)⋅≠⋅∅  \\
& ⋅⋅⇒⋅⋅(F′)^{-1}(c\pA)=(F′)^{-1}(c\pB)⋅,⋅\text{and} \end{maneq} \nt
\begin{maneq} \text{[c]}
& ∪_{c′∈C′}(F′)^{-1}(c′) = (F′)^{-1}(C′). \end{maneq} \notag
\zz
\end{gather} To show [a], take $c′$.  By the bijectivity of $δ$, $δ^{-1}(c′)⋅∈⋅C$.  Thus by [P3], $F^{-1}(δ^{-1}(c′))$ $≠$ $∅$.  Thus $τ(F^{-1}(δ^{-1}(c′)))⋅≠⋅∅$.  Thus by Claim~\ref{6966}, $(F′)^{-1}(c′)⋅≠⋅∅$.  To show [b], suppose that [b] were false.  Then there would be $c\pA$ and $c\pB$ such that $(F′)^{-1}(c\pA)$ and $(F′)^{-1}(c\pB)$ intersect and are unequal.  Hence by Claim~\ref{6966}, $τ(F^{-1}(δ^{-1}(c\pA)))$ and \linebreak $τ(F^{-1}(δ^{-1}(c\pA)))$ intersect and are unequal.  Hence by the bijectivity of $τ$, \linebreak $F^{-1}(δ^{-1}(c\pA))$ and $F^{-1}(δ^{-1}(c\pB))$ intersect and are unequal.  This contradicts [P3] because both $δ^{-1}(c\pA)$ and $δ^{-1}(c\pB)$ belong to $C$ by the bijectivity of $δ$.  Finally, [c] holds by definition (recall the last sentence of note~\ref{8128}). 

\end{cllist}

\lstep{(b)}.  This paragraph shows that $[Π,Π′,τ,δ]$ is a morphism.  $Π$ is a preform by assumption and $Π′$ is a preform by part (a).  [PM1] and [PM2] hold by assumption (a fortiori).  [PM3] holds with equality by the definition of $⊗′$.

Finally, SP Theorem~3.7 implies that $[Π,Π′,τ,δ]$ is an isomorphism because [a] it is a morphism by the previous paragraph and [b] $τ$ and $δ$ are bijective by assumption. \end{pf}

\pagebreak\begin{lemma}\label{8058} Suppose $Φ = (I,T,(C_i)_{i∈I},⊗)$ is an \ct{NCF} form.  Also suppose $ι{:}I→I′$, $τ{:}T→T′$, and $δ{:}∪_{i∈I}C_i→C′$ are bijections.  Define $⊗′$ by surjectivity and $⊗′\gr = ⎨(τ(t),δ(c),τ(t\sh)|(t,c,t\sh)∈⊗\gr⎬$.  Also define $(C′_{i′})_{i′∈I′}$ at each $i′$ by $C′_{i′} = δ(C_{ι^{-1}(i′)})$.  Also define $Φ′ = (I′,T′,(C′_{i′})_{i′∈I′},⊗′)$.  Then (a) $Φ′$ is an \ct{NCF} form and (b) $[Φ,Φ′,ι,τ,δ]$ is an \ct{NCF} isomorphism. \end{lemma}

\begin{pf} Define $C = ∪_{i∈I}C_i$.  Define $Π = (T,C,⊗)$.  Define $Π′ = (T′,C′,⊗′)$. 

\begin{cllist} \yl{8059} {\em (a)} $Π′$ {\em is an \ct{NCP} preform and (b)} $[Π,Π′,τ,δ]$ {\em is an \ct{NCP} isomorphism.}  Consider Lemma~\ref{6946}.  The assumptions of Lemma~\ref{6946} are met because [i] $Π$ is an \ct{NCP} preform by [F1], [ii] $τ{:}T→T′$ is a bijection by assumption, and [iii] $δ{:}C→C′$ is a bijection because $C = ∪_{i∈I}C_i$ by definition and $δ{:}∪_{i∈I}C_i→C′$ is a bijection by assumption.  Further, Lemma~\ref{6946}'s definitions of $⊗′$ and $Π′$ coincide with the present definitions of $⊗′$ and $Π′$.  Thus Lemma~\ref{6946} implies this claim's two conclusions.

\yl{8060} $C′ = ∪_{i′∈I′}C′_{i′}$.  I argue, in four steps, that $C′$ by the bijectivity of $δ$ equals $δ(∪_{i∈I}C_i)$, which by rearrangement equals $∪_{i∈I}δ(C_i)$, which by the bijectivity of $ι$ equals $∪_{i′∈I′}δ(C_{ι^{-1}(i′)})$, which by the definition of $(C′_{i′})_{i′∈I′}$ equals $∪_{i′∈I′}C′_{i′}$.

\yl{8129} $Φ′$ {\em satisfies [F1]}. It must be shown that $(T′,C\st,⊗′)$ is a preform where $C\st$ is defined as $∪_{i′∈I′}C′_{i′}$.  Claim~\ref{8060} implies that $C\st = C′$.  Hence $Π′ = (T′,C\st,⊗′)$.  Hence Claim~\ref{8059}(a) implies that $(T′,C\st,⊗′)$ is a preform.

\yl{8061} $Φ′$ {\em satisfies [F2]}. Take $i′⋅∈⋅I′$ and $j′⋅∈⋅I′⧷⎨i′⎬$.  The bijectivity of $ι$ implies $ι^{-1}(i′)⋅∈⋅I$ and $ι^{-1}(j′)⋅∈⋅I⧷⎨ι^{-1}(i′)⎬$.  Thus [F2] for $Φ$ implies \linebreak $C_{ι^{-1}(i′)}\,∩\,C_{ι^{-1}(j′)} = ∅$.  Hence the bijectivity of $δ$ implies $δ(C_{ι^{-1}(i′)})\,∩\,δ(C_{ι^{-1}(j′)}) = ∅$.  Hence the definition of $(C′_{i′})_{i′∈I′}$ implies $C′_{i′}∩C′_{j′} = ∅$.

\yl{8062} $Φ′$ {\em satisfies [F3]}. Take $t′⋅∈⋅T′$.  The bijectivity of $τ$ implies $τ^{-1}(t′)⋅∈⋅T$.  Hence [F3] for $Φ$ implies there is $i⋅∈⋅I$ such that $F(τ^{-1}(t′))⋅⊆⋅C_i$.  Hence the bijectivity of $ι$ implies there is $i′⋅∈⋅I′$ such that \ilc{ff11} $F(τ^{-1}(t′))⋅⊆⋅C_{ι^{-1}(i′)}$.  Also, I show \il{ff12} $F(τ^{-1}(t′)) = δ^{-1}(F′(t′))$ by arguing, in steps, that $F(τ^{-1}(t′))$ by rearrangement equals $⎨c∈C|(τ^{-1}(t′),c)∈F\gr⎬$, which by Claim~\ref{8059}(b) and SP Proposition~3.8(c) equals $⎨c∈C|(t′,δ(c))∈F′\gr⎬$, which by the bijectivity of $δ$ equals $⎨c|(∃c′∈C′)$ \linebreak $c{=}δ^{-1}(c′),\,(t′,δ(c))∈F′\gr⎬$, which by rearrangement equals $⎨δ^{-1}(c′)|(∃c′∈C′)$ \linebreak $(t′,c′)∈F′\gr⎬$, which by rearrangement equals $δ^{-1}(⎨c′∈C′|(t′,c′)∈F′\gr⎬)$, which by rearrangement equals $δ^{-1}(F′(t′))$.  \ref{ff11} and \ref{ff12} imply $δ^{-1}(F′(t′))⋅⊆⋅C_{ι^{-1}(i′)}$.  Hence the bijectivity of $δ$ implies $F′(t′)⋅⊆⋅δ(C_{ι^{-1}(i′)})$.  Hence the definition of $C′_{i′}$ implies $F′(t′)⋅⊆⋅C′_{i′}$.

\yl{8063} $Φ′$ {\em is an \ct{NCF} form}.  This follows from Claims \ref{8129}--\ref{8062}.

\yl{8064} $[Φ,Φ′,ι,τ,δ]$ {\em is an \ct{NCF} morphism}. $Φ$ is an \ct{NCF} form by assumption, and $Φ′$ is an \ct{NCF} form by Claim~\ref{8063}.  [FM1] holds because $[Π,Π′,τ,δ]$ is an \ct{NCP} morphism a fortiori by Claim~\ref{8059}(b).  [FM2] holds by assumption.  For [FM3], take $i⋅∈⋅I$.  I argue, in two steps, that $δ(C_i)$ by the bijectivity of $ι$ equals $δ(C_{ι^{-1}○ι(i)})$, which by definition of $C′_{ι(i)}$ equals $C′_{ι(i)}$.  

\yl{8065} $[Φ,Φ′,ι,τ,δ]$ {\em is an \ct{NCF} isomorphism}.  This follows from the reverse direction of Corollary~\ref{6333} because [a] $[Φ,Φ′,ι,τ,δ]$ is an \ct{NCF} morphism by Claim~\ref{8064}, [b] $[Π,Π′,τ,δ]$ is an \ct{NCP} isomorphism by Claim~\ref{8059}(b), and [c] $ι$ is a bijection by assumption. \end{cllist}

\lstep{Conclusion}.  The lemma's conclusions follow from Claims~\ref{8063} and \ref{8065}. \end{pf}

\section{\ct{CsqF}} \markb{\sc Appendix B. \ct{CsqF}}

\begin{npf}[for Proposition~\ref{7720}]\label{7720p} The \xin{%
\begin{picture}(0,0)
\put(-100,38){\color{white} \rule{100ex}{3ex}}
\put(-35,42){\sc \SMALL Appendix B. \ct{CsqF}}
\end{picture}}%%%%%%%%%%%%%%%
proposition follows from Claims~\ref{7721p}--\ref{7733p} and \ref{8110p}--\ref{8111p}. \begin{cllist}

\yl{7721p} {\em (\ref{7721}) holds.}  Suppose [a] $t^o⋅≠⋅⎨⎬$.  [Csq1] states [b] $⎨⎬⋅∈⋅T$.  [a] and [b] imply $⎨⎬⋅∈⋅T⧷⎨t^o⎬$.  Thus by [P1], there are $t⋅∈⋅T$ and $c⋅∈⋅C$ such that $(t,c,⎨⎬)⋅∈⋅⊗\gr$.  Thus by [Csq2], $p(⎨⎬)±(c)$ $= ⎨⎬$.  This is impossible because the left-hand sequence has positive length and the right-hand sequence has zero length.

\yl{7728p} {\em (\ref{7728}) holds.}  Take $t\sh⋅∈⋅T⧷⎨⎨⎬⎬$.  \ptrf{7721} implies $t\sh⋅∈⋅T⧷⎨t^o⎬$.  Thus the reverse direction of SP Proposition~3.1(a) implies $(p(t\sh),q(t\sh),t\sh)$ $∈$ $⊗$.  Thus [Csq2] implies $p(t\sh)±(q(t\sh)) = t\sh$.  Thus $p(t\sh) = {_1t\sh_{|t\sh|-1}}$ and $q(t\sh) = t\sh_{|t\sh|}$.

\yl{7726p} {\em (\ref{7726}) holds.}  Assume $(t,c,t\sh)⋅∈⋅⊗\gr$.  Then [P1] yields $(t,c,t\sh)⋅∈⋅T×C×T$, and [Csq2] yields $t±(c) = t\sh$.  Conversely, suppose \lic{7764} $(t,c,t\sh)⋅∈⋅T×C×T$ and \li{7765} $t±(c) = t\sh$.  \ref{7765} implies \li{7766} $t = {_1t\sh_{|t\sh|-1}}$ and \li{7767} $c = t\sh_{|t\sh|}$.  Further, \ref{7767} implies $t\sh⋅≠⋅⎨⎬$.  This and \ref{7764} implies \li{7768} $t\sh⋅∈⋅T⧷⎨⎨⎬⎬$.  \ref{7768} and \ptrf{7728} imply \li{7769} $p(t\sh) = {_1t\sh_{|t\sh|-1}}$ and \li{7770} $q(t\sh) = t\sh_{|t\sh|}$.  \ref{7766} and \ref{7769} imply \li{7772} $t = p(t\sh)$.  \ref{7767} and \ref{7770} imply \li{7773} $c = q(t\sh)$.   Further, \ref{7768} and \ptrf{7721} imply $t\sh⋅≠⋅t^o$, and thus SP Proposition~3.1(a) implies \li{7771} $(p(t\sh),q(t\sh),t\sh)⋅∈⋅⊗\gr$.  \ref{7772}--\ref{7771} imply $(t,c,t\sh)⋅∈⋅⊗\gr$.

\yl{7727p} {\em (\ref{7727}) holds.} By [P1], $F⋅⊆⋅T×C$.  Thus it suffices to show $(∀t∈T,c∈C)$ $(t,c)⋅∈⋅F\gr$ iff $t±c⋅∈⋅T$.  Suppose $(t,c)⋅∈⋅F\gr$.  Then [P1] implies there is \lic{7777} $t\sh⋅∈⋅T$ such that \li{7778} $(t,c,t\sh)⋅∈⋅⊗\gr$.  \ref{7778} and \ptrf{7726} imply $t±(c) = t\sh$.  This and \ref{7777} imply $t±(c)⋅∈⋅T$.  Conversely, suppose $t±(c)⋅∈⋅T$.  There there is $t\sh⋅∈⋅T$ such that $t±(c) = t\sh$.  Thus \ptrf{7726} implies $(t,c,t\sh)⋅∈⋅⊗\gr$.  This and [P1] imply $(t,c)⋅∈⋅F\gr$.

\yl{7729p} {\em (\ref{7729}) holds.} Take $t⋅∈⋅T$.  I will use induction on $m⋅∈⋅⎨0,1,...\,|t|⎬$.  For the initial step, assume $m = 0$.  Then $p^0(t) = t = {_1t_{|t|}} = {_1t_{|t|-0}} = {_1t_{|t|-m}}$ by inspection.  For the inductive step, assume $m > 0$.  Note $m⋅≤⋅|t|$ implies $|t|{-}m⋅≥⋅0$, which implies $|t|{-}(m{-}1) > 0$, which implies \lic{7741} $_1t_{|t|-(m-1)}⋅≠⋅⎨⎬$.  I then argue, in steps, that $p^m(t)$ by $m > 0$ equals $p(p^{m-1}(t))$, which by the inductive hypothesis equals $p(_1t_{|t|-(m-1)})$, which by \ref{7741} and \ptrf{7728} at $t\sh = {_1t_{|t|-(m-1)}}$ equals $_1t_{|t|-(m-1)-1}$, which by rearrangement equals $_1t_{|t|-m}$.

\yl{7722p} {\em (\ref{7722}) holds.} Take $t⋅∈⋅T$.  I show $p^{|t|}(t) = t^0$ by arguing, in steps, that $p^{|t|}(t)$ by \ptrf{7729} at $m = |t|$ equals $_1t_{|t|-|t|}$, which equals $_1t_0$, which equals $⎨⎬$, which by \ptrf{7721} equals $t^o$.  This and the definition of $k$ imply $k(t) = |t|$.

\yl{7723p} {\em (\ref{7723}) holds.} Take $t⋅∈⋅T$.  By inspection, the result is equivalent to $(∀ℓ∈⎨1,2,...\,|t|⎬)$ $t_ℓ = q○p^{|t|-ℓ}(t)$.  On the one hand, take $t = ⎨⎬$.  Then $|t| = 0$ so the result is vacuously true.  On the other hand, take $t⋅≠⋅⎨⎬$.  Then take \lic{7739} $ℓ⋅∈⋅⎨1,2,...\,|t|⎬$.  First I show \li{7738} $p^{|t|-ℓ}(t) = {_1t_ℓ}$ by arguing, in steps, that $p^{|t|-ℓ}(t)$ by \ptrf{7729} at $m = |t|{-}ℓ$ equals $_1t_{|t|-(|t|-ℓ)}$, which by rearrangement equals $_1t_ℓ$.  Then I argue, in steps, that $q○p^{|t|-ℓ}(t)$ by \ref{7738} equals $q(_1t_ℓ)$, which by \ref{7739} and \ptrf{7728} equals $t_ℓ$.

\yl{7733p} {\em (\ref{7733}) holds.} Suppose $c⋅∈⋅C$.  This and [P3] imply $F^{-1}(c)⋅≠⋅∅$.  Thus there is $t^\star⋅∈⋅T$ such that $(t^\star,c)⋅∈⋅F\gr$.  This and \ptrf{7727} imply $t^*±(c)⋅∈⋅T$.  Thus $c⋅∈⋅R(t^*±(c))⋅⊆⋅∪⎨R(t)|t∈T⎬$.  Conversely, suppose $b⋅∈⋅∪⎨R(t)|t∈T⎬$.  There there is \lic{7775} $t^*⋅∈⋅T$ such that \li{7776} $b⋅∈⋅R(t^*)$.  \ref{7775} and \ptrf{7723} imply that $t^* = (q○p^{|t^*|-ℓ}(t^*))^{|t^*|}_{ℓ=1}$.  This and \ref{7776} imply there is $ℓ^*⋅∈⋅⎨1,2,...\,|t^*|⎬$ such that $b = q○p^{|t^*|-ℓ^*}(t^*)$.  This implies $b⋅∈⋅C$ since the codomain of $q$ is $C$ by the definition of $q$.

\yl{7719o} $(∀t^A∈T,t^B∈T)$ $(|t^A|\,{<}\,|t^B|⋅\text{and}⋅t^A\,{=}\,{_1t^B_{|t^A|}})$ {\em iff} $t^A⋅⊊⋅t^B$.  Take $t^A⋅∈⋅T$ and $t^B⋅∈⋅T$.  First, suppose \lic{7748} $|t^A| < |t^B|$ and \li{7742} $t^A = {_1t^B_{|t^A|}}$.  \ref{7748} and the definition of $_1t^B_{|t^A|}$ imply \li{7743} $_1t^B_{|t^A|}⋅⊊⋅t^B$.  \ref{7742} and \ref{7743} imply $t^A⋅⊊⋅t^B$.  Conversely, suppose \li{7744} $t^A⋅⊊⋅t^B$.  [Csq1] implies \li{7745} $t^A = ⎨(1,t^A_1),(2,t^A_2),...\,(|t^A|,t^A_{|t^A|})⎬$ and \li{7746} $t^B = ⎨(1,t^B_1),(2,t^B_2),...\,(|t^B|,t^B_{|t^B|})⎬$.  By inspection, \ref{7744}--\ref{7746} imply $|t^A| < |t^B|$ and $t^A = {_1t^B_{|t^A|}}$. 

\newcommand{\noteburrito}{\footnote{Claim~\ref{7747o} says that one sequence is an initial segment of another sequence iff the former is a restriction of the latter.  This may appear implausible.  For example, $⎨(\f2,\ff)⎬$ is not an initial sequence of $t^* = $ $⎨(1,\fg),(2,\ff),(3,\ff)⎬$ even though $⎨(\f2,\ff)⎬$ is a restriction of $t^*$.  This is consistent with Claim \ref{7747o}, because $⎨(2,\ff)⎬$ is not a sequence and thus not an element of $T$ by [Csq1].} }

\yl{7747o} $(∀t^A∈T,t^B∈T)$ $(|t^A|\,≤\,|t^B|⋅\text{and}⋅t^A\,{=}\,{_1t^B_{|t^A|}})$ {\em iff} $t^A⋅⊆⋅t^B$.\noteburrito  In the proof of Claim~{7719o}, change $<$ to $≤$, and $⊊$ to $⊆$.

\yl{7724o} $(∀t^A∈T,t^B∈T)⋅t^A⋅≺⋅t^B$ {\em iff} $t^A⋅⊊⋅t^B$.  Take $t^A⋅∈⋅T$ and $t^B⋅∈⋅T$.  First, suppose $t^A⋅≺⋅t^B$.  This and the definition of $≺$ imply there is \lic{7754} $m⋅∈⋅⎨1,2,...\,k(t^B)⎬$ such that \li{7757} $t^A = p^m(t^B)$.  \ref{7754} and \ptrf{7722} imply \li{7758} $m⋅∈⋅⎨1,2,...\,|t^B|⎬$.  Finally, I argue, in steps, that $t^A$ by \ref{7757} equals $p^m(t^B)$, which by \ref{7758} and \ptrf{7729} equals $ {_1t^B_{|t^B|-m}}$, which by \ref{7758} is a strict subset of $_1t^B_{|t^B|}$, which by inspection equals $t^B$.

Conversely, suppose $t^A⋅⊊⋅t^B$.  This and Claim~{7719o} imply \lic{7759} $|t^A| < |t^B|$ and \li{7760} $t^A = {_1t^B_{|t^A|}}$.  For convenience, let \li{7762} $m = |t^B|{-}|t^A|$.  Note \ref{7759} and $|t^A|⋅≥⋅0$ imply \li{7763} $m⋅∈⋅⎨1,2,...\,|t^B|⎬$.  I now show \li{7788} $t^A = p^m(t^B)$ by arguing, in steps, that $t^A$ by \ref{7760} equals $_1t^B_{|t^A|}$, which by \ref{7762} equals $_1t^B_{|t^B|-m}$, which by \ref{7763} and \ptrf{7729} at $t = t^B$ equals $p^m(t^B)$.  Finally, \ref{7788}, \ref{7763}, and the definition of $≺$ imply $t^A⋅≺⋅t^B$.

\yl{7725o} $(∀t^A∈T,t^B∈T)⋅t^A⋅≼⋅t^B$ {\em iff} $t^A⋅⊆⋅t^B$.  Take $t^A⋅∈⋅T$ and $t^B⋅∈⋅T$.  First, suppose $t^A⋅≼⋅t^B$.  Then by the definition of $≼$, either $t^A⋅≺⋅t^B$ or $t^A = t^B$.  In the first case, Claim~{7724o} implies $t^A⋅⊊⋅t^B$.  Thus $t^A⋅⊆⋅t^B$ in both cases.  Conversely, suppose $t^A⋅⊆⋅t^B$.  Then either $t^A⋅⊊⋅t^B$ or $t^A = t^B$.  In the first case, Claim~\ref{7724o} implies $t^A⋅≺⋅t^B$.  Thus the definition of $≼$ implies $t^A⋅≼⋅t^B$ in both cases. 

\yl{8110p} {\em (\ref{8110}) holds.}  Combine Claims~\ref{7719o} and \ref{7724o}.

\yl{8111p} {\em (\ref{8111}) holds.}  Combine Claims~\ref{7747o} and \ref{7725o}.  \end{cllist}\qedup\end{npf}

\begin{lemma}\label{8035} Suppose $(T,C,⊗)$ is a node-and-choice preform with its $t^o$, $p$, and $q$.  Let $\dT = ⎨\,(q○p^{k(t)-ℓ}(t))^{k(t)}_{ℓ=1}\,|\,t∈T\,⎬$.  Then\begin{gather}
\zz
T⋅∋⋅t⋅\mapsto⋅(q○p^{k(t)-ℓ}(t))^{k(t)}_{ℓ=1}⋅∈⋅\dT \notag
\zz
\end{gather} is a well-defined bijection.  Its inverse is \begin{gather}
\zz
T⋅∋⋅((...((t^o⊗\dt_1)⊗\dt_2)\,\dots\,)⊗\dt_{|\dt|-1})⊗\dt_{|\dt|}⋅\mapsfrom⋅\dt⋅∈⋅\dT \notag
\zz
\end{gather} (to be clear, $T⋅∋⋅t^o⋅\mapsfrom⋅⎨⎬⋅∈⋅\dT$). \end{lemma}

\begin{pf} Let $α$ be the function from $T$ to $\dT$, and conversely, let $β$ be the function to $T$ from $\dT$.  

This paragraph shows that $β○α$ is the identity function on $T$.  The composition is well-defined because [1] the domain of $β$ is $\dT$ and [2] the range of $α$ is $\dT$ by the definition of $\dT$.  Thus it suffices to show $(∀t∈T)⋅β○α(t) = t$.  Toward that end, take $t⋅∈⋅T$.  First, suppose $k(t) = 0$.  I argue, in steps, that $β○α(t)$ by the definition of $α$ equals $β(⎨⎬)$, which by the definition of $β$ equals $t^o$, which by $k(t) = 0$ equals $t$.  Second, suppose $k(t) = 1$.  I argue, in steps, that $β○α(t)$ by the definition of $α$ equals $β[(q(t))]$, which by the definition of $β$ equals $t^o⊗q(t)$, which by $k(t) = 1$ equals $p(t)⊗q(t)$, which by SP Proposition~3.1(b) equals $t$.  Third and finally, suppose $k(t)⋅≥⋅2$.  I will argue \begin{align}
\zz
β○α(t) =&⋅β(⋅(q○p^{k(t)-ℓ}(t))^{k(t)}_{ℓ=1}⋅) \notag \\[-1mm]
=&⋅[[...[[t^o⊗q○p^{k(t)-1}(t)]⊗q○p^{k(t)-2}(t)]\,...⋅]⊗q○p(t)]⊗q(t) \notag \\[-1mm]
=&⋅[[...[[p^{k(t)}(t)⊗q○p^{k(t)-1}(t)]⊗q○p^{k(t)-2}(t)]\,...⋅]⊗q○p(t)]⊗q(t) \notag \\[1mm]
=&⋅[[...[[p○p^{k(t)-1}(t)⊗q○p^{k(t)-1}(t)]⊗q○p^{k(t)-2}(t)]\,...⋅]⊗q○p(t)]⊗q(t) \notag\\[-1mm]
=&⋅[[...[p^{k(t)-1}(t)⊗q○p^{k(t)-2}]\,...⋅]⊗q○p(t)]⊗q(t) \notag \\[1mm]
 &⋅·⋅·⋅· \nt
=&⋅p(t)⊗q(t) = t. \notag
\zz
\end{align} The first equality holds by the definition of $α$, the second by the definition of $β$, and the third by the definition of $k$.  The fourth and fifth equalities hold by a rearrangement and SP Proposition 3.1(b).  The sixth equality holds by $k(t){-}2$ similar applications of SP Proposition 3.1(b), and the final equality holds by a final application of SP Proposition 3.1(b). 

This paragraph shows that $α○β$ is the identity function on $\dT$.  The composition is well-defined because [a] the domain of $α$ is $T$ and [b] each value of $β$ is a value of $⊗$ and the codomain of $⊗$ is a subset of $T$.  Thus it suffices to show $(∀\dt∈\dT)⋅α○β(\dt) = \dt$.  Toward that end, take $\dt$.  First, suppose $\dt = ⎨⎬$.  I argue, in steps, that $α○β(⎨⎬)$ by the definition of $β$ equals $α(t^o)$, which by the definition of $α$ equals $⎨⎬$.  Second, suppose $\dt⋅≠⋅⎨⎬$.  Then it suffices to show that $(∀\dt∈\dT)$ $(∀ℓ∈⎨1,2,...|\dt|⎬)$ $(α○β(\dt))_ℓ = \dt_ℓ$.  Toward this end, take $\dt$ and $ℓ$.  [i] First assume $ℓ < |\dt|$.  I will argue \begin{align}
\zz
(α○β(\dt))_ℓ =&⋅q○p^{k(β(\dt))-ℓ}(β(\dt)) \nt
=&⋅q○p^{|\dt|-ℓ}(β(\dt)) \nt
=&⋅q○p^{|\dt|-ℓ}[⋅((...((t^o⊗\dt_1)⊗\dt_2)\,\dots\,)⊗\dt_{|\dt|-1})⊗\dt_{|\dt|}⋅] \nt
=&⋅q○p^{|\dt|-ℓ-1}[⋅((...((t^o⊗\dt_1)⊗\dt_2)\,\dots\,)⊗\dt_{|\dt|-2})⊗\dt_{|\dt|-1}⋅] \nt
 &⋅·⋅·⋅· \nt
=&⋅q○p^{|\dt|-ℓ-(|\dt|-ℓ)}[⋅((...((t^o⊗\dt_1)⊗\dt_2)\,\dots\,)⊗\dt_{|\dt|-(|\dt|-ℓ)-1})⊗\dt_{|\dt|-(|\dt|-ℓ)}⋅] \nt
=&⋅q○p^0[⋅((...((t^o⊗\dt_1)⊗\dt_2)\,\dots\,)⊗\dt_{ℓ-1})⊗\dt_ℓ⋅] = \dt_ℓ. \notag
\zz
\end{align} The first equality holds by the definition of $α$.  The second equality holds because $k(β(\dt)) = |\dt|$ by inspecting the definitions of $k$ and $β$.  The third holds by the definition of $β$.  The fourth holds by the definition of $p$. The fifth holds by $|t|{-}ℓ{-}1$ similar applications of the definition of $p$.  The sixth is a rearrangement.  The seventh holds by the definition of $q$.  [ii] Second assume $ℓ = |\dt|$.  Then I will argue \begin{gather}
\zz
(α○β(\dt))_{|\dt|} = q○p^{k(β(\dt))-|\dt|}(β(\dt)) \nt
= q○p^{|\dt|-|\dt|}(β(\dt)) = q(β(\dt)) = \dt_{|\dt|}⋅, \notag
\zz
\end{gather} The first equality holds by the definition of $α$ and $ℓ = |\dt|$.  The second equality holds because $k(β(\dt)) = |\dt|$ by inspecting the definitions of $k$ and $β$.  The third is trivial.  The fourth holds by the definitions of $q$ and $β$.  \end{pf}

\begin{npf}[for Theorem~\ref{8066}]\label{8066p} {\em (a)}.  Lemma~\ref{8035} implies $\bar{τ}{:}T→\dT$ is a bijection.  Thus the assumptions of Lemma~\ref{6946} are met at $T′ = \dT$, $C′ = C$, and $δ = \id_C$.  Further, the definition of $\bar{⊗}$ here coincides with the definition of $⊗′$ in Lemma~\ref{6946}.  Therefore Lemma~\ref{6946} implies that $(\dT,C,\bar{⊗})$ is an \ct{NCP} preform, and that $[(T,C,⊗),(\dT,C,\bar{⊗}),\bar{τ},\id_C]$ is an \ct{NCP} isomorphism. Thus [Csq1] and [Csq2] remain to be shown.  

For [Csq1], note that the definition of $\dT$ implies that $\dT$ is a collection of finite sequences.  Further, since $t^o⋅∈⋅T$ by [P1], the definition of $\dT$ implies \linebreak that $(q○p^{k(t^o)-ℓ}(t^o))^{k(t^o)}_{ℓ=1}⋅∈⋅\dT$.  Thus, since $k(t^o) = 0$ by the definition of $k$, \linebreak $(q○p^{0-ℓ}(t^o))^0_{ℓ=1}⋅∈⋅\dT$.  Thus $⎨⎬⋅∈⋅\dT$. 

For [Csq2], take $(\dt,c,\dt\sh)⋅∈⋅\bar{⊗}$.  Then by the definition of $\bar{⊗}$, there are $t⋅∈⋅T$ and $t\sh⋅∈⋅T$ such that \ilc{rr22} $\bar{τ}(t) = \dt$, \il{rr23} $\bar{τ}(t\sh) = \dt\sh$, and \il{rr24} $(t,c,t\sh)⋅∈⋅⊗\gr$.  \ref{rr22}, \ref{rr23}, and the definition of $\bar{τ}$ imply \il{rr31} $\dt = (q○p^{k(t)-ℓ}(t))^{k(t)}_{ℓ=1}$ and \il{rr32} $\dt\sh = (q○p^{k(t\sh)-ℓ}(t\sh))^{k(t\sh)}_{ℓ=1}$.  Also \ref{rr24} and SP Proposition~3.1(b) imply \il{rr41} $t = p(t\sh)$ and \il{rr42} $c = q(t\sh)$.  \ref{rr41} and the definition of $k$ imply \il{rr43} $k(t) = k(t\sh){-}1$.   Finally, I argue, in steps, that $\dt±(c)$ by \ref{rr31} equals $(q○p^{k(t)-ℓ}(t))^{k(t)}_{ℓ=1}±(c)$, which by \ref{rr41}--\ref{rr43} equals $(q○p^{k(t\sh)-1-ℓ}○p(t\sh))^{k(t\sh)-1}_{ℓ=1}±(q(t\sh))$, which by rearrangement equals $(q○p^{k(t\sh)-ℓ}(t\sh))^{k(t\sh)-1}_{ℓ=1}±(q(t\sh))$, which by rearrangement equals $(q○p^{k(t\sh)-ℓ}(p(t\sh))^{k(t\sh)}_{ℓ=1}$, which by \ref{rr32} equals $\dt\sh$.

\lstep{(b)}.  By assumption, $(I,T,(C_i)_{i∈I},⊗)$ is an \ct{NCF} form.  Thus [F1] implies $(T,∪_{i∈I}C_i,⊗)$ is an \ct{NCP} preform.  Further, part (b) defines $\dT$, $\bar{τ}$, and $\bar{⊗}$ as part (a) did.  Thus part (a) implies \lic{uu33} $(\dT,∪_{i∈I}C_i,\bar{⊗})$ is a \ct{CsqP} preform.

Meanwhile, Lemma~\ref{8035} implies $\bar{τ}{:}T→\dT$ is a bijection.  Thus the assumptions of Lemma~\ref{8058} are met at $I′ = I$, $ι = \id_I$, $T′ = \dT$, $C′ = ∪_{i∈I}C_i$, and $δ = \id_{∪_{i∈I}C_i}$.  Further, the definition of $\bar{⊗}$ here coincides with the definition of $⊗′$ in Lemma~\ref{8058}.  Also, the transparent definitions of $ι$ and $δ$ here, and the definition of $(C′_{i′})_{i′∈I′}$ in Lemma~\ref{8058}, imply that $(C′_{i′})_{i′∈I′} = (C_i)_{i∈I}$.  Hence Lemma~\ref{8058} implies that \lic{uu11} $(I,\dT,(C_i)_{i∈I},\bar{⊗})$ is an \ct{NCF} form, and \li{uu12} $[(I,T,(C_i)_{i∈I},⊗),$ $\!(I,\dT,(C_i)_{i∈I},\bar{⊗}),$ $\!\id_I,\bar{τ},$ $\!\id_{∪_{i∈I}C_i}]$ is an \ct{NCF} isomorphism.

\ref{uu11} and \ref{uu33} imply that $(I,\dT,(C_i)_{i∈I},\bar{⊗})$ is a \ct{CsqF} form.  This and \ref{uu12} are part (b)'s conclusions. \end{npf}

\begin{lemma}\label{7969}  Suppose $(T,C,⊗)$ is a \ct{CsqP} preform.  Then the following are equivalent. \begin{tlist}
\yl{7981} $(T,C,⊗)$ has no absentmindedness.
\yl{7982} $(∄H∈\HH,t∈H,ℓ{<}|t|)⋅_1t_ℓ⋅∈⋅H$.
\yl{7983} $(∀t∈T,H∈\HH)⋅|⎨\,ℓ\,{:}\,1≤ℓ≤|t|,\,t_ℓ∈F(H)\,⎬|⋅≤⋅1$.\footnote{The ``\,{:}\,'' replaces ``$\,|\,$'' for clarity.}
\yl{7984} $(∀t∈T)⋅|R(t)| = |t|$.
\yl{7985} $R|_T$ is injective. 
\end{tlist} \end{lemma} 

\begin{pf} The lemma follows from Claims \ref{7998}--\ref{8000}. \begin{cllist} 

\yl{7979} {\em Not (\ref{7981}) $⇒$ not (\ref{7982})}.  Assume absentmindedness.  Then there are \linebreak \ilc{cb90} $H⋅∈⋅\HH$, \il{cb91} $t⋅∈⋅H$, and \il{7986} $s⋅∈⋅H$ such that \il{7980} $s⋅≺⋅t$.  \ref{7980} and Proposition \ref{7720}(\ref{8110}) imply \il{7988} $|s| < |t|$ and \il{7989} $s = {_1t_{|s|}}$.  \ref{7989} and \ref{7986} imply \il{8029} ${_1t_{|s|}}⋅∈⋅H$.  \ref{cb90}, \ref{cb91}, \ref{7988}, and \ref{8029} show property (\ref{7982}) is violated at $ℓ = |s|$.

\yl{7991} {\em Not (\ref{7982}) $⇒$ not (\ref{7983})}. Assume not (\ref{7982}).  Then there is $H⋅∈⋅\HH$, \ilc{7993} $t⋅∈⋅H$, and \il{7994} $ℓ < |t|$ such that \il{7995} $_1t_ℓ⋅∈⋅H$.  \ref{7994} implies $t_{ℓ+1}$ is well-defined; thus the definition of $F$ implies $t_{ℓ+1}⋅∈⋅F({_1t_ℓ})$; and thus \ref{7995} implies \il{8001} $t_{ℓ+1}⋅∈⋅F(H)$.  \ref{8001} and \ref{7993} imply $t_{ℓ+1}⋅∈⋅F(t)$; and thus $t^* = t±(t_{ℓ+1})$ is a member of $T$.  Finally, I argue, in steps, that  $|⎨ℓ′{:}1≤ℓ′≤|t^*|,t^*_{ℓ′}∈F(H)⎬|$ by \ref{8001} is at least $|⎨ℓ′{:}1≤ℓ′≤|t^*|,t^*_{ℓ′}{=}t_{ℓ+1}⎬|$, which by the construction of $t^*$ is at least $|⎨ℓ{+}1,|t|{+}1⎬|$, which by \ref{7994} is 2.

\yl{7996} {\em Not (\ref{7983}) $⇒$ not (\ref{7984})}. Assume not (\ref{7983}).  There there are $t⋅∈⋅T$ and $H⋅∈⋅\HH$ such that $|⎨ℓ{:}1≤ℓ≤|t|,t_ℓ∈F(H)⎬|⋅≥⋅2$.  Hence there are $ℓ$ and $ℓ′$ such that \ilc{8002} $ℓ < ℓ′$, \il{8003} $t_{ℓ}⋅∈⋅F(H)$, and \il{8004} $t_{ℓ′}⋅∈⋅F(H)$.  I argue in three steps that \ref{8004} by Proposition~\ref{7720}(\ref{7728}) implies $q({_1t_{ℓ′}})⋅∈⋅F(H)$, which by Proposition~\ref{7646}(\ref{7648}) implies $p({_1t_{ℓ′}})⋅∈⋅H$, which by Proposition~\ref{7720}(\ref{7728}) implies \il{8006} $_1t_{ℓ′-1}⋅∈⋅H$.  \ref{8003} and \ref{8006} imply $t_ℓ⋅∈⋅F(_1t_{t′-1})$.  Thus $t^* = {_1t_{ℓ′-1}}±(t_ℓ)$ is a member of $T$.  Further, \ref{8002} implies $t^*_ℓ$ is well-defined and equal to $t_ℓ$.  Thus $t^*_ℓ = t^*_{ℓ′}$.  This and \ref{8002} again imply $|R(t^*)| < |t^*|$. 

\yl{7997} {\em Not (\ref{7984}) $⇒$ not (\ref{7981})}. Assume not (\ref{7984}).  There there is $t⋅∈⋅T$ such that $|R(t)|⋅≠⋅|t|$.  Since $|R(t)| > |t|$ is inconceivable, $|R(t)| < |t|$.  Thus there are $ℓ$ and $ℓ′$ in $⎨1,2,...\,|t|⎬$ such that \ilc{8007} $ℓ < ℓ′$ and \il{8008} $t_ℓ = t_{ℓ′}$.  The definition of $F$ implies \il{8009} ${_1t_{ℓ-1}}⋅∈⋅F^{-1}(t_ℓ)$ and \il{8010} ${_1t_{ℓ′-1}}⋅∈⋅F^{-1}(t_{ℓ′})$.  \ref{8008} and \ref{8010} imply \il{8013} ${_1t_{ℓ′-1}}⋅∈⋅F^{-1}(t_ℓ)$.   [P3] implies \il{8011} $F^{-1}(t_ℓ)⋅∈⋅\HH$.  Finally, \ref{8007} implies $ℓ{-}1 < ℓ′{-}1$; thus Proposition~\ref{7720}(\ref{8110}) implies \il{8012} ${_1t_{ℓ-1}}⋅≺⋅{_1t_{ℓ′-1}}$.  \ref{8011}, \ref{8009}, \ref{8013}, and \ref{8012} imply absentmindedness.

\yl{7998} {\em (\ref{7981}), (\ref{7982}), (\ref{7983}), and (\ref{7984}) are equivalent.}  This follows from Claims~\ref{7979}--\ref{7997}.

\yl{7999} {\em Not (\ref{7984}) $⇒$ not (\ref{7985}).} Assume not (\ref{7984}).  Then there is $t⋅∈⋅T$ such that $|R(t)|⋅≠⋅|t|$.  Thus since $|R(t)| > |t|$ is inconceivable, $|R(t)| < |t|$.  Thus there are $ℓ$ and $ℓ′$ in $⎨1,2,...\,|t|⎬$ such that \ilc{8014} $ℓ < ℓ′$ and \il{8015} $t_ℓ = t_{ℓ′}$.  \ref{8014} and \ref{8015} imply $R({_1t_{ℓ′-1}}) = R({_1t_{ℓ′}})$.  Thus $R|_T$ is not injective.

\yl{8000} {\em Not (\ref{7985}) $⇒$ not (\ref{7984}).} Assume not (\ref{7985}).  Then $R|_T$ is not injective.  Then there are $s$ and $t$ in $T$ such that \ilc{8016} $s⋅≠⋅t$ and \il{8017} $R(s) = R(t)$.

On the one hand, suppose there is not an $ℓ$ in $⎨1,2,...\,\text{min}⎨|s|,|t|⎬⎬$ such that $s_ℓ⋅≠⋅t_ℓ$.  Then \il{8030} $_1s_{\text{max}⎨|s|,|t|⎬} = {_1t_{\text{max}⎨|s|,|t|⎬}}$.  Thus \ref{8016} implies $|s|⋅≠⋅|t|$.  Hence $|s| < |t|$ or $|t| < |s|$.  Without loss of generality assume \il{8019} $|s| < |t|$.  Hence \ref{8030} implies \il{8018} $s = {_1t_{|s|}}$. \ref{8019} implies $t_{|s|+1}$ exists.  Thus \ref{8017} implies $s⋅≠⋅⎨⎬$ and there is \il{8021} $ℓ⋅∈⋅⎨1,2,...\,|s|⎬$ such that \il{8020} $s_ℓ = t_{|s|+1}$.  But \ref{8018} implies $s_ℓ = t_ℓ$, and thus \ref{8020} implies $t_ℓ = t_{|s|+1}$.  This and \ref{8021} imply $|R(t)| < |t|$.  In other words, property (\ref{7984}) is violated. 

On the other hand, suppose there is an $ℓ$ in $⎨1,2,...\,\text{min}⎨|s|,|t|⎬⎬$ such that $s_ℓ⋅≠⋅t_ℓ$.  Let $j$ be the smallest such $ℓ$.  Then \il{8022} $_1s_{j-1} = {_1t_{j-1}}$ and \il{8023} $s_j⋅≠⋅t_j$.  The definition of $F$ implies $s_j⋅∈⋅F({_1s_{j-1}})$ and $t_j⋅∈⋅F({_1t_{j-1}})$, and thus, \ref{8022} implies \il{8024} $⎨s_j,t_j⎬⋅⊆⋅F({_1s_{j-1}})$.  A fortiori \ref{8024} and [P3] imply there is $H⋅∈⋅\HH$ such that ${_1s_{j-1}}⋅∈⋅H$.  Hence \ref{8024} also implies \il{8026} $⎨s_j,t_j⎬⋅⊆⋅F(H)$. Further, \ref{8017} and \ref{8023} imply there is $j^*⋅∈⋅⎨1,2,...\,|s|⎬$ such that \il{8027} $j^*⋅≠⋅j$ and \il{8025} $s_{j^*} = t_j$.  \ref{8025} and \ref{8026} imply \il{8028} $⎨s_j,s_{j^*}⎬⋅⊆⋅F(H)$.  Finally, I argue that $|⎨\,ℓ\,{:}\,1≤ℓ≤|s|,s_ℓ∈F(H)\,⎬|$ is at least as great as $|⎨j,j^*⎬|$ by \ref{8028}, which is 2 by \ref{8027}.  Thus the proposition's property (\ref{7983}) is violated.  So Claim~\ref{7998}((\ref{7983})$⟺$(\ref{7984})) implies property (\ref{7984}) is violated.\end{cllist}\qedup\end{pf}

\section{\ct{CsetF}} \markb{\sc Appendix C. \ct{CsetF}}

\begin{lemma}\label{8103} Suppose $C$ is a set, $t\,⊆\,C$, $c\,∈\,C$, and $t\sh\,⊆\,C$.  Then the following are equivalent.  (a) $c\,∉\,t$ and $t∪⎨c⎬\,{=}\,t\sh$.  (b) $t\,≠\,t\sh$ and $t∪⎨c⎬\,{=}\,t\sh$.  (c) $t\,≠\,t\sh$ and $t\,{=}\,t\sh⧷⎨c⎬$.  (d) $t\,⊆\,t\sh$ and $⎨c⎬\,{=}\,t\sh⧷t$. \end{lemma} 

\begin{pf} {\em (a)$⟺$(b)}.  It suffices to show that if $t∪⎨c⎬ = t\sh$, then $c⋅∉⋅t$ and $t⋅≠⋅t\sh$ are equivalent.  Toward that end, assume $t∪⎨c⎬ = t\sh$.  Then both directions of the equivalence hold by inspection.

{\em (b)$⟺$(c)}.  It suffices to show that if $t⋅≠⋅t\sh$, then $t∪⎨c⎬ = t\sh$ and $t = t\sh⧷⎨c⎬$ are equivalent.  Toward that end, assume \lic{ca00} $t⋅≠⋅t\sh$.  For the forward direction, assume \li{ca01} $t∪⎨c⎬ = t\sh$.  \ref{ca00}, \ref{ca01} imply \li{ca02} $c⋅∉⋅t$.  \ref{ca01} implies $(t∪⎨c⎬)⧷⎨c⎬ = t\sh⧷⎨c⎬$, and \ref{ca02} implies the left-hand side is $t$.  For the reverse direction, assume \li{ca11} $t = t\sh⧷⎨c⎬$.  \ref{ca00} and \ref{ca11} imply \li{ca12} $c⋅∈⋅t\sh$.  \ref{ca11} implies $t∪⎨c⎬ = (t\sh⧷⎨c⎬)∪⎨c⎬$, and \ref{ca12} implies the right-hand side is $t\sh$.

{\em (a)$⟺$(d)}.  Assume (a).  That is, assume \ilc{cb01} $c⋅∉⋅t$ and \il{cb02} $t∪⎨c⎬ = t\sh$.  \ref{cb02} implies $t⋅⊆⋅t\sh$.  Further, \ref{cb02} implies $(t∪⎨c⎬)⧷t = t\sh⧷t$, and \ref{cb01} implies that the left-hand side is $⎨c⎬$.  Conversely, assume (d).  That is, assume \il{cb03} $t⋅⊆⋅t\sh$ and \il{cb04} $⎨c⎬ = t\sh⧷t$.  \ref{cb04} implies $c⋅∉⋅t$.  Further, \ref{cb04} implies $t∪⎨c⎬ = t∪(t\sh⧷t)$, and \ref{cb03} implies the right-hand side is $t\sh$. \end{pf}

\begin{npf}[for Proposition~\ref{7568}]\label{7568p} The proposition if proved by Claims \ref{7511p}, \ref{7592p}, \ref{7594p}, \ref{7596p}, \ref{7644p}, \ref{7915p}, \ref{7572p}, \ref{7645p}, \ref{7632p}, \ref{7660p}, \ref{7661p}, \ref{7567p}, and \ref{7667p}. \begin{cllist}

\yl{7511p} \pt{7511}{$t^o = ⎨⎬$.}  Suppose [a] $t^o⋅≠⋅⎨⎬$.  [Cset1] states [b] $⎨⎬⋅∈⋅T$.  [a] and [b] imply $⎨⎬⋅∈⋅T⧷⎨t^o⎬$.  Thus by [P1], there is $t⋅∈⋅T$ and $c⋅∈⋅C$ such that $(t,c,⎨⎬)⋅∈⋅⊗\gr$.  Thus by [Cset2], $t∪⎨c⎬$ $= ⎨⎬$.  This implies $c⋅∈⋅⎨⎬$, which is impossible.

\yl{7566o} $⊗⋅⊆⋅⎨\,(t,c,t\sh)∈T×C×T\,|\,c∉t,\,t∪⎨c⎬{=}t\sh\,⎬$.  Take $(t,c,t\sh)⋅∈⋅⊗\gr$.  [P1] yields \ilc{7668} $(t,c,t\sh)⋅∈⋅T×C×T$.  [P2] yields \il{7669} $t = p(t\sh)$.  Remark [ii] in the paragraph following SP equation (1) yields \il{7670} $p(t\sh)⋅≠⋅t\sh$.  \ref{7669} and \ref{7670} imply \il{7671} $t⋅≠⋅t\sh$.  [Cset2] yields \il{7672} $t∪⎨c⎬{=}t\sh$. \ref{7671} and \ref{7672} yield \il{7673} $c⋅∉⋅t$. \ref{7668}, \ref{7673}, and \ref{7672} are the desired results. 

\yl{7592p} \pt{7592}{$(∀t\sh∈T⧷⎨⎨⎬⎬)$ $q(t\sh)⋅∉⋅t\sh$ {\em and} $p(t\sh)∪⎨q(t\sh)⎬ = t\sh$.}  Take $t\sh⋅∈⋅T⧷⎨⎨⎬⎬$.  \ptrf{7511} implies that $t\sh⋅∈⋅T⧷⎨t^o⎬$.  Thus SP Proposition~3.1(a) implies that $(p(t\sh),q(t\sh),t\sh)$ $∈$ $⊗$.  Thus Claim~\ref{7566o} implies that $q(t\sh)⋅∉⋅p(t\sh)$ and $p(t\sh)∪⎨q(t\sh)⎬ = t\sh$.

\yl{7595o} $(∀t∈T,∀m∈⎨0,1,2,...\,k(t)⎬)$ $|p^m(t)| = |t|\,{-}\,m$.  Note by inspection, that \ptrf{7592} implies \ilc{7676} $(∀t\sh∈T⧷⎨⎨⎬⎬)$ $|p(t\sh)| = |t\sh|{-}1$.  To prove the present claim, take $t⋅∈⋅T$.  I will show $(∀m∈⎨0,1,2,...\,k(t)⎬)$ $|p^m(t)| = |t|\,{-}\,m$ by induction.  For the initial step ($m{=}0$), $|p^0(t)| = |t| = |t|{-}0 = |t|{-}m$ by inspection.  For the inductive step ($m≥1$), I first note that by assumption $m⋅≤⋅k(t)$, which trivially implies $m{-}1 < k(t)$, which by the definition of $k$ implies $p^{m-1}(t)⋅≠⋅t^o$, which by \ptrf{7511} implies \il{7677} $p^{m-1}(t)⋅≠⋅⎨⎬$.  I then argue, in steps, that $|p^m(t)|$ by rearrangement equals $|p○p^{m-1}(t)|$, which by \ref{7677} and \ref{7676} at $t\sh = p^{m-1}(t)$ equals $|p^{m-1}(t)| - 1$, which by the inductive hypothesis equals $(|t|{-}(m{-}1)) - 1$, which by rearrangement equals $|t|-m$.

\yl{7594p} \pt{7594}{$(∀t∈T)$ $k(t) = |t|$.}  Take $t⋅∈⋅T$.  Note [a] $|p^{k(t)}(t)| = |t|\,{-}\,k(t)$ by Claim~\ref{7595o} at $m\,{=}\,k(t)$.  Also note [b] $|p^{k(t)}(t)| = |t^o| = |⎨⎬| = 0$ by the definition of $k(t)$ and by \ptrf{7511}.   [a] and [b] imply $|t|\,{-}\,k(t) = 0$.  Hence $|t| = k(t)$.

\yl{7596p} \pt{7596}{$(∀t∈T⧷⎨⎨⎬⎬,m∈⎨0,1,...\,|t|⎬)$ $p^m(t)\,⊆\,t$ {\em and} $t⧷p^m(t)$ $=$ $⎨\,q○p^n(t)\,|\,\!$ $m{>}n≥0\,⎬$.}  Take $t⋅∈⋅T$.  I will use induction on $m⋅∈⋅⎨0,1,...\,|t|⎬$.  For the initial step ($m{=}0$), $p^m(t) = p^0(t) = t$ and $t⧷p^m(t) = t⧷p^0(t) = t⧷t = ⎨⎬ = ⎨\,q○p^n(t)\,|\,0{>}n≥0\,⎬ = ⎨\,q○p^n(t)\,|\,m{>}n≥0\,⎬$.  For the inductive step ($m≥1$), note $m⋅≤⋅|t|$ by assumption; which implies $m{-}1 < |t|$; which implies $m{-}1 < k(t)$ by \ptrf{7594}; which implies $p^{m-1}(t)⋅≠⋅t^o$ by the definition of $k$; which implies $p^{m-1}(t)⋅≠⋅⎨⎬$ by \ptrf{7511}.  Hence, \ptrf{7592} at $t\sh = p^{m-1}(t)$ implies $q○p^{m-1}(t)⋅∉⋅p○p^{m-1}(t)$ and $p○p^{m-1}(t)∪⎨q○p^{m-1}(t)⎬ = p^{m-1}(t)$.  By \linebreak Lemma~\ref{8103}(a)$⟺$(d), this is equivalent to $p○p^{m-1}(t)⋅⊆⋅p^{m-1}(t)$ and \linebreak
$p^{m-1}(t)⧷p○p^{m-1}(t) = ⎨q○p^{m-1}(t)⎬$.  By a small rearrangement, this is equivalent to \il{7597} $p^m(t)⋅⊆⋅p^{m-1}(t)$ and \il{7598} $p^{m-1}(t)⧷p^m(t)$ $=$ $⎨q○p^{m-1}(t)⎬$.  Meanwhile, the inductive hypothesis is \il{7599} $p^{m-1}(t)⋅⊆⋅t$ and \il{7600} $t⧷p^{m-1}(t)$ $=$ $⎨\,q○p^n(t)\,|\,m{-}1{>}n≥0\,⎬$.  \ref{7597} and \ref{7599} imply \il{7601} $p^m(t)⋅⊆⋅t$.  \ref{7597} and \ref{7599} also imply \il{7602} $t⧷p^m(t)$ $=$ \linebreak $(t⧷p^{m-1}(t))⋅∪⋅(p^{m-1}(t)⧷p^m(t))$.  \ref{7602}, \ref{7600}, and \ref{7598} imply \il{7603} $t⧷p^m(t)$ $=$ \linebreak $⎨\,q○p^n(t)\,|\,m{-}1{>}n≥0\,⎬$ $∪$ $⎨q○p^{m-1}(t)⎬$.  The right-hand side of \ref{7603} is equal to \linebreak $⎨\,q○p^n(t)\,|\,m{-}1≥n≥0\,⎬$, which is equal to $⎨\,q○p^n(t)\,|\,m{>}n≥0\,⎬$.  Hence \ref{7603} is equivalent to \il{wxyz} $t⧷p^m(t)$ $=$ $⎨\,q○p^n(t)\,|\,m{>}n≥0\,⎬$.  \ref{7601} and \ref{wxyz} are the desired results.  

\yl{7581o} $(∀t^A∈T,t^B∈T)$ $t^A⋅≺⋅t^B$ $⇒$ $t^A⋅⊊⋅t^B$. Suppose $t^A⋅≺⋅t^B$.  Then the definitions of $≺$ and $k$ imply there is \ilc{7679} $m⋅∈⋅⎨1,2,...\,k(t^B)⎬$ such that \il{7680} $t^A = p^m(t^B)$.  \ref{7679} and \ptrf{7594} imply $m⋅∈⋅⎨1,2,...\,|t^B|⎬$.  Thus \ptrf{7596} at $t = t^B$ implies \il{7681} $p^m(t^B)⋅⊆⋅t^B$ and \il{7682} $t^B⧷p^m(t^B) = ⎨\,q○p^n(t^B)\,|\,m{>}n≥0\,⎬$.  Since $m⋅≥⋅1$ by \ref{7679}, $⎨\,q○p^n(t^B)\,|\,m{>}n≥0\,⎬$ is nonempty.  Thus \ref{7681} and \ref{7682} imply $p^m(t^B)⋅⊊⋅t^B$. Thus \ref{7680} implies $t^A⋅⊊⋅t^B$.  

\yl{7918o} $(∀t^A∈T,t^B∈T)$ $t^A⋅≼⋅t^B$ $⇒$ $t^A⋅⊆⋅t^B$. Suppose $t^A⋅≼⋅t^B$.  Then the definition of $≼$ implies $t^A⋅≺⋅t^B$ or $t^A = t^B$.  The first implies $t^A⋅⊆⋅t^B$ by Claim~\ref{7581o}.  The second implies $t^A⋅⊆⋅t^B$ trivially.

\yl{7644p} \pt{7644}{$(∀t∈T)⋅t = ⎨\,q○p^n(t)\,|\,|t|{>}n≥0\,⎬$.}  Take $t⋅∈⋅T$.  I argue, in steps, that $t$ trivially equals $t⧷⎨⎬$, which by \ptrf{7511} equals $t⧷t^o$, which by the definition of $k$ equals $t⧷p^{k(t)}(t)$, which by \ptrf{7594} equals $t⧷p^{|t|}(t)$, which by \ptrf{7596} at $m = |t|$ equals $⎨\,q○p^n(t)\,|\,|t|{>}n≥0\,⎬$

\yl{7915p} \pt{7915}{$C = ∪T$.} Forward direction.  Take $c⋅∈⋅C$.  By [P3], $F^{-1}(c)$ is a member of a partition, and thus, it is nonempty.  Take $t^*⋅∈⋅F^{-1}(c)$.  By [P1], $t^*⊗c⋅∈⋅T$.  Thus by [Cset2], $t^*∪⎨c⎬⋅∈⋅T$.  Thus $c$ belongs to an element of $T$. Reverse direction. Take any $t$.  [Cset1] implies that $t$ is a set.  Take $b⋅∈⋅t$.  By \ptrf{7644}, there is $n$ such that $b = q○p^n(t)$.  Thus, since the codomain of $q$ is $C$, $b⋅∈⋅C$.

\yl{7572p} \pt{7572}{{\em $(T,C,⊗)$ has no-absentmindedness.}}  Suppose there were $H⋅∈⋅\HH$, \ilc{7574} $t^A⋅∈⋅H$, and \il{7575} $t^B⋅∈⋅H$ such that \il{7576} $t^A⋅≺⋅t^B$.  \ref{7576} and the definition of $≺$ imply there is \il{7577} $m > 1$ such that \il{7578} $t^A = p^m(t^B)$.  \ref{7577} and \ref{7578} imply $t^A = p○p^{m-1}(t^B)$.  Thus [P2]'s definition of $p$ implies there is $c⋅∈⋅C$ such that \il{7573} $(t^A,c,p^{m-1}(t^B))⋅∈⋅⊗\gr$.  Thus the definition of $F$ implies $c⋅∈⋅F(t^A)$.  This, \ref{7574}, \ref{7575}, and SP Proposition~3.2(16a) imply $c⋅∈⋅F(t^B)$.  Thus the definition of $F$ implies there is $t\st⋅∈⋅T$ such that $(t^B,c,t\st)⋅∈⋅⊗\gr$.  This and Claim~\ref{7566o} implies \il{7590} $c⋅∉⋅t^B$.  But, \ref{7573} and Claim~\ref{7566o} imply \il{7591} $c⋅∈⋅p^{m-1}(t^B)$.  And, the definition of $≼$ implies $p^{m-1}(t^B)⋅≼⋅t^B$, and thus Claim~\ref{7918o} implies \il{7685} $p^{m-1}(t^B)⋅⊆⋅t^B$.  \ref{7591} and \ref{7685} imply $c⋅∈⋅t^B$, which contradicts \ref{7590}.

\yl{7645p} \pt{7645}{$(∀t∈T,H∈\HH)$ $|t∩F(H)|⋅≤⋅1$.} Suppose $|t∩F(H)|⋅≥⋅2$.  Then by \ptrf{7644}, there exist distinct $m′$ and $m$ such that $⎨⋅q○p^{m′}(t),⋅q○p^m(t)⋅⎬$ $⊆$ $F(H)$.  Thus by Lemma~\ref{7646}(\ref{7648}), \ilc{7656} $⎨⋅p^{m′+1}(t),⋅p^{m+1}(t)⋅⎬$ $⊆⋅H$.  Without loss of generality assume $m′ > m$.  Then $p^{m′+1}(t) = p^{m′-m}○p^{m+1}(t)$.  Hence \il{7657} $p^{m′+1}(t)⋅≺⋅p^{m+1}(t)$ by the definition of $≺$.  \ref{7656} and \ref{7657} show there is absentmindedness, which contradicts \ptrf{7572}.

\yl{7633o} $(∀t^A∈T,t^B∈T)$ $t^A⋅⊊⋅t^B$ {\em implies} $(∀m∈⎨0,1,...\,|t^A|⎬)$ $p^m(t^A)$ $=$ \linebreak $p^{m+|t^B|-|t^A|}(t^B)$.   Suppose \lic{7634} $t^A⋅⊊⋅t^B$.  I will use induction on $m$.
  
For the initial step, assume \li{we11} $m = |t^A|$.  I argue, in steps, that 
$p^m(t^A)$ by \ref{we11} equals 
$p^{|t^A|}(t^A)$, which by \ptrf{7594} equals 
$p^{k(t^A)}(t^A)$, which by the definition of $k$ equals 
$t^o$, which by the definition of $k$ again equals 
$p^{k(t^B)}(t^B)$, which by \ptrf{7594} again equals 
$p^{|t^B|}(t^B)$, which by manipulation equals 
$p^{|t^A|+|t^B|-|t^A|}(t^B)$, which by \ref{we11} again equals
$p^{m+|t^B|-|t^A|}(t^B)$.  

For the inductive step, assume \li{we12} $m < |t^A|$.  (The next two sentences concern $p^m(t^A)$ alone.)  \ref{we12} and \ptrf{7594} imply $m < k(t^A)$, which by the definition of $k$ implies $p^m(t^A)⋅≠⋅t^o$.  This and SP Proposition~3.1(a) at $t\sh = p^m(t^A)$ yield 
\zz
\ci\li{7636} $p^{m+1}(t^A)⋅⊗⋅q○p^m(t^A) = p^m(t^A)$. \co
\zz
(The next three sentences concern $p^{m+|t^B|-|t^A|}(t^B)$ alone.)  \ref{we12} and manipulation imply $m{+}|t^B|{-}|t^A| < |t^A|{+}|t^B|{-}|t^A| = |t^B|$, which by \ptrf{7594} implies $m{+}|t^B|{-}|t^A| < k(t^B)$, which by the definition of $k$ implies $p^{m{+}|t^B|{-}|t^A|}(t^B)⋅≠⋅t^o$.  This and SP Proposition~3.1(a) at $t\sh = p^{m{+}|t^B|{-}|t^A|}(t^B)$ yield
\zz
\ci\li{7637} $p^{m+1+|t^B|-|t^A|}(t^B)⋅⊗⋅q○p^{m+|t^B|-|t^A|}(t^B) = p^{m+|t^B|-|t^A|}(t^B)$. \co
\zz
Since the inductive hypothesis is $p^{m+1}(t^A) = p^{m+1+|t^B|-|t^A|}(t^B)$, \ref{7637} yields
\zz
\ci\li{7638} $p^{m+1}(t^A)⋅⊗⋅q○p^{m+|t^B|-|t^A|}(t^B) = p^{m+|t^B|-|t^A|}(t^B)$.\co
\zz

\ref{7636}, \ref{7638}, and the definition of $F$ yield
\li{7639} $⎨⋅q○p^m(t^A),⋅q○p^{m+|t^B|-|t^A|}(t^B)⋅⎬$ $⊆$ $F(p^{m+1}(t^A))$.
Also, \ptrf{7644} and \ref{7634} yield \li{7640} $q○p^m(t^A)⋅∈⋅t^A⋅⊆⋅t^B$.  
Also, \ptrf{7644} yields \li{7641} $q○p^{m+|t^B|-|t^A|}(t^B)⋅∈⋅t^B$.  
\ref{7639}, \ref{7640}, \ref{7641}, and \ptrf{7645} imply \li{7642} $q○p^m(t^A) = q○p^{m+|t^B|-|t^A|}(t^B)$.

Finally, I argue, in steps, that 
$p^m(t^A)$ by \ref{7636} equals 
$p^{m+1}(t^A)⊗q○p^m(t^A)$, which by \ref{7642} equals
$p^{m+1}(t^A)$ $⊗$ $q○p^{m+|t^B|-|t^A|}(t^B)$, which by \ref{7638} equals
$p^{m+|t^B|-|t^A|}(t^B)$. 

\yl{7632p} \pt{7632}{$(∀t^A∈T,t^B∈T)$ $t^A⋅⊊⋅t^B$ {\em implies} $t^A = p^{|t^B|-|t^A|}(t^B)$.} This follows from Claim~\ref{7633o} at $m = 0$.

\yl{7660p} \pt{7660}{$(∀t^A∈T,t^B∈T)$ $t^A⋅≺⋅t^B$ {\em iff} $t^A⋅⊊⋅t^B$.}  Because of Claim~\ref{7581o}, it suffices to show the reverse direction.  Toward that end, suppose \lic{qq01} $t^A⋅⊊⋅t^B$.  \ref{qq01} and \ptrf{7632} imply $t^A = p^{|t^B|-|t^A|}(t^B)$.  Further, \ref{qq01} and [Cset1] imply $|t^B|{-}|t^A| > 0$.  The last two sentences and the definition of $≺$ imply $t^A⋅≺⋅t^B$.  

\yl{7661p} \pt{7661}{$(∀t^A∈T,t^B∈T)$ $t^⋅≼⋅t^B$ {\em iff} $t^A⋅⊆⋅t^B$.}  Because of Claim~\ref{7918o}, it suffices to show the reverse direction.  Toward that end, suppose $t^A⋅⊆⋅t^B$.  Then either $t^A⋅⊊⋅t^B$ or $t^A = t^B$.  In the first case, \ptrf{7660} implies $t^A⋅≼⋅t^B$.  In the second case, $t^A⋅≼⋅t^B$ holds trivially by the definition of $≼$.

\yl{7593o} $(∀t∈T,c∈C,t\sh∈T)$ ($c\,∉\,t$ {\em and} $t∪⎨c⎬\,{=}\,t\sh$) $⇒$ $(t,c)\,{=}\,(p(t\sh),q(t\sh))$.  Suppose \ilc{7607} $c⋅∉⋅t$ and \il{7608} $t∪⎨c⎬ = t\sh$.  \ref{7607} and \ref{7608} imply \il{7662} $t⋅⊊⋅t\sh$ and \il{7663} $|t\sh|{-}|t| = 1$.  \ref{7662} and \ptrf{7632} at $(t^A,t^B) = (t,t\sh)$ imply $t = p^{|t\sh|-|t|}(t\sh)$.  This and \ref{7663} imply \il{7664} $t = p(t\sh)$.  Further, \ref{7662} implies \il{7686} $t\sh⋅≠⋅⎨⎬$.  I then argue, in steps, that 
$⎨c⎬$ by \ref{7607}--\ref{7608} equals 
$t\sh⧷t$, which by \ref{7664} equals 
$t\sh⧷p(t\sh)$, which by \ref{7686} and \ptrf{7592} equals
$⎨q(t\sh)⎬$.  Thus \il{7666} $c = q(t\sh)$.  \ref{7664} and \ref{7666} are the required results.

\yl{7567p} \pt{7567}{$⊗ = ⎨\,(t,c,t\sh)∈T×C×T\,|\,c∉t,\,t∪⎨c⎬{=}t\sh\,⎬$.}  By Claim~\ref{7566o}, it suffices to show the reverse direction.  Toward that end, suppose [a] $c⋅∉⋅t$ and [b] $t∪⎨c⎬ = t\sh$.  [b] implies $t\sh⋅≠⋅⎨⎬$.  Thus \ptrf{7511} implies $t\sh⋅≠⋅t^o$.  Thus SP Proposition 3.1(a) implies [c] $p(t\sh)⊗q(t\sh) = t\sh$.  Also, [a], [b], and Claim~\ref{7593o} imply [d] $(t,c) = (p(t\sh),q(t\sh))$.  [c] and [d] imply $t⊗c = t\sh$.

\yl{7667p} \pt{7667}{$F = ⎨\,(t,c)∈T×C\,|\,c∉t,\,t∪⎨c⎬∈T\,⎬$.}  I argue, in three steps, that $(t,c)⋅∈⋅F\gr$ by [P1] is equivalent to [a] $(t,c)∈T×C$ and [b] $(∃t\sh∈T)\,(t,c,t\sh)∈⊗\gr$, which by \ptrf{7567} is equivalent to [a] and [b$′$] $(∃t\sh∈T)\,c\,∉\,t$ and $t∪⎨c⎬\,{=}\,t\sh$, which by rearrangement is equivalent to [a] and [b$″$] $c\,∉\,t$ and $t∪⎨c⎬\,∈\,T$.  \end{cllist}\qedup\end{npf}

\begin{npf}[for Theorem~\ref{7967}]\label{7967p} {\em (a)}. Lemma~\ref{7969}[(\ref{7981})$⇒$(\ref{7985})] implies $R|_{\dT}{:}\dT→T$ is a bijection.  Thus the assumptions of Lemma~\ref{6946} are met at [1] its $(T,C,⊗)$ equal to $(\dT,\dC,\bar{⊗})$ here, [2] its $τ{:}T→T′$ equal to $R|_{\dT}{:}\dT→T$ here, and [3] its $δ{:}C→C′$ equal to $\id_{\dC}{:}\dC→\dC$ here.  Further, the definition of $⊗′$ in the lemma coincides with the definition of $⊗$ here.  Therefore the lemma implies that $(T,\dC,⊗)$ is an \ct{NCP} preform, and that $[(\dT,\dC,\bar{⊗}),(T,\dC,⊗),R|_{\dT},\id_{\dC}]$ is an \ct{NCP} isomorphism.  Thus it remains to show that $(T,\dC,⊗)$ is a \ct{CsetP} preform.  By definition, it suffices to show [Cset1] and [Cset2]. 

For [Cset1], first note that $\dT$ is a collection of (finite) sequences by assumption.  Hence $T$ is a collection of finite sets by the definitions of $T$ and $R$.  Further, $⎨⎬$ belongs to $\dT$ by [Csq1].  Hence $R(⎨⎬) = ⎨⎬$ belongs to $T$.

For [Cset2], take $(t,\dcc,t\sh)⋅∈⋅⊗\gr$.  Then by the definition of $⊗$, there are $\dt⋅∈⋅\dT$ and $\dt\sh⋅∈⋅\dT$ such that \ilc{xy11} $R(\dt) = t$, \il{xy12} $R(\dt\sh) = t\sh$, and \il{xy13} $(\dt,\dcc,\dt\sh)⋅∈⋅\bar{⊗}$.  \ref{xy13} and [Csq2] implies \il{xy21} $\dt±(\dcc) = \dt\sh$.  Finally, I argue, in steps, that $t∪⎨\dcc⎬$ by \ref{xy11} equals $R(\dt)∪⎨\dcc⎬$, which by inspection equals $R(\dt±(\dcc))$, which by (\ref{xy21}) equals $R(\dt\sh)$, which by \ref{xy12} equals $t\sh$.

\lstep{(b)}. Lemma~\ref{7969}[(\ref{7981})$⇒$(\ref{7985})] implies $R|_{\dT}{:}\dT→T$ is a bijection.  Thus the assumptions of Lemma~\ref{8058} are met at [1] its $(I,T,(C_i)_{i∈I},⊗)$ equal to $(\dI,\dT,(\dC_\di)_{\di∈\dI},\bar{⊗})$ here, [2] its $ι{:}I→I′$ equal to $\id_{\dI}{:}\dI→\dI$ here, [3] its $τ{:}T→T′$ equal to $R|_{\dT}{:}\dT→T$ here, and [4] its $δ{:}∪_{i∈I}C_i→C′$ equal to $\id_{∪_{\di∈\dI}\dC_\di}{:}∪_{\di∈\dI}\dC_\di→∪_{\di∈\dI}\dC_\di$ here.  Also, the definition of $⊗′$ in Lemma~\ref{8058} coincides with the definition of $⊗$ here.  Also, the transparent definitions of $δ$ and $ι$ here, and the definition of $(C′_{i′})_{i′∈I′}$ in Lemma~\ref{8058}, imply $(C′_{i′})_{i′∈I′}$ there equals $(\dC_\di)_{\di∈\dI}$ here.  Hence Lemma~\ref{8058} implies that $(\dI,T,(\dC_\di)_{\di∈\dI},⊗)$ is an \ct{NCF} form, and that $[(\dI,\dT,(\dC_\di)_{\di∈\dI},\bar{⊗}),$ $\!(\dI,T,(\dC_\di)_{\di∈\dI},⊗),$ $\!\id_{\dI},R|_{\dT},\id_{∪_{\di∈\dI}C_\di}]$ is an \ct{NCF} isomorphism. 

It remains to show that $(\dI,T,(\dC_\di)_{\di∈\dI},⊗)$ is a \ct{CsetF} form.  Since the previous paragraph showed that it is an \ct{NCF} form, it suffices to show that its preform $(T,∪_{\di∈\dI}\dC_\di,⊗)$ is an \ct{CsetP} preform.  By assumption, $(\dT,∪_{\di∈\dI}\dC_\di,\bar{⊗})$ is a \ct{CsqP_\ga} preform.  Thus the assumption of part (a) is met at $(\dT,\dC,\bar{⊗}) = (\dT,∪_{\di∈\dI}\dC_\di,\bar{⊗})$.  Further, part (b) defines $T$ and $⊗$ just as part (a) does.  Hence part (a) implies $(T,∪_{\di∈\dI}\dC_\di,⊗)$ is an \ct{CsetP} preform. \end{npf}

\markb{\sc References}

\showbib \end{document}